\documentclass[fleqn,usenatbib]{mnras}

\usepackage[T1]{fontenc}

\DeclareRobustCommand{\VAN}[3]{#2}
\let\VANthebibliography\thebibliography
\def\thebibliography{\DeclareRobustCommand{\VAN}[3]{##3}\VANthebibliography}

\usepackage{graphicx}	
\usepackage{amsmath}	
\usepackage{amssymb}	
\usepackage{longtable}
\usepackage{array}

\usepackage{soul}       
\soulregister\cite7
\soulregister\ref7
\soulregister\figref7
\soulregister\tabref7
\soulregister\eqref7
\soulregister\cite7
\soulregister\citep7
\soulregister\citet7
\usepackage[dvipsnames]{xcolor} 
\usepackage{newtxtext,newtxmath}

\title[MR-SNe: a nucleosynthetic analysis]{Magnetorotational supernovae: a nucleosynthetic analysis of sophisticated 3D models.}

\author[M. Reichert]{
M. Reichert$^{1}$\thanks{E-mail: moritz.reichert@uv.es},
M. Obergaulinger$^{1}$,
M. \'A. Aloy$^{1}$,
M. Gabler$^{1}$,
A. Arcones$^{2,3}$ and
F. K. Thielemann$^{3,4}$
\\
$^{1}$Departament d'Astonomia i Astrof\'{\i}sca, Universitat de Val\`encia, Edifici d'Investigatci\'{o} Jeroni Munyoz, C/Dr. Moliner, 50, E-46100 Burjassot (Val\`encia), Spain\\
$^{2}$Institut f\"ur Kernphysik, Technische Universit\"at Darmstadt, Schlossgartenstr. 2, D-64289
Darmstadt, Germany\\
$^{3}$GSI Helmholtzzentrum f\"ur Schwerionenforschung GmbH, Planckstr. 1, D-64291 Darmstadt, Germany\\
$^{4}$ Department of Physics, University of Basel, Klingelbergstrasse 82, CH-4056 Basel, Switzerland
}

\date{Accepted XXX. Received YYY; in original form ZZZ}

\pubyear{2015}

\defcitealias{Aloy2021}{AO21}
\defcitealias{Obergaulinger2021}{OA21}
\defcitealias{Obergaulinger2022}{OA22}

\begin{document}
\label{firstpage} 
\pagerange{\pageref{firstpage}--\pageref{lastpage}}
\maketitle

\begin{abstract}
Magnetorotational supernovae are a rare type of core-collapse supernovae where the magnetic field and rotation play a central role in the dynamics of the explosion. We present the post-processed nucleosynthesis of state-of-the-art neutrino-MHD supernova models that follow the post explosion evolution for few seconds. We find three different dynamical mechanisms to produce heavy r-process elements: (i) a prompt ejection of matter right after core bounce, (ii) neutron-rich matter that is ejected at late times due to a reconfiguration of the protoneutronstar shape, (iii) small amount of mass ejected with high entropies in the centre of the jet. We investigate total ejecta yields, including the ones of unstable nuclei such as $^{26}$Al, $^{44}$Ti, $^{56}$Ni, and $^{60}$Fe. The obtained $^{56}$Ni masses vary between $0.01 - 1\,\mathrm{M_\odot}$. The latter maximum is compatible with hypernova observations. Furthermore, all of our models synthesize Zn masses in agreement with observations of old metal-poor stars. We calculate simplified light curves to investigate whether our models can be candidates for superluminous supernovae. The peak luminosities obtained from taking into account only nuclear heating reach up to a few $\sim 10^{43} \,\mathrm{erg\,s^{-1}}$. Under certain conditions, we find a significant impact of the $^{66}$Ni decay chain that can raise the peak luminosity up to $\sim 38\%$ compared to models including only the $^{56}$Ni decay chain. This work reinforces the theoretical evidence on the critical role of magnetorotational supernovae to understand the occurrence of hypernovae, superluminous supernovae, and the synthesis of heavy elements.
\end{abstract}


\section{Introduction}
The advent of time-domain astronomy has unveiled a multitude of luminous transients of stellar origin. Among them, supernovae (SNe) are an exceptional example marking the death of massive stars. Besides ordinary SNe releasing energies $\sim 10^{51}\,$erg with peak luminosity $\sim 10^{41}\,\text{erg\,s}^{-1}$ \citep{Nomoto2013}, superluminous supernovae  \citep[SLSNe;][]{Gal-Yam2012,Gal-Yam2019, Nicholl2014, Moriya_et_al__2018__ssr__SuperluminousSupernovae} and hypernovae \citep[HNe;][]{Iwamoto_1998Natur.395..672_SN1998bw} excel by their extreme luminosity and energy. Theoretically, these extreme SNe are promising nurseries for the nucleosynthesis of the heaviest chemical elements in the Universe \citep[][]{Nishimura2006,Nishimura2015,Nishimura2017,Winteler2012,Reichert2021a}. Furthermore, there is overwhelming observational evidence connecting extreme SNe with other rarer, extremely powerful events, namely, long gamma-ray bursts \citep[lGRB;][]{Woosley_Bloom__2006__araa__The_Supernova_Gamma-Ray_Burst_Connection}. Their extreme properties indicate the presence of distinctive and somewhat extraordinary conditions in the stellar progenitors and/or in their circumstellar environment, from which SLSNe and HNe result. Rotation and magnetic fields may be the differential factors bringing exceptionally powerful or energetic magnetorotationally driven supernovae (MR-SNe).
With respect to the expression MR-SNe we include in this category all events where at the end of stellar evolution core collapse with rotation and magnetic fields plays an essential role. This includes events that end with a magnetized neutron star (magnetar) and a specific supernova explosion, as well as events that lead to central black holes (BHs), where the later evolution beyond that point can cause black-hole accretion disc outflows and long-duration gamma-ray bursts. The simulations in the present paper do not yet include the second phase of the evolution of the latter models. In absence of unambiguous detections of MR-SNe only theoretical/numerical models may provide clues on indirect observational signatures beyond their direct detection. 

The first theoretical models of MR-SNe date back to 1970s \citep{LeBlanc1970,Bisnovatyi-Kogan1976,Meier1976,Mueller1979,Symbalisty1984}. These pioneering works have been extended and improved during the years reaching some consensus on the fact that MR-SNe tend to produce collimated ejecta (jets) along the stellar rotational axis. In the first 2D axisymmetric hydrodynamical models that became available \citep[e.g.,][]{Maeda2003, Nishimura2006, Burrows2007, Tominaga2007} the developing jets were often injected artificially. Until today, there still exist uncertainties in the models, such as the magnitude of the magnetic field. This magnetic field can be inherited from the stellar evolution models \citep[e.g.,][]{Maeder2003,Maeder2004,Maeder2005,woosley-heger2006,Braithwaite2008}, and further amplified by the magneto-rotational instability (MRI; \citealt[][]{Obergaulinger2009,Masada2012,Moesta2015,Rembiasz_2016MNRAS.456.3782,Nishimura2017}). 
Currently, the computational frontier include (all or most of) the following elements: three dimensions, general relativity, sophisticated neutrino transport, and detailed microphysics \citep[recent MHD CC-SN simulations are presented in, e.g.,][\citealt{Obergaulinger2021}, \citetalias{Obergaulinger2021} hereafter]{Moesta2015,Mueller2020b,Kuroda2020,Bugli2021,Matsumoto2022,Varma2022}. 
In parallel to the (magneto-)hydrodynamical modelling, frameworks to investigate nuclear processes were developed. Only more recently, magnetohydrodynamic (MHD) simulations and large nuclear reaction networks were combined to calculate the nucleosynthesis of MR-SNe \citep{Nishimura2006,Winteler2012,Nishimura2015,Nishimura2017,Halevi2018,Moesta2018,Reichert2021a}. 

Existing neutrino-driven supernova models can hardly explain the extreme energies and ejected Nickel masses in HNe. However, state-of-the-art 3D MHD simulations usually last a couple of hundreds of milliseconds, resulting into ejected Nickel masses of the order of $10^{-2}\, \mathrm{M_\odot}$ \citep[e.g.,][]{Winteler2012,Nishimura2015,Nishimura2017,Moesta2018,Reichert2021a}, a value much lower than the (model dependent) Nickel ejecta mass in observed HNe \citep[around $\gtrsim10^{-1}\,\mathrm{M_\odot}$][]{Nomoto2013}. We shall show here that longer simulation times may yield larger Nickel masses \citep[c.f.,][]{Witt2021}, broadly compatible with observational models.

The high luminosities of SLSNe are hard to explain with the radioactive decay of Nickel only. Thus, two other ingredients could be relevant to explain the observed luminosities. First, a central engine (e.g., a just born magnetar or protomagnetar) may transfer energy to the ejecta \citep[][\citealt{Obergaulinger2022}; \citetalias{Obergaulinger2022} hereafter]{Kasen2010,Woosley2010,Dessart_et_al__2012__mnras__Superluminoussupernovae:$56$Nipowerversusmagnetarradiation,Chatzopoulos2013,Inserra2013b,Nicholl2014,Nicholl2015,Metzger2015,Soker2017}. Secondly, a potential contribution to the light curve due to interactions with circumstellar matter \citep[e.g.,][and references therein]{Jerkstrand2020}. MR-SNe may not yield enough Nickel mass to explain the high luminosity of SLSNe from radioactive sources, but have the merit that they may produce protomagnetars, whose contribution to the peak luminosity can be dominant \citepalias[e.g.][]{Obergaulinger2022}. 

Whether MR-SNe are able to provide the necessary conditions for the r-process depends on many processes during the evolution, which can only be addressed with realistic MHD simulations of a variety of progenitors. In 3D simulations a kink instability may develop that will lead to less neutron-rich conditions suppressing the synthesis of heavy nuclei \citep{Moesta2014,Moesta2018,Kuroda2020}. Whether this happens is still an open question and may depend on physical conditions such as the strength and geometry of the magnetic field as well as numerical ones like the grid resolution.
Furthermore, neutrino reactions may lead to more proton-rich conditions \citep{Nishimura2017,Reichert2021a}. 

Here we present the first nucleosynthesis calculations based on 3D simulations with sophisticated neutrino transport. Our results advance our understanding of the following key questions:
\begin{itemize}
    \item Can MR-SNe synthesize the heaviest nuclei known in our Universe? Are the necessary magnetic field configurations and strengths realistic?
    \item Which radioactive nuclei are synthesized in MR-SNe? How does the explosion energy correlate with the Nickel mass synthesized in this special type of SNe? Can MR-SNe reproduce the typical imprints of elements in old metal-poor stars?
    \item What are the dominant radioactive nuclei, and what are the resulting peak luminosities?
\end{itemize}
To answer these questions, we structure the paper as follows: in sect.~\ref{sct:methods} we describe the neutrino-MHD, Eulerian models, and the method to extract Lagrangian tracer particles out of them. Also we report on our procedure to extrapolate in time the SN ejecta conditions to estimate the nucleosynthetic yields on time-scales much longer than the computed ones in our neutrino-MHD simulations. Results are discussed in sect.~\ref{sct:results}, which contains the final ejecta yields (sect.~\ref{ssct:ej_comp}), a brief comparison of the yields of 2D and respective 3D models (sect.~\ref{sct:2D_3D_comp}), and an analysis of the conditions that are necessary for the r-process (sect.~\ref{sct:r-process}). Additionally, we analyse the yields, conditions, and spatial distribution of radioactive nuclei such as $^{26}$Al, $^{44}$Ti, $^{56}$Ni, $^{60}$Fe in sect.~\ref{sct:radioactive_isotopes}. Afterwards in sect.~\ref{ssct:zinc}, we investigate the amount of produced Zinc. A simplified light curve model is presented in sect.~\ref{ssct:lightcurve}. Discussions and conclusions are given in sect.~\ref{sct:conclusion}.

\section{Simulations}
\label{sct:methods}
\subsection{Neutrino-MHD models}
\label{sct:mhd_models}
We investigate four models in full 3D \citepalias[][]{Obergaulinger2021} and two long-time axisymmetric 2D simulations \citep[][\citet{Aloy2021}; \citetalias{Aloy2021} hereafter]{ObergaulingerAloy2017} of the same Wolf-Rayet progenitor star with $35\,\mathrm{M_\odot}$ zero age main sequence mass (ZAMS), a pre-collapse mass of $28.1\,\mathrm{M_\odot}$, and a metallicity of 1/10th of solar metallicity (model 35OC of \citealt[][]{woosley-heger2006}). Hence, the progenitor for all of our models is the same and they only differ in the parametrization and strength of the magnetic field (see magnetic energies in Tab~\ref{tab:sim_properties}).
All six models were initialized using the rotational profile given by the stellar evolution calculations. In one of the 3D models, model O, we also use the magnetic field given by \citealt[][]{woosley-heger2006}. Also model P in 3D and the equivalent model in 2D (35OC-Rp3) take the estimated magnetic field at the pre-supernova link, but slightly increase the poloidal component strength. In the other three models, the magnetic field is set up as a combination of a large-scale dipole field and a toroidal component, following the prescription of \citep[][]{Suwa2007}. We changed its normalization such as to increase or decrease the field strength w.r.t.~the original field obtained by  \citealt[][]{woosley-heger2006}. We note that, magnetic fields are only estimated from the saturation of the Tayler-Spruit dynamo process in stellar evolution. Thus, neither the strength, nor the topology of the fields are exactly known. Hence, there is room for some variation of these quantities without modifying any essential property of the pre-supernova model (see Tab.~\ref{tab:sim_properties} and \citetalias[][]{Aloy2021}; \citealt{Maeder2012,Wheeler2015,Keszthelyi2019,Mueller2020b,Varma2021,Griffiths_2022arXiv220400016}).
Before core bounce, all models are calculated in 2D axisymmetry and then mapped into full 3D.

Our models O, W, and S are 3D versions of axisymmetric models for which we already computed the nucleosynthetic yields \citep{Reichert2021a}. Model P is another 3D model with an initial field strength between the original and the very strong field of models O and S, respectively (Tab.~\ref{tab:sim_properties}). We include its axisymmetric version, model 35OC-Rp3, in the present analysis. 
The 2D equivalent of model S (35OC-Rs) was already presented in \cite{Reichert2021a}. There it was run until a final time of $t_{\mathrm{f}} = 0.9 \, \mathrm{s}$. Model 35OC-Rs$_\mathrm{N}$ has the same physical initial conditions and a slightly different numerical setup, which allows us to evolve for a much longer time, $t_{\mathrm{f, pb}} = 2.53 \, \mathrm{s}$ \citepalias{Aloy2021}.

The models were calculated using the neutrino-MHD code \textsc{Aenus-Alcar} \citep[][]{Just2015,ObergaulingerAloy2017}. We refer to \citet{ObergaulingerAloy2017} and \citetalias{Aloy2021} for a detailed overview of the simulation setup, equation of states (EOS), neutrino transport, and general relativistic corrections on the Newtonian gravitational potential. 
The resolution differs among our models, having $300-480$ zones unevenly spaced (linear and logarithmically) in radial direction, $64-128$ zones in $\theta$, and for the case of 3D, $128$ zones in $\phi$ (Tab.~\ref{tab:sim_properties}). For the neutrino transport M1 scheme \citep[][]{Just2015}, neutrinos are binned into $8-10$ logarithmically spaced energy bins. The applied spatial resolution is comparable with the one used in \citet[][]{Nishimura2015} and slightly less compared to the long-time simulations presented in \citet[][]{Nakamura2019}.
We note that, the numerical grids are not completely analogous among the 2D and 3D models. While the impact of, e.g., grid resolution remains to be explored, the main differences between 2D and 3D models may come from the assumption or relaxation of the artificial condition of axisymmetry.

\begin{figure*}
 \includegraphics[width=2\columnwidth]{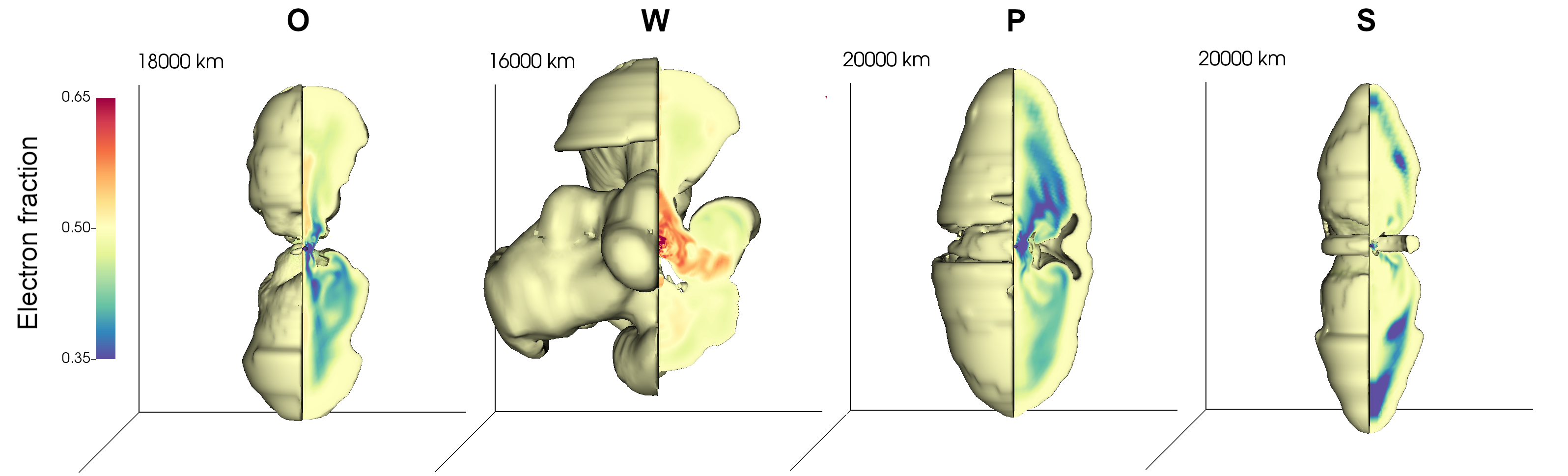}\\
  \includegraphics[width=2\columnwidth]{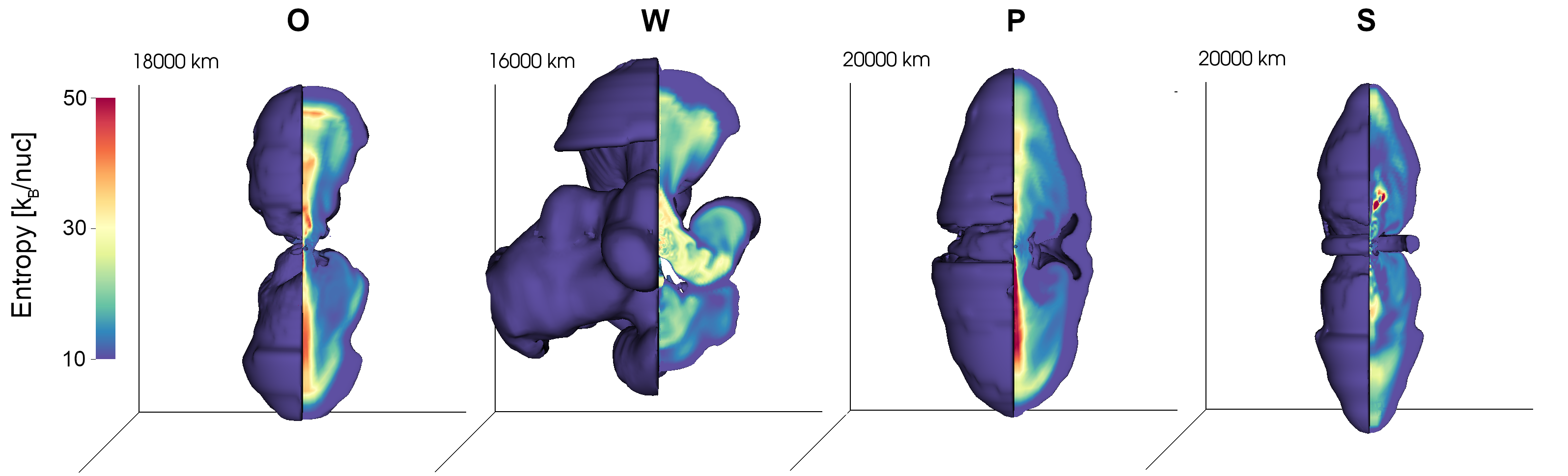}
  \caption{
  Electron fractions (upper panels) and entropies per baryon (lower panels) of the unbound matter in 3D models. O and W are shown at the final simulation time ($t_\mathrm{pb}=0.80\,\mathrm{s}$ and $1.13\,\mathrm{s}$, respectively) while P and S are shown at earlier times at similar maximum shock radii ($t_{\rm pb}=0.63\,\mathrm{s}$ and $0.40\,\mathrm{s}$, respectively).
  }
 \label{fig:entropy_3D}
\end{figure*}

The models span a wide variety of SN explosion conditions by only varying the magnetic field setup. They include a neutrino-rotational driven case, which behaves in many ways like a classical CC-SN without magnetic fields (model W), a prompt explosion without any stall of the shock (model S and 35OC-Rs$_\mathrm{N}$), models that contain a high entropy outflow (P and 35OC-Rp3), and an intermediate case as inherited by the progenitor magnetic field properties (model O). We shortly summarize the dynamics of the models. Details can be found in \citet{ObergaulingerAloy2017}, \citetalias{Aloy2021}, and \citetalias{Obergaulinger2021}. 

\textbf{Model O} bears a moderate magnetization, with $b^{\mathrm{pol; tor}} \approx 1.7\times 10^{10};\, 1.7\times 10^{11}\,\mathrm{G}$ for the poloidal and toroidal component at the centre of the star, respectively. Due to these field strengths, the magnetic field has already a dominant influence on the shock revival and the explosion is magnetorotational driven (though neutrinos still have some role). After shock revival, the model develops two jets that reach a radius of $r_\mathrm{shock}\sim 8.3 \times 10^{3} \,\mathrm{km}$ at the end of the simulation ($0.8\,\mathrm{s}$ post-bounce, Tab.~\ref{tab:sim_properties}). The morphology of the ejecta (left-hand panels of Fig.~\ref{fig:entropy_3D}) corresponds to that of typical bipolar jets, namely a central high-entropy beam or spine, where $Y_e\gtrsim 0.5$, surrounded by a double-lobed cavity. Flanking the jet, we find denser and lower entropy ejecta, where the electron fraction is smaller (with minimum values $Y_e\sim 0.32$). The cavity is limited by the unbound shocked matter, where, owed to its larger density, the entropy per baryon is lower. The neutron-rich ejecta is more pronounced in the southern (downward) direction (Fig.~\ref{fig:entropy_3D}). On the other hand, also slightly neutron-deficient matter ($Y_e > 0.5$) is ejected in the centre of the jet. The entropy is typically higher in the centre of the jet and at the shock front than elsewhere (bottom left-hand panel of Fig.~\ref{fig:entropy_3D}). At the end of the simulation the diagnostic energy reached \mbox{$4.9\times 10^{50}\,\mathrm{erg}$}.\footnote{\label{foot:energy}The smaller values of the diagnostic explosion energy computed here compared to \citetalias{Obergaulinger2021} arise from a more restrictive criterion for estimating that a computational cell contributes to the ejecta. Here we also request that the radial velocity of an unbound cell must be positive.} 

\textbf{Model W} differs from model O in two aspects. First, the original topology of the magnetic field is changed to that of a large scale dipole. Secondly, it has a factor $10$ weaker magnetic field, compared to model O \mbox{($b^{\mathrm{pol; tor}} \approx 10^{10}\,\mathrm{G}$)}. This leads to a neutrino-rotational explosion, as the magnetic field becomes dynamically irrelevant for the shock revival. The model develops rather spherical after shock revival\footnote{We note that this is a major difference compared to the 2D version, model 35OC-Rw, which develops an aspherical explosion \citep[][]{ObergaulingerAloy2017}} and the shock expands slower compared to model O.  It reaches $\sim 3.7\times 10^{3}\,\mathrm{km}$ at $0.8\,\mathrm{s}$ post-bounce. In contrast to model O, the model dominantly ejects symmetric and proton-rich material ($Y_e \gtrsim 0.5$). Due to different simulation times, the diagnostic energy at the end of the simulation is slightly higher ($5.2\times 10^{50}\,\mathrm{erg}$) compared to model O, even though the explosion occurs less violently.

\textbf{Model S} is the model with the strongest (large-scale, dipolar) magnetic field \mbox{($b^{\mathrm{pol; tor}} =10^{12}\,\mathrm{G}$)}. The magnetorotationally driven explosion happens promptly after core bounce, without shock stagnation. This leads to a neutron-rich cocoon at the shock front, more pronounced around the southern jet (top right-hand panel of Fig.~\ref{fig:entropy_3D}). The expansion of the shock happens extremely quickly, and it reaches already $r_\mathrm{shock}\sim 3.0 \times 10^{4} \,\mathrm{km}$ after $0.8\,\mathrm{s}$ post-bounce. The diagnostic energy reaches $1.28\times 10^{52}\,\mathrm{erg}$ at the end of the simulation ($t_\mathrm{f, pb}=1.17\,\mathrm{s}$). Such a large energy is already sufficient to account for the large explosion energies that are observed in HNe. Noteworthy, this model most likely yields an strongly magnetized proto-neutron star as compact remnant.

\textbf{Model 35OC-Rs}$_\mathrm{N}$, the 2D counterpart of the 3D model S, 
explodes promptly. At first, the shock expands faster than in model S, but later on it falls behind the shock in the latter and reaches  $r_s\sim 2.6\times 10^{4}\,\mathrm{km}$ at $0.8\,\mathrm{s}$ after bounce. Nevertheless, the cocoon around the jet is dominated by even more neutron-rich material, which is also more neutron-rich than the 3D version. The explosion happens less violently, and the diagnostic energy at the end of the simulation ($t_\mathrm{f, pb}=2.53\,\mathrm{s}$) is $8.96\times 10^{51}\,\mathrm{erg}$. 

\textbf{Model P} has the same toroidal field as model O, but a three times stronger poloidal component. Its magnetization is between models O and S.  The shock expands slower compared to models S and 35OC-Rs$_\mathrm{N}$ and reaches $\sim 1.5\times 10^{4}\,\mathrm{km}$ $0.8\,\mathrm{s}$ after bounce. Neutron-rich material is ejected within a cone of half-opening angle of $\sim 45^\circ$ around the rotational axis. Small regions in the jet beam reach high entropies of  $S>100\,\mathrm{k_B/nuc}$ ($\rm k_B$ is the Boltzmann constant). The explosion energy is with $2\times 10^{51}\,\mathrm{erg}$ at the end of the simulation ($t_\mathrm{f, pb}=1.8\,\mathrm{s}$) thus fairly large. 

\textbf{Model 35OC-Rp3} develops very similarly to its 3D version, model P. The shock expands marginally faster in 2D, reaching $r_s\sim 1.9\times 10^{4}\,\mathrm{km}$ at $0.8\,\mathrm{s}$ after bounce. This model was calculated for a long time ($\simeq 9\,$s post bounce), reaching a maximum shock radius of $4.1\times 10^{5}\,\mathrm{km}$. The outflow can be moderately neutron-rich ($Y_e \sim 0.25$), though less than in model S. Compared to model P, the maximum ejecta entropy is slightly lower, however, also reaching values $S> 100\,\mathrm{k_B/nuc}$ and still larger than the maximum values of all other models. The explosion energy is with $6.6\times 10^{51}\,\mathrm{erg}$ at the end of the simulation also fairly high, as expected from the strong magnetization in combination with the long simulation times. 

\begin{table*}
 \caption{Main properties of the simulations. The first two columns list the model name and their spatial dimensionality. The next columns show 
 the pre-collapse energies of the poloidal and toroidal magnetic field components, $E_{\mathrm{mag},0}^{\mathrm{pol,tor}}$, (for comparison, the initial rotational energy of all models is $E_{\mathrm{rot},0} \approx 1.7 \times 10^{50} \, \mathrm{erg}$). We note that the magnetic energies are integrated over the whole numerical grid. As the outer radii of the 2D and 3D models are slightly different, this also results in a slightly different magnetic energy.
 The final simulation time after bounce, $t_{\rm f, pb}$, 
 and the ejected mass, as well as the diagnostic explosion energy$^{\ref{foot:energy}}$ at $t=t_{\rm f}$, the total amount of tracers set and total amount of tracers in which a detailed nucleosynthetic calculation is performed. The last four columns give the number of cells in the radial, $\theta$, and $\phi$ directions, and  the number of neutrino energy bins.  
 }
 \label{tab:sim_properties}
 \begin{tabular}{lccccccrrrrr}
  \hline
  Model & D 
  & $E_{\mathrm{mag},0}^{\mathrm{pol}}$ & $E_{\mathrm{mag},0}^{\mathrm{tor}}$
  & $t_{\rm f, pb}$ 
  & Ejected mass & Energy & \#Tracers & $N_r$ & $N_\theta$ & $N_\phi$ & $N_\mathrm{E_\nu}$ \\
  & & [$\mathrm{erg}$] 
  & $[\mathrm{erg}]$ & $[\mathrm{s}]$
  & [$M_{\sun}$] & [$10^{51}\,\mathrm{erg}$] &  Tot./Calc.& & &\\
  \hline
  35OC-Rp3 & 2 & $5.0 \times 10^{48}$ & $1.3 \times 10^{49}$ & $8.96$ & $1.58$ & $6.60$  &   $41980$/$41980$& 480 &128 &  1 & 10\\
  35OC-Rs$_\mathrm{N}$  & 2 & $7.5 \times 10^{47}$ & $2.6 \times 10^{49}$ & $2.53$ & $2.56$ & $8.96$  &   $74521$/$3451$ & 480 &128 &  1 & 8\\
  W      & 3 & $1.1 \times 10^{44}$ & $1.3 \times 10^{45}$ & $1.13$ & $0.20$ & $0.52$  &  $952412$/$2618$ & 300 & 64 & 128& 10\\
  O      & 3 & $2.6 \times 10^{47}$ & $6.6 \times 10^{48}$ & $0.80$ & $0.15$ & $0.49$  &  $838984$/$2032$ & 300 & 64 & 128& 10\\
  P     & 3 & $2.4 \times 10^{48}$ & $6.6 \times 10^{48}$ & $1.80$ & $0.66$ & $2.06$  & $1286322$/$9641$ & 320 & 64 & 128& 10\\
  S      & 3 & $1.1 \times 10^{48}$ & $1.3 \times 10^{49}$ & $1.17$ & $1.61$ & $12.8$  & $1486316$/$3459$ & 300 & 64 & 128& 10\\
 \end{tabular}
 
\end{table*}

\subsection{Tracer particles}
\label{sec:tracer} 
The nucleosynthesis calculations are based on Lagrangian tracer particles representing fluid elements of the unbound ejecta. We integrate their equation of motion $\partial_t \vec{X} = \vec v(\vec X)$, where $\vec X$ is the position of a tracer particle and $\vec v$ the velocity field on the simulation grid. Our strategy differs from similar analyses in that we carry it out after the MHD models rather than at run-time and perform the integration backward in time instead of forward. This approach has the advantage that it allows us to insert the tracers at the final time of the MHD simulation directly in the gravitationally unbound regions. That way, we achieve a fine coverage of the ejecta without wasting resources on fluid elements that in the end are not ejected. The drawback of the method is that we have to store the velocity field with a sufficiently high cadence (in our case, every 1 ms), thus consuming a large amount of disc space \citep[see][for a similar approach]{Wanajo2018,Sieverding2020,Witt2021}. The reliability of this method is tested in Appendix~\ref{app:tracer_integration}.

\begin{figure}
 \includegraphics[width=\columnwidth]{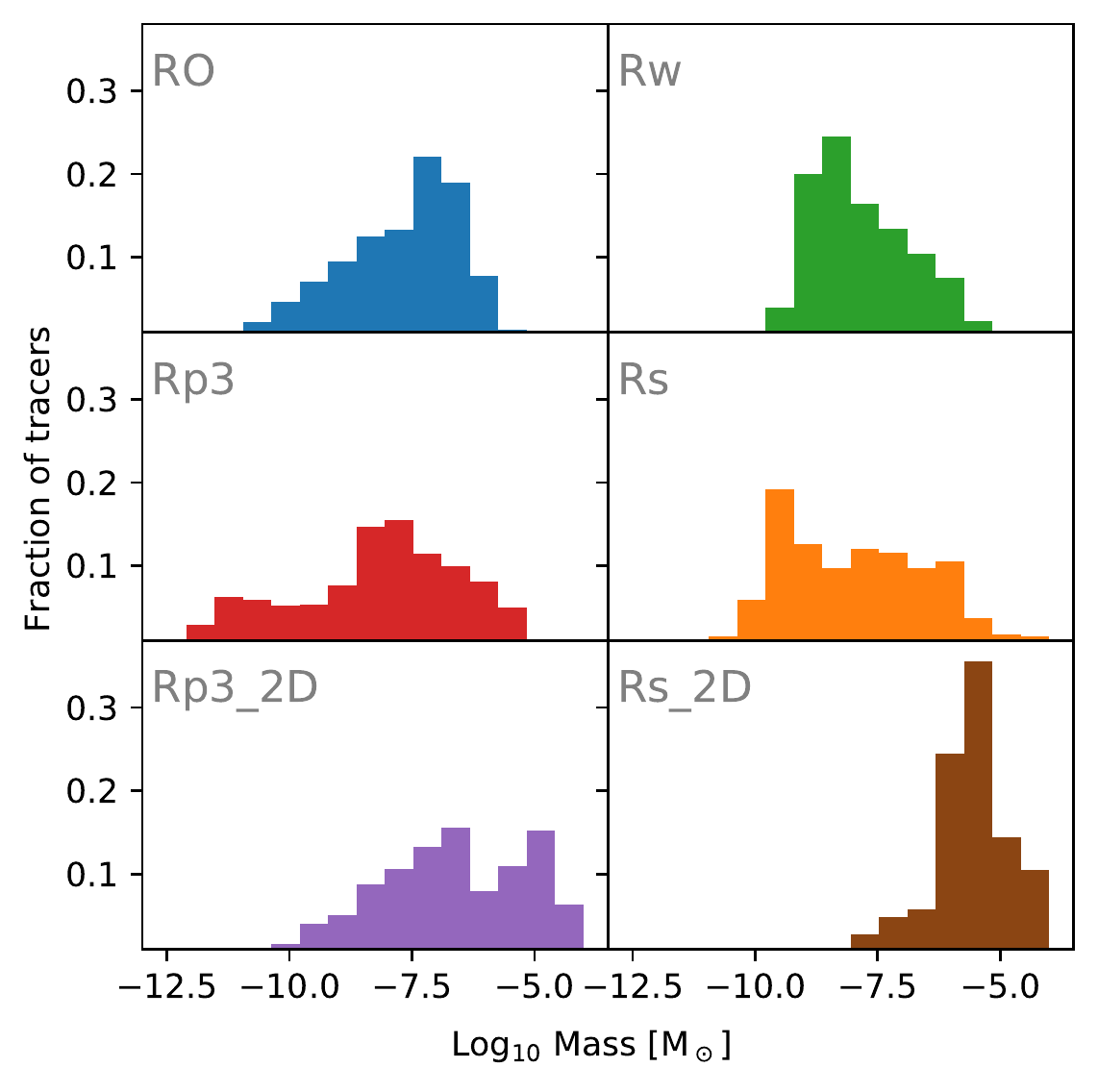}
 \caption{Histogram of the fraction of tracer particles as a function of their masses.}
 \label{fig:mass_distr_tracer}
\end{figure}

Due to the large computational domain (c.f., \mbox{$r_\mathrm{max, domain}\sim 7.7\times 10^5\,\mathrm{km}$} with a stellar radius $r_\ast \sim 5.3\times 10^5\,\mathrm{km}$), no unbound matter has left the domain at the end of the simulation. Therefore, we do not omit any ejected mass when placing the tracer particles at the last available time in each unbound cell. We say that matter in a computational cell is unbound when two conditions hold: that both the total energy (i.e., internal, kinetic, magnetic, plus gravitational) and the radial velocity are positive. While the second condition (positive radial velocity) is not a necessary condition for unbound ejecta, it filters out fluid elements that fall back onto the central object.$^{\ref{foot:energy}}$
We distribute the total mass of each cell, i.e., the product of its density and volume, $M_{\mathrm{cell}} = \rho_{\mathrm{cell}} V_{\mathrm{cell}}$, flagged as unbound among a number $n_{\mathrm{ptc0}}$ of tracer particles placed at random positions in the cell \citep[see e.g.,][for a discussion of uncertainties arising from different initial placements of tracer particles]{Bovard2017}. By construction, each tracer represents a mass of $M_{\mathrm{ptc0}} = M_{\mathrm{cell}} / n_{\mathrm{ptc0}}$. The number $n_{\mathrm{ptc0}}$ can vary between cells.  It is determined by two conditions: each unbound cell has to contain at least two tracers and the mass of each tracer is limited to $M_{\mathrm{ptc0}} \le 10^{-4}\, \mathrm{M_\odot}$.
In all models, most of the tracer particles have a mass less than $10^{-6}\, \mathrm{M_\odot}$ (Fig.~\ref{fig:mass_distr_tracer}). \cite{Nishimura2015} have shown that this resolution is sufficient to obtain a converged result of the nucleosynthesis.
For 35OC-Rp3 we set additional tracers into a five degree solid angle around the symmetry axis (i.e., in the jet) to cover better this low density region. The amount of tracer particles is shown in Tab.~\ref{tab:sim_properties}. 

We can vastly reduce the computational cost of the nucleosynthesis calculation by applying a binning procedure and only calculate representative tracers. For this we differentiate between hot ($T_\mathrm{peak}\geq 7\,\mathrm{GK}$) and cold ($T_\mathrm{peak}< 7\,\mathrm{GK}$) tracers. The bins of hot tracers are based on either the electron fraction and entropy at $7\,\mathrm{GK}$. For cold tracers they are based on their peak temperature and density. Our approach works because similar hydrodynamic conditions will end up in similar final abundances (see Appendix~\ref{app:tracer_selection} for further details and tests) and we catch all nucleosynthesis relevant conditions. We note that we still use the spatial information of all tracer particles to enable a detailed analysis of the spatial distribution of the elements.

\subsection{Nucleosynthesis}
To calculate the nucleosynthetic yields, we employ an upgraded version of the nuclear reaction network \textsc{WinNet} \citep[][]{Winteler2012} as in \citet{Reichert2021a}. We included $6545$ nuclei up to $Z=111$. We used reaction rates from the JINA Reaclib database \citep[][]{Cyburt2010}. Furthermore, we include fission reactions as well as fragment distributions from \citet{Panov2005,Panov2010} and theoretical $\beta$-decay rates and electron/positron captures at stellar conditions \citep[][]{Langanke2001b}. The latter reaction rates are exchanged for experimentally known ones that are included in the JINA Reaclib below the temperature tabulation of \citet{Langanke2001b} at $T=0.01\,\mathrm{GK}$. Neutrino reactions on nucleons are included as in \citet[][]{Froehlich2006b}, using the rate tabulation of \citet[][]{Langanke2001a}.

\subsection{Finding estimates of the final yields}
\label{sect:upper_limit}
\begin{figure}
 \includegraphics[width=\columnwidth]{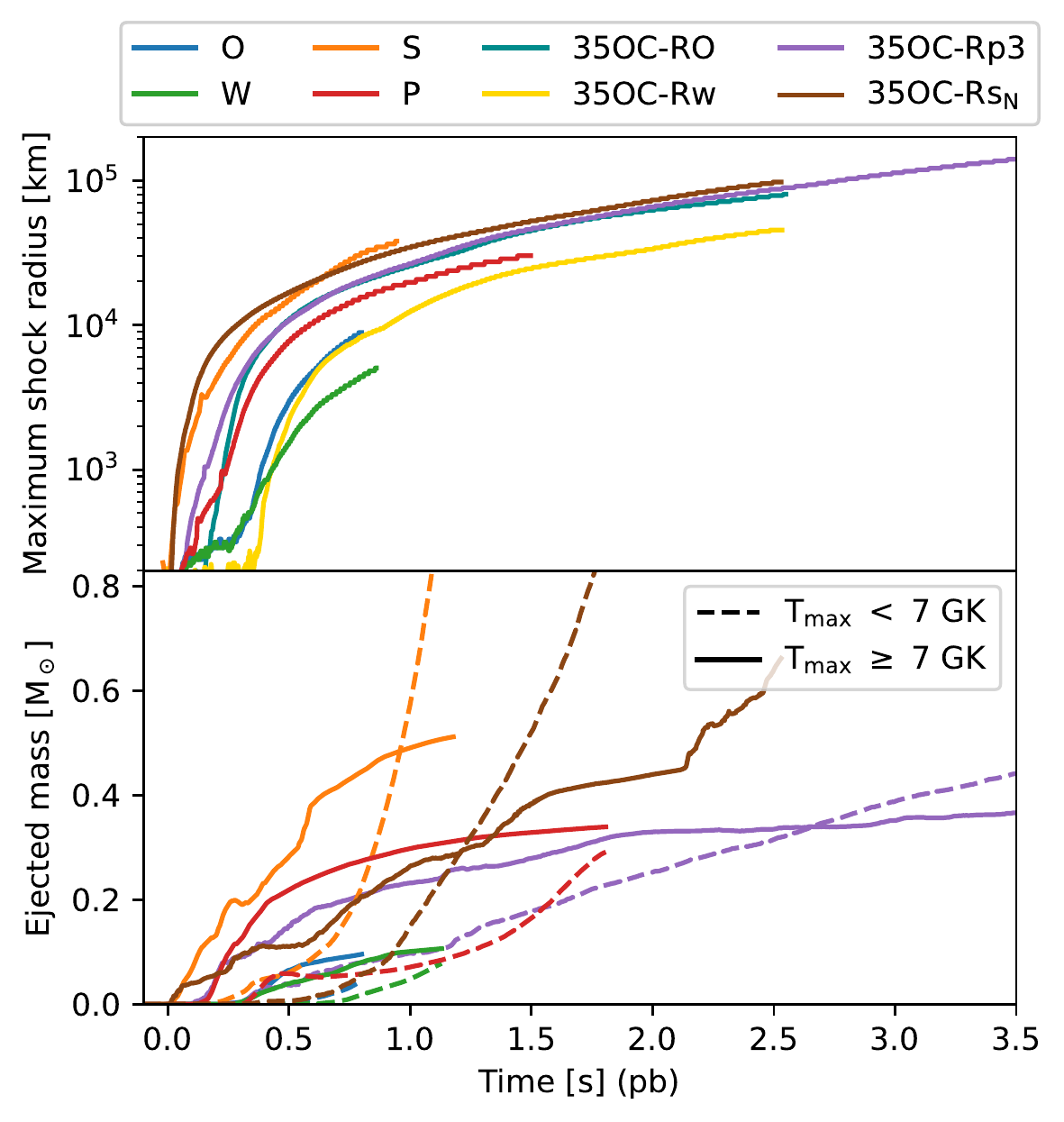}
 \caption{Upper panel: Maximum shock radii of different models versus time post-bounce. In addition to the models discussed within this work, also the 2D versions of model O and W are shown \citep[][]{ObergaulingerAloy2017,Reichert2021a}. It is visible that 2D models tend to develop a larger maximum shock radius at similar times after bounce. Lower panel: Ejected mass of the different models (except models 35OC-RO and 35OC-Rw) as a function of the time post-bounce, distinguishing for each model matter whose maximum temperature is \mbox{$T_{\rm max}\ge 7\,$GK} (solid lines) or  \mbox{$T_{\rm max}< 7\,$GK} (dashed lines). The ejected mass of both components is growing until the end of the simulation.}
 \label{fig:ej_mass_temp}
\end{figure}

It is challenging and, to date, not possible to calculate sophisticated 3D neutrino-MHD models for long enough (of the order of few tens of seconds) that nucleosynthesis has effectively finished in a SN explosion. For the time computed in this work, the calculated ejected mass (Tab.~\ref{tab:sim_properties}) is only a fraction of the total foreseeable ejected mass of the event \citep[see e.g.,][for an overview of uncertainties in state-of-the-art nucleosynthesis calculations]{Harris2017}. To get a more complete picture of the expected total yields of the explosions, it is useful to split the ejected mass into a hot and a cold component, depending on the maximum temperature reached during its evolution, $T_{\rm max}$ (Fig.~\ref{fig:ej_mass_temp}).

\begin{figure}
 \includegraphics[width=\columnwidth]{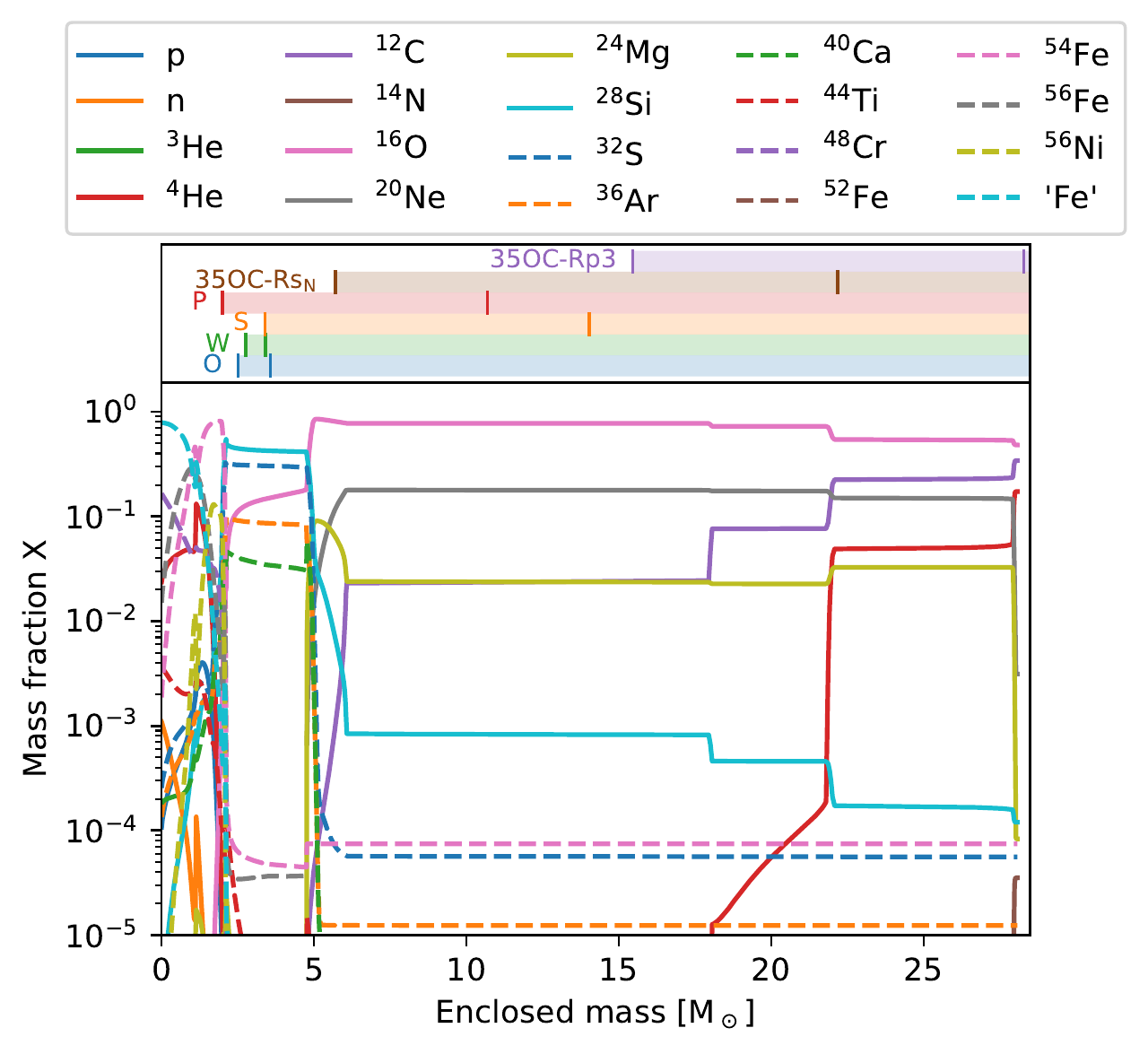}
 \caption{Lower panel: Composition of the progenitor (35OC) from \citet{woosley-heger2006}. Upper panel: Position of the ejecta outermost and innermost shock wave in the individual models. The two vertical lines indicate the minimum and maximum shock positions at the end of the simulations. Coloured regions ahead of the shock positions correspond to progenitor stellar matter, part of which will be ejected.}
 \label{fig:progenitor}
\end{figure}

Extrapolating the behaviour found in Fig.~\ref{fig:ej_mass_temp}, most of the subsequent additions to the ejecta mass will not reach $7\,\mathrm{GK}$ and therefore its composition will be dominated by light elements (i.e., lighter than iron, see also Fig.~\ref{fig:progenitor}). However, there is also a fraction of hot matter still getting ejected until the end of the simulation even for model 35OC-Rp3, which was simulated up to $t_\mathrm{f, pb}=8.96\,\mathrm{s}$. Since this hot matter will still contribute significantly to create lighter heavy elements such as Fe, Ni, Zn, or even Sr, we shall estimate the contribution of this fraction of the ejecta too.

In order for unbound matter to attain $T_{\rm max}\ge 7\,$GK, we anticipate two possibilities. When the shock moves outward, the post-shock temperature decreases from above this threshold to lower values. Thus, progenitor matter can only be shock heated to $T_{\rm max}\ge 7\,$GK while the shock is still deep inside the core. At later times, such high temperatures can be achieved if matter gets closer to the hot PNS, from where it may be re-ejected. 
To estimate the final yields, it is useful to recall that the progenitor is a rapidly rotating star. Hence, at a finite distance from the centre, the rotating stellar layers posses sufficiently high specific angular momentum to circularize as they hit a centrifugal barrier in their nearly free falling (collapse) trajectory.
\citetalias{Aloy2021} estimated this orbit to be located at $M_\mathrm{D}\approx 7.5\,\mathrm{M_\odot}$ and $j\sim 10^{17}\,\mathrm{cm^2\,s^{-1}}$ at the equator (for model 35OC, the progenitor of all models used here). This can be used to find a crude estimate of the outermost mass shell in the star that may contribute to the hot ejecta. Inside $M_\mathrm{D}$, matter has insufficient angular momentum to form an accretion disc. Thus, (a fraction of) these stellar mass shells may be ejected after having fallen close to the central object. The part of the star beyond $M_\mathrm{D}$ might fall onto the central object, after assembling an accretion disc. However, our models develop successful (and powerful) SN explosions, which make uncertain what fraction of the mass above $M_\mathrm{D}$ may finally end up accreted onto the central object (and hence, not contributing to the ejecta). A more restrictive bound can be obtained by the shock itself. We estimate if the shock is able to drive matter outwards by the mass flux ahead and behind the shock,
\begin{align}
    \dot{M}_\mathrm{pre}  = 4\pi r^2\rho_\mathrm{pre} v_\mathrm{r,pre} \quad \mathrm{and} \quad \dot{M}_\mathrm{post} =  4\pi r^2 \rho_\mathrm{post} v_\mathrm{r,post},
\end{align}
with the density $\rho$ and radial velocity $v_\mathrm{r}$ ahead (pre) and behind the shock (post). If $\dot{M}_\mathrm{pre}+\dot{M}_\mathrm{post}>0$, matter will be pushed out by the shock, even when it may not get unbound immediately. 
For each $\theta$ and $\phi$ direction we define three radii, $r_j(\theta,\phi)$, $r_s(\theta,\phi)$, and $r_u(\theta,\phi)$. The radius $r_j$ is defined as the radius at which $j = 10^{17}\,\mathrm{cm^2/s}$ holds and $r_s$ is the shock radius. 
We define the radius $r_u$ as the radius at which matter will most likely not be able to fall onto the central object, $r_u=\text{min}(r_j,r_s)$. Thus, $r_u=r_j$ if $r_j<r_s$ or if the shock is not able to move matter outwards in the sense as defined above. Otherwise, $r_u$ is set to the shock radius. Summarized, our ejecta is split into several groups:
\begin{itemize}
    \item \textbf{Group 1:} Ejecta that are unbound already during the computed neutrino-MHD evolution, i.e., have positive energy and radial velocity. Matter in this group was followed by the tracer particles and constitutes the minimum amount of material that will be ejected.
    \item \textbf{Group 2:} Bound matter (i.e., negative total energy or radial velocity) at the end of the simulation outside of the central object but inside $r<r_u$. We assume that this matter heats up to at least $7\,\mathrm{GK}$, thus contributing to the hot ejecta.
    \item \textbf{Group 3:} Bound matter that is located at radii $r>r_u$ at the final time of the computed neutrino-MHD evolution. We assume that this matter will be shocked and ejected.  
    \item \textbf{Group 4:} Stellar wind that was ejected prior to the explosion. The mass is given as the difference between the stellar mass at ZAMS ($35\,\text{M}_\odot$) and at collapse (28.1 M$_\odot$). 
\end{itemize}
We stress that it is not guaranteed that the entirety of group 2 and group 3 are ejected. A growing mass of the central object could swallow large parts of the mass of the groups, especially if a BH forms. Furthermore, we did not account for the gravitational binding energy of the outer shells from the total energy of each tracer for the definition of group~1 and parts of this group may therefore be not ejected \citep[see][]{Bruenn2016}. This effect may, however, be comparably small. Our extrapolation, assuming that all gas in groups 2 and 3 is ejected, should therefore be seen as a very optimistic case if the entirety of groups get ejected.

For group 1, we already obtained the nuclear composition via the tracer particles in the neutrino-MHD simulation. 

\begin{figure}
 \includegraphics[width=\columnwidth]{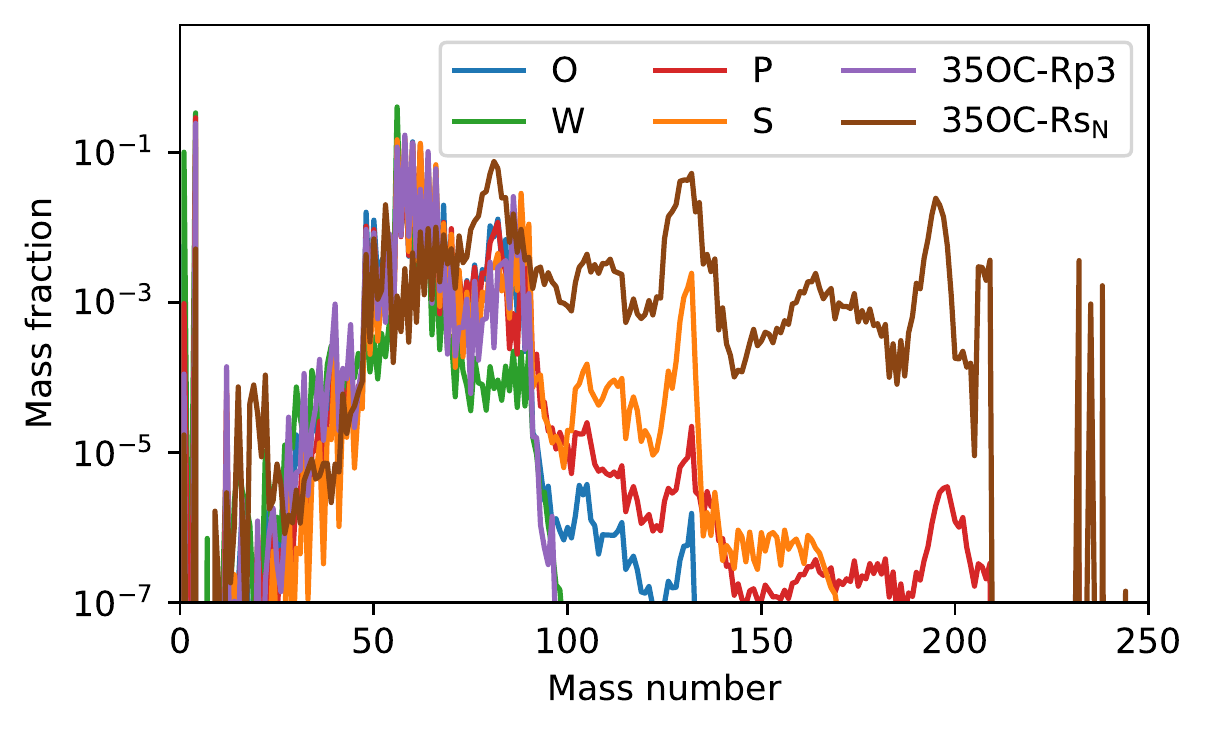}
 \caption{Mass fraction versus mass number for matter ejected during the last $100\,\mathrm{ms}$ before the neutrino-MHD simulation ends.}
 \label{fig:last100ms}
\end{figure}

To obtain the composition of group 2, we assume that this matter will eventually be ejected similarly to hot matter that got ejected during the last 100$\,$ms before the simulations end (Fig.~\ref{fig:last100ms}). This assumption obviously has many weaknesses. For example, it cannot be reliably applied to model 35OC-Rs$_\mathrm{N}$, because the neutron-rich ejecta at the end of the simulation is not expected to continue for a long time (see the r-process pattern in Fig.~\ref{fig:last100ms}). Additionally, model 35OC-Rp3 develops an almost stable downflow configuration which exists until the end of the simulation. As a result, the estimated amount of significantly heated material is huge ($\approx 7.3\,\mathrm{M}_\odot$) and, most likely, an artifact of the symmetry assumption of this 2D-axisymmetric model \citep[see also][]{Witt2021}.

For material that is located at radii that exceed $r_u$ (group 3), we calculate the shock temperatures and densities following \citet[][]{Nadyozhin2002}. We assume the shock temperature to follow \citep[e.g.,][]{Nadyozhin2002,Woosley2002}
\begin{equation}
    T_\mathrm{s} =  2.37\times 10^9 E_{51}^{0.25} \times R_{09}^{-0.75}\,\mathrm{K},
\end{equation}
with the final explosion energy $E_{51}$ in units of $10^{51}\,\mathrm{erg}$ and pre shock radius $R_{09}$ in $10^9 \,\mathrm{cm}$. Furthermore, the shock properties evolve significantly on time-scales \citep[][]{Nadyozhin2002},
\begin{equation}
    t_u = 3.83\times10^{-3} \rho_0^{0.5} \times E_{51}^{-0.5} \times R_{09}^{2.5} \, \mathrm{s},
\end{equation}
where $\rho_0$ is the density before shock arrival. For a Lagrangian layer crossed by the expanding shock, \citet[][]{Nadyozhin2002} obtained a temperature, density, and radius evolution that follows

\begin{align}
    T(t) &= \frac{T_p}{1+\epsilon_T\times t/t_u}, \quad T_p=\epsilon_p T_s \label{eq:Ts}\\
    \rho(t) &= \rho_p \left( \frac{T}{T_p}\right)^3, \quad \rho_p = 7\rho_0\\
    R(t) &= R_0\times (1+\epsilon_r t/t_u ), \label{eq:Rs}
\end{align}
with $\epsilon_p$, $\epsilon_T$, and $\epsilon_r$ being parameters that we calibrate for each model individually. Precisely, we fit the evolution of the temperature, density and radial location of each ejected tracer particle that has been shocked by the expanding SN shock during the neutrino-MHD computed time to the functional forms of Eqs.\,\eqref{eq:Ts}-\eqref{eq:Rs} (the mean value and standard deviation of all fits is given in Tab.~\ref{tab:fit_pars}). We confirmed that matter is indeed radiation dominated for most of the density range and $\gamma = 4/3$ and consequently $\rho_p=7\rho_0$ is a reasonable approximation. Furthermore, we tested an extrapolation using $\gamma = 5/3$ and only find minor differences in the final abundances. With these Lagrangian evolutions we account for the mass located outside the shock by calculating a typical evolution for each radius, using the initial composition in mass coordinates of the progenitor. For $E_{51}$ we take the diagnostic explosion energy at the end of the neutrino-MHD simulation as given in Tab.~\ref{tab:sim_properties}. 
\begin{table}
 \caption{Post-shock calibration parameters of the different models. The error is defined as the standard deviation from the parameter from all fitted tracer particles.}
 \label{tab:fit_pars}
 \begin{tabular}{cccc}
  \hline
  Model &$\epsilon_p$& $\epsilon_T$ & $\epsilon_R$ \\
  \hline
  W & $1.04\pm 0.09$ & $0.40\pm 0.17$ & $0.95\pm0.30$\\
  O & $1.14\pm 0.09$ & $1.08\pm 0.33$ & $1.12\pm0.40$\\
  P & $0.81\pm 0.14$ & $0.44\pm 0.44$ & $0.42\pm0.34$\\
  S & $0.55\pm 0.11$ & $0.19\pm 0.11$ & $0.31\pm0.20$\\
  35OC-Rp3           & $0.59\pm 0.11$ & $0.15\pm 0.14$ & $0.27\pm0.19$\\
  35OC-Rs$_\mathrm{N}$ & $0.54\pm 0.09$ & $0.19\pm 0.40$ & $0.16\pm0.15$\\
  \hline
 \end{tabular}
\end{table}

To test our extrapolation, we calculated total ejected yields as described above for model 35OC-Rp3 after different times from $1\,\mathrm{s}$ to $9\,\mathrm{s}$ in $0.5\,\mathrm{s}$ steps using the same fitting parameters as in Tab.~\ref{tab:fit_pars}. For elements dominantly ejected within group 3, namely the $\alpha$-elements such as $^{16}$O, $^{20}$Ne, or $^{24}$Mg the extrapolated values agree within $\sim 30$\% for all tested times. For extrapolations after $3\,\mathrm{s}$ post-bounce, they even agree within $\sim 3$\%. The extrapolated yields of $^{44}$Ca and $^{56}$Fe that are mainly produced within group 2 agree within a factor of $\sim 2$. Heavier elements with $90\lesssim A \lesssim 130$ are converged within a factor of $10$. Again, the estimate gets significantly better for $t>3\,\mathrm{s}$ after which the extrapolated values agree within a factor of $2$. For these heavier elements, we tend to estimate higher ejected masses for extrapolations at earlier times. This is expected as the electron fraction of the ejected matter moves to more symmetric conditions at later simulation times while our estimate is based on the distribution of the last $100\,\mathrm{ms}$. This tends to favor more neutron-rich conditions in the extrapolation. Elements heavier than $A \gtrsim 130$ show a bad agreement within $2$ mag only.

We are aware that our estimates are very crude. They are aimed to stress that the amounts of nucleosynthetic yields computed until $t=t_{\rm f}$ are lower bounds of the final products in our models. Our treatment completely neglects the presence of reverse shocks. Furthermore, the total yields strongly depend on the final fate of the central object. In the case of models O and W, the PNS may collapse to a BH as this happens in the 2D version of O and even when not happened so far, it is also expected in the 2D version of W \citepalias[see][for a discussion of BH formation]{Aloy2021}. In this case, a major fraction of the inflowing matter will be accreted by the central BH. Contrary, more material from a later forming accretion disc may get unbound (e.g. by viscous effects) and might also synthesize heavier elements \citep[the ejecta of a so-called collapsar, for which a still ongoing discussion exists,][]{MacFadyen_1999ApJ...524..262,Surman2004,McLaughlin2005,Surman2006, Fujimoto2008,Siegel2019,Miller2020,Siegel2021,Just_2022arXiv220514158}. Within the scope of our work, we can not estimate the conditions of material from a collapsar as detailed simulations would be required.

Nevertheless, our treatment gives a rough estimate of the potential of the models to still synthesize material as, e.g., $^{56}$Ni and other unstable isotopes further discussed in sect.~\ref{sct:radioactive_isotopes}. Such an extrapolation is important to shed light into the question whether MR-SNe can be candidates for HNe or SL-SNe.

\section{Results}
\label{sct:results}
\subsection{Ejecta composition}
\label{ssct:ej_comp}
\begin{figure}
 \includegraphics[width=\columnwidth]{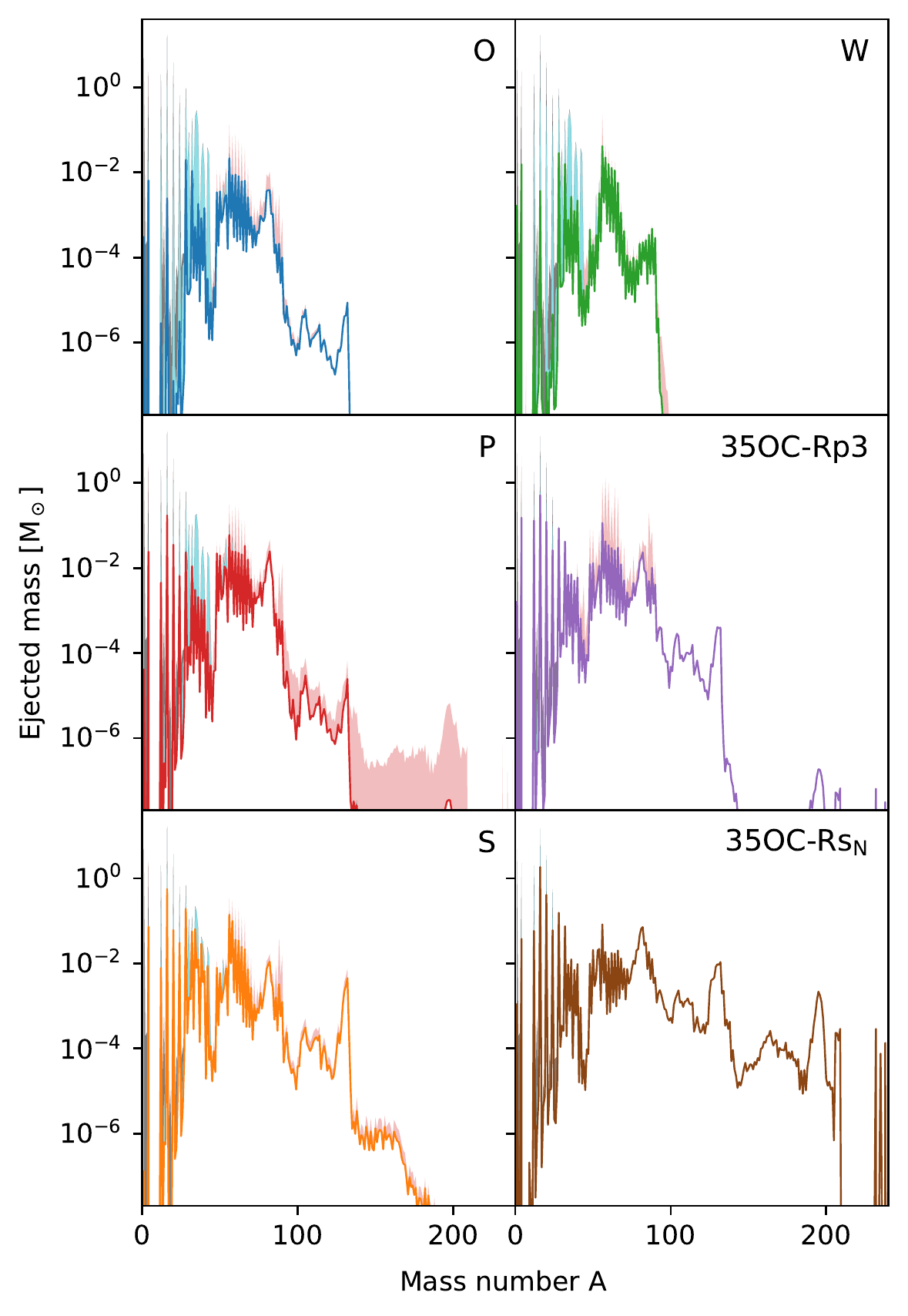}
 \caption{Final ejected masses as a function of mass number for the individual models. Lower lines represent the ejecta mass contained in the Lagrangian tracer particles (group 1, see Sect.~\ref{sect:upper_limit}). Shaded red regions indicate the potential contribution of significantly heated material after the simulation has ended (group 2, except for model 35OC-Rs$_\mathrm{N}$, see text). Shaded cyan regions show the contribution from later shocked material (group 3). The contribution of the stellar wind is indicated as grey region.}
 \label{fig:final_masses}
\end{figure}

The yields, as shown in the final composition of the ejecta in Fig.~\ref{fig:final_masses}, vary considerably among our models, with stronger initial magnetic fields leading to heavier elements. The most magnetized models P, \mbox{35OC-Rs$_\mathrm{N}$}, \mbox{35OC-Rp3}, and to some extends also S,  synthesize elements up to the third r-process peak ($A\sim 200$), while models with weaker magnetization only reach the second r-process peak (O) or the first peak (W). 

\begin{figure}
 \includegraphics[width=\columnwidth]{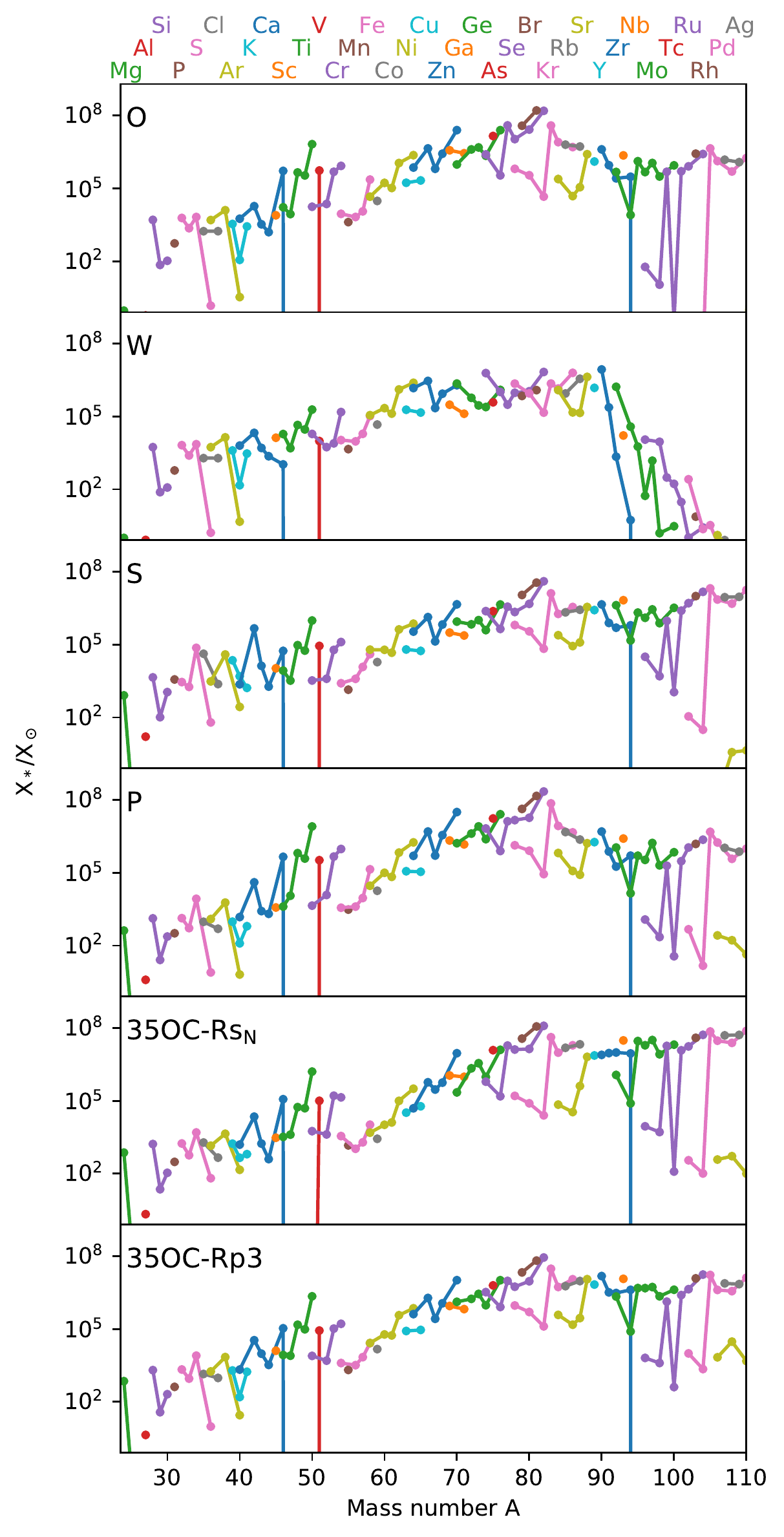} 
 \caption{Isotopic ratios relative to the sun for ejecta as obtained from our tracer particles (i.e., group 1, see Sect.~\ref{sect:upper_limit}). Isotopes of the same element are illustrated as a chain of the same colour. Solar values are taken from \citet{Lodders2009}. The upper labels correspond to the elements represented (according to their atomic number from left to right).}
 \label{fig:isotopes}
\end{figure}

Figure~\ref{fig:isotopes} shows the overproduction factor of isotopes, $X_*/X_\odot$, i.e., the ratio between the mass fraction in one of our models, $X_*$, and the solar value, $X_\odot$. The overproduction factor is shown for the lower bound of the ejecta (i.e., group 1, see sect.~\ref{sect:upper_limit}). Individual isotopes of one element are connected by lines. Positive/negative slopes of the lines indicate that a model favors the production of more neutron-/proton-rich isotopes compared to the Sun. Most models populate more neutron-rich isotopes (e.g., Ti); however, model W is an exception. There, the isotopic ratios are more similar to the sun (i.e., flatter, e.g., Kr). Furthermore, for elements heavier than Sr such as, e.g., Zr, Mo, and Ru more proton-rich isotopes are synthesized for this model. This overproduction factor is caused by more proton-rich conditions compared to the other models (c.f., Fig.~\ref{fig:entropy_3D}).
Complete tables of the ejecta composition can be found in Appendix~\ref{app:yields}.

\subsection{Comparison of 2D and 3D models}
\label{sct:2D_3D_comp}
\begin{figure}
 \includegraphics[width=\columnwidth]{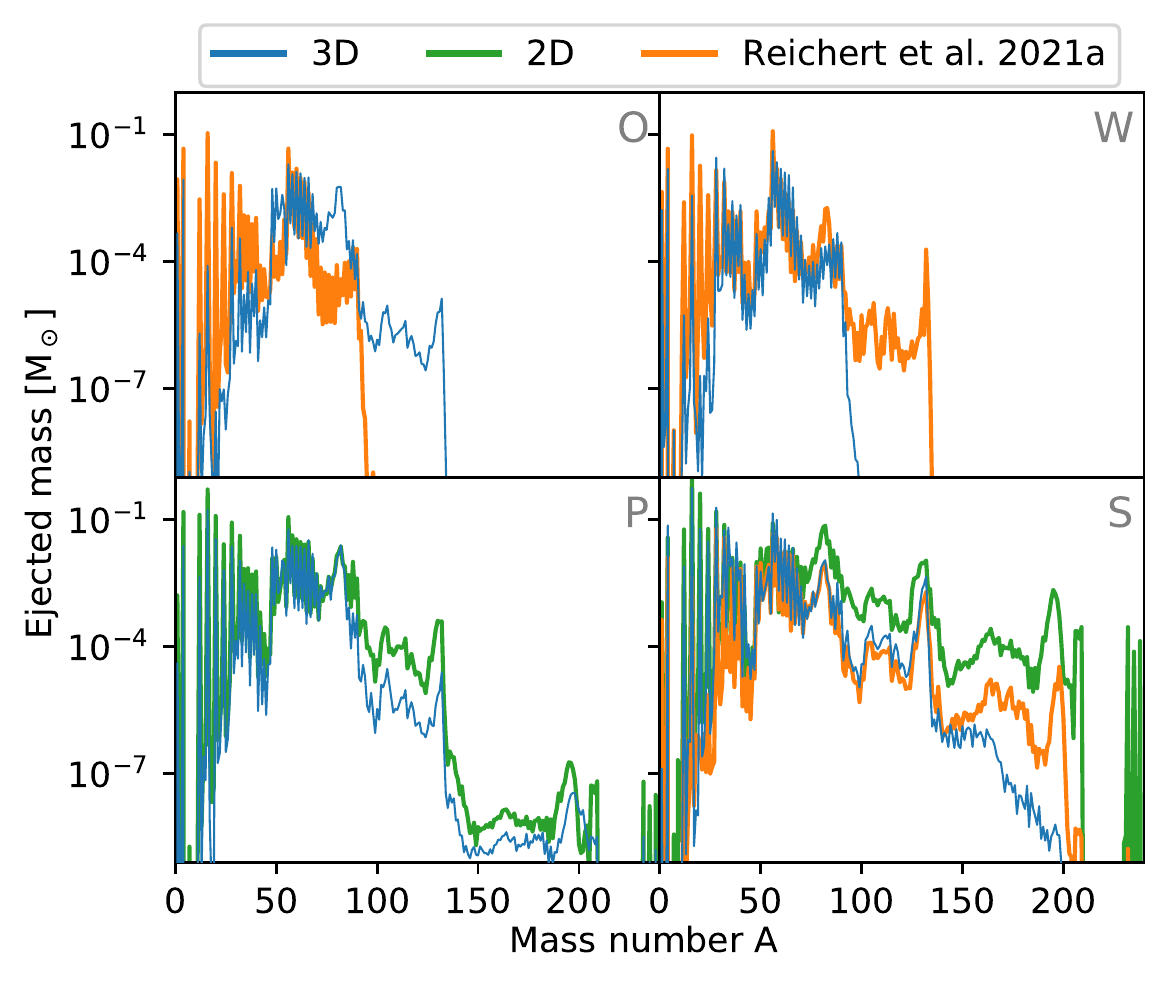}
 \caption{Comparison between the nucleosynthetic yields of 2D axisymmetric and full 3D versions of different models as indicated in each panel. Blue and green lines are models calculated within this work, whereas orange lines refer to 2D models presented in \citet{Reichert2021a}. The yields are shown for our lower limit, i.e., the yields obtained by the tracer particles (group 1, sect.~\ref{sect:upper_limit}).}
\label{fig:2d_3d_nucleosynth}
\end{figure}

Based on Fig.~\ref{fig:2d_3d_nucleosynth}, we discuss the differences in the nucleosynthesis between 2D and 3D models. 
Most of our 3D models show a signature of saturation of the explosion energies already by the final simulation times (and note that 3D models are computed for shorter post-bounce times than 2D ones). Contrarily, the explosion energy within the 2D models grows continuously with simulation time. Eventually, we expect that 2D models will reach a higher explosion energy. Model S is an exception to this behaviour as it does not show any saturation in 3D and the explosion energy grows faster than in its 2D counterpart.
For a deeper discussion of the impact of the dimensionality on the dynamical evolution, see \citet[][]{Obergaulinger2020,Obergaulinger2021}, \citet{Bugli2021}, and \citetalias{Aloy2021}. We want to stress that the differences between 2D and 3D models discussed here are not only influenced by the assumption of axisymmetry, but also by different resolutions in the simulation setups (see Sect.\ref{sct:mhd_models}). We expect the effects of different resolutions to be relatively minor, but a more detailed study will be necessary in the future.

\begin{figure}
 \includegraphics[width=\columnwidth]{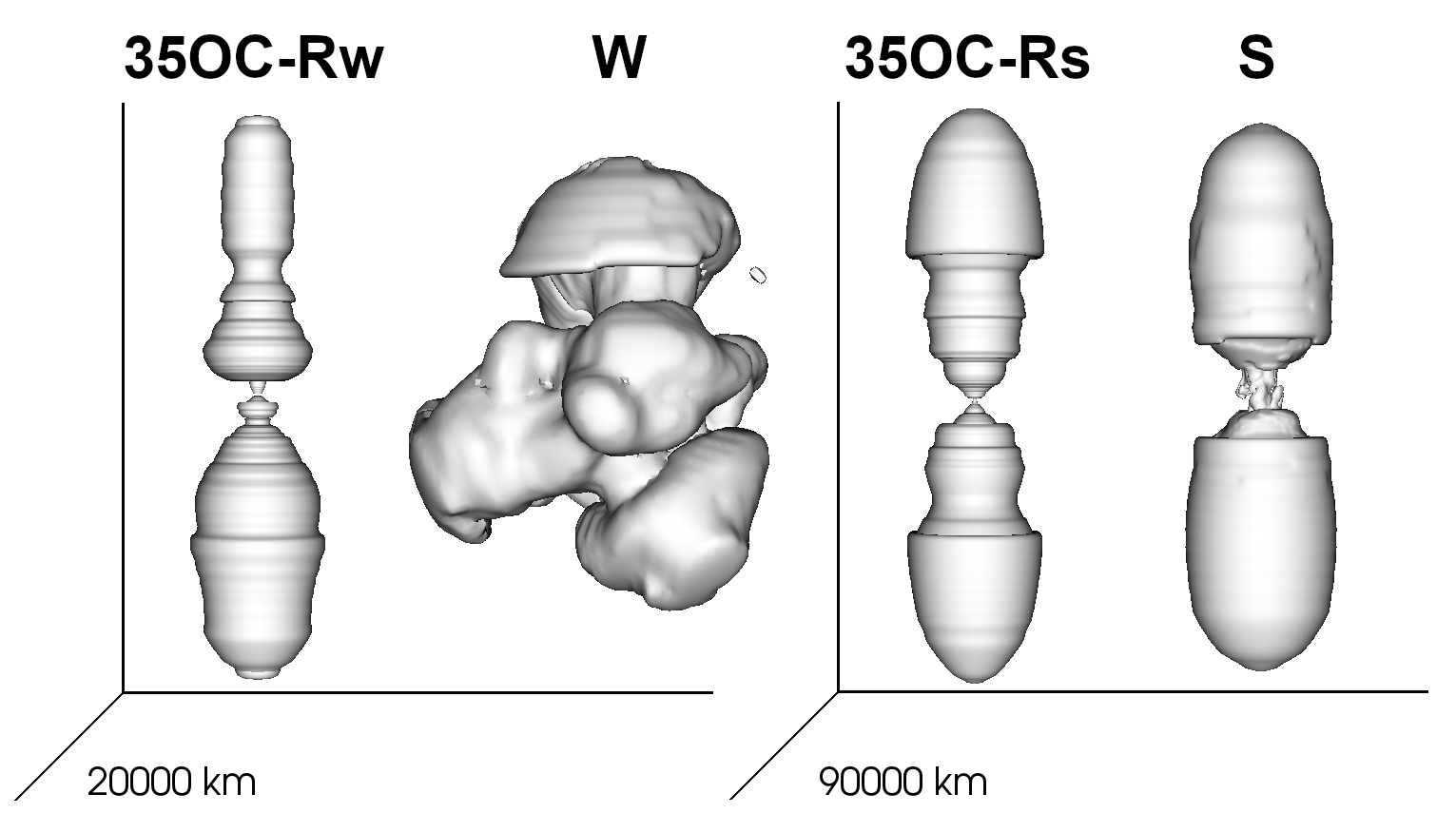}
 \caption{Surface containing the unbound ejecta of models 35OC-Rw, 35OC-Rs (2D axisymmetric), W, and S (full 3D). The models are shown at $0.46\,\mathrm{s}$, $1.45\,\mathrm{s}$, $1.13\,\mathrm{s}$, and $1.17\,\mathrm{s}$ post bounce, respectively. These times were chosen to obtain similar outermost shock positions of 2D and 3D models. The 2D and 3D version of model W develop a different geometry, while the strongly magnetized model S is similar to its 2D counterpart.}
\label{fig:comp_2D_3D}
\end{figure}

\begin{figure*}
 \includegraphics[width=0.5\columnwidth]{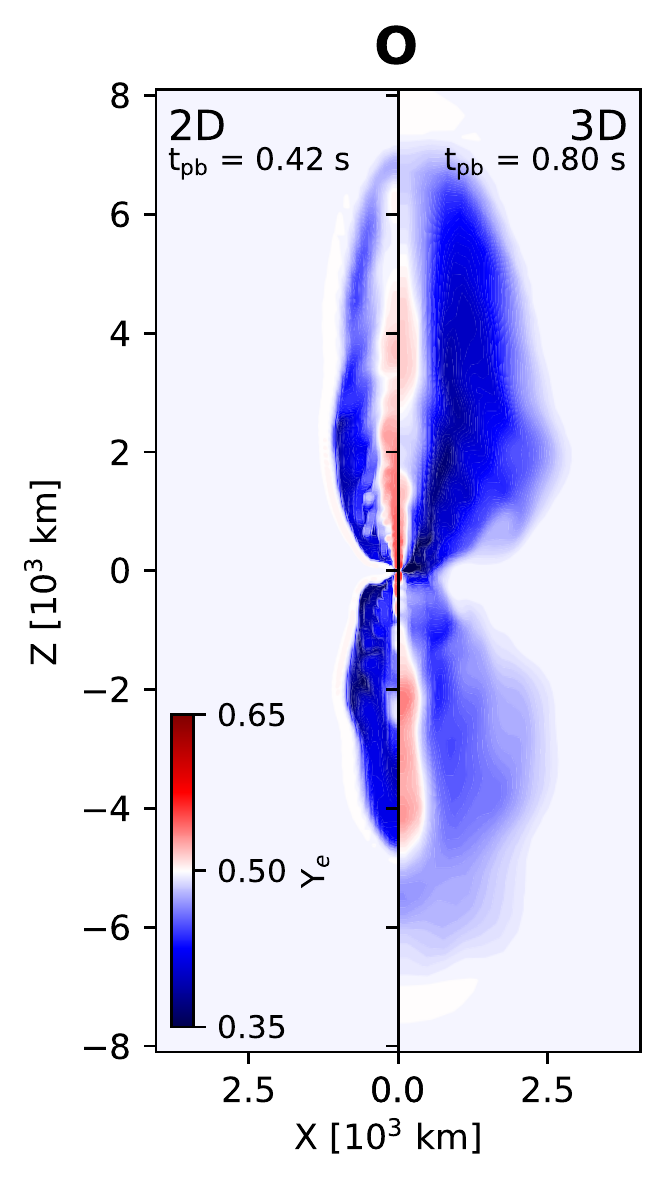}%
 \includegraphics[width=0.5\columnwidth]{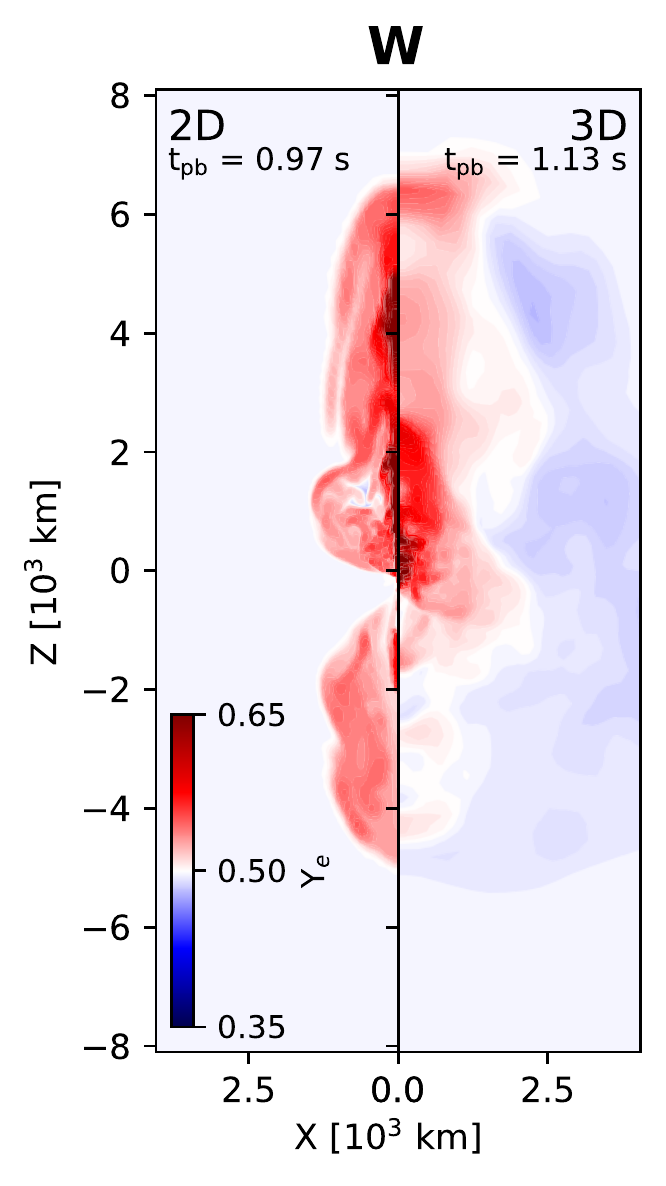}%
  \includegraphics[width=0.5\columnwidth]{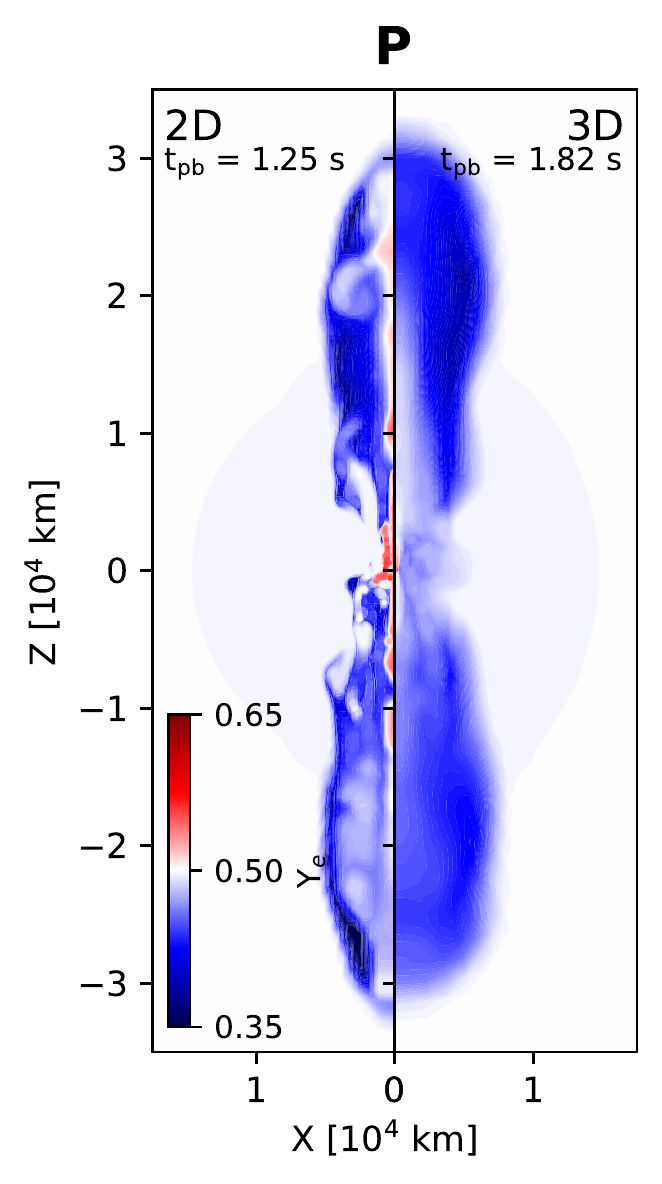}%
  \includegraphics[width=0.5\columnwidth]{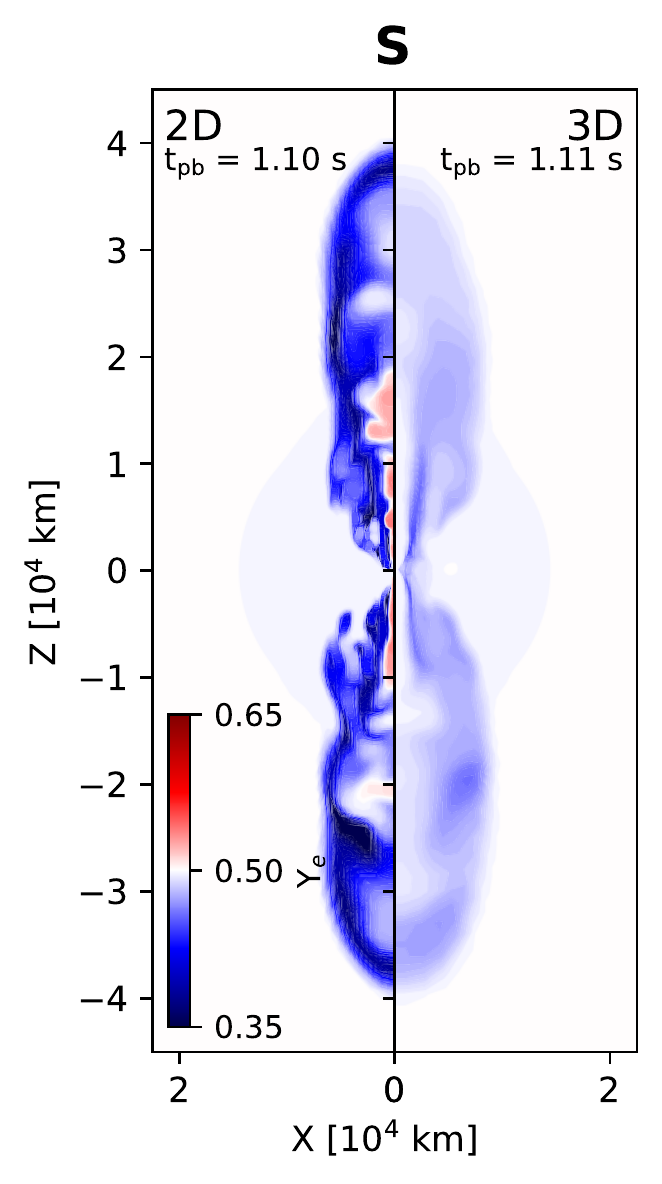}%
 \caption{Comparison of the electron fraction between 2D and 3D models. The left-hand panels show a snapshot of the 2D models, the right-hand panels an average of the electron fraction over all $\phi$ angles in the 3D models. The time of each of the snapshots is selected so that the 2D and 3D versions of the same model reach similar maximum shock radii. We note that an average in the case of the 3D model will blur the minimum and maximum electron fractions.}
\label{fig:2d_3d_hydro_ye}
\end{figure*}

Within our models, we observe general features in between 2D and 3D models. For instance, 2D models hosts a much more collimated jet cavity compared to 3D models (Fig.~\ref{fig:comp_2D_3D} and~\ref{fig:2d_3d_hydro_ye}). The difference is larger for a weaker magnetization and becomes less for stronger magnetizations. Connected to this, the maximum shock radius develops, as a tendency, faster in 2D (Fig.~\ref{fig:ej_mass_temp}). The ejected mass, however, is lower at times with similar maximum shock radii. Additionally, 2D models seem to produce conditions for a more proton-rich jet. When analysing this phenomena in more detail, we discovered a potential correlation between the shape of the PNS and the proton-rich jet. A more oblate PNS leads to an increased neutrino flux on the rotational axis and, as a consequence, to larger electron fractions. Even though not impossible in full 3D models, this effect is more common and amplified within 2D axisymmetric models (Fig.~\ref{fig:2d_3d_hydro_ye}). 

The weakly magnetized model W explodes more spherical in 3D compared to its 2D counterpart (Fig.~\ref{fig:comp_2D_3D}). Most strikingly, compared to the 3D model, a lower electron fraction component ($Y_e<0.5$) is present. We note that this component is not located in the jet that is in both cases (2D and 3D) proton-rich (see Fig.~\ref{fig:2d_3d_hydro_ye} for a snapshot at similar maximum explosion radii). From a nucleosynthesis perspective, the differences between 2D and 3D are small for elements around iron. However, there are significant differences in the amount of lighter elements ($A\lesssim 40$). The lower yields in 3D are related to two reasons. The first is the shorter evolution time of the 3D model, ending at a final time $t_{\mathrm{f, pb}} = 1.13 \, \mathrm{s}$ compared to the 2D version ($t_{\mathrm{f, pb}} = 2.5 \, \mathrm{s}$), i.e., while the ejection of these elements is still ongoing. A second possible reason are different conditions for explosive nucleosynthesis owing to the different shapes of the shock \citepalias[left-hand panel of Fig.~\ref{fig:comp_2D_3D}, see also][]{Obergaulinger2021}.

Heavier elements ($A>100$) are exclusively synthesized in the 2D model 35OC-Rw \citep[Fig.~\ref{fig:2d_3d_nucleosynth}, see also][]{Reichert2021a}, but not the 3D model W. As discussed in \citealt[][]{Reichert2021a} and \citetalias[][]{Aloy2021}, the synthesis of these nuclei is related to a change in the shape of the PNS due to angular momentum transport of the magnetic field at very late times ($\sim 2\,\mathrm{s}$). The simulation of the 3D model (W) only reached a much earlier time. Although a similar effect cannot be excluded after the end of the simulations, the evolutionary path to it are too specific to deem it very likely. 

Similarly to the case of models W and 35OC-Rw, the shorter simulation time of model O causes an underproduction of lighter nuclei compared to the axisymmetric version 35OC-RO. In 3D, lower neutrino luminosities and shorter ejection time-scales lead to more neutron-rich ejecta containing a component ejected early on that has an electron fraction $Y_e \approx 0.4$ when dropping out of NSE at $T = 7\,\mathrm{GK}$.

The strongly magnetized models 35OC-Rp3 and P evolve fairly similarly to each other. There are slight differences and the 2D model hosts more extreme (low and high) electron fractions. Also this model shows a slightly larger collimation in 2D compared to 3D (Fig.~\ref{fig:2d_3d_hydro_ye}). In the southern hemisphere, the 3D model develops a significantly wider beam of proton-rich matter, nearly absent in the 2D model. The combined effects of resolution and enforced axial symmetry drive a more intermittent jet beam in 2D than in 3D. This intermittency results from the strong pinching that the toroidal field drives in axial symmetry. However, the nucleosynthetic fingerprint is qualitatively almost identical. In both models, heavier elements up to the second r-process peak are synthesized by slightly neutron-rich material with $Y_e \sim 0.4$. Both models reach maximum entropies of the order of $\gtrsim 100 \,\mathrm{k_B}$ for a small portion of matter ($\sim 10^{-4}\,\mathrm{M_\odot}$). 

\begin{figure}
 \includegraphics[width=\columnwidth]{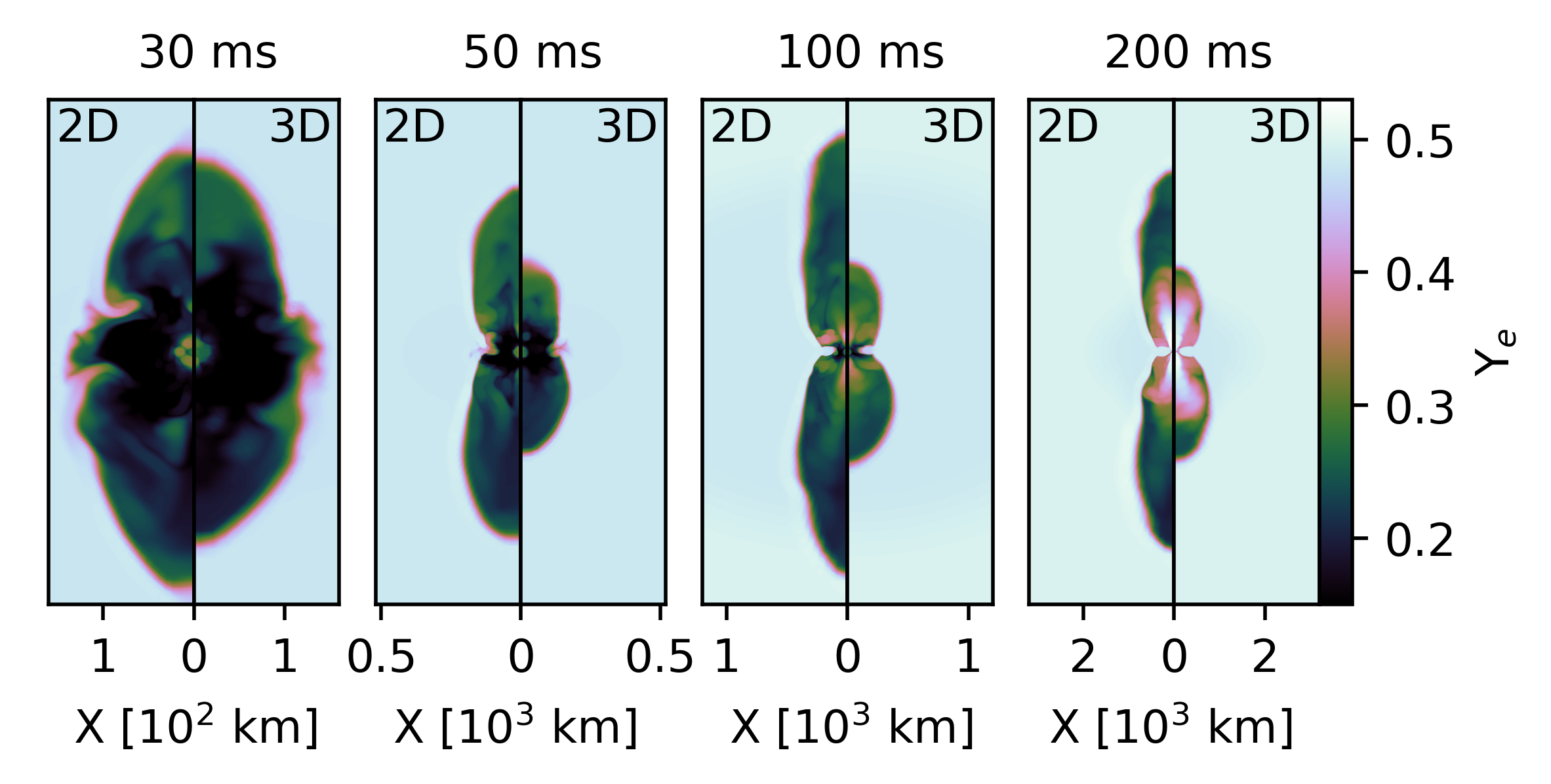}
 \caption{Electron fraction of model 35OC-Rs (left-hand panels) and of a vertical slice of model S (right-hand panels) at different times after bounce. Model 35OC-Rs expands faster and therefore maintains more neutron-rich conditions compared to model S.}
 \label{fig:2d_3d_expansion}
\end{figure}

Among the three versions of the strongest magnetized model, the 3D model S and the 2D model 35OC-Rs from \citet{ObergaulingerAloy2017}, \citetalias{Aloy2021}, and \citetalias{Obergaulinger2021} agree well for elements up to the second r-process peak, while abundances of heavier elements are lower in the 3D model. In both models, these elements are located in the jets, whose dynamics does not differ much between 2D and 3D. In particular, we do not observe the development of non-axisymmetric kink instabilities which have the potential to reduce the neutron richness of the jet \citep{Moesta2014,Kuroda2020}. 
However, we also observe a slower shock expansion in 3D compared to 2D (Fig.~\ref{fig:2d_3d_expansion}, see also \citealt{Obergaulinger2020}). The faster expansion velocity will drive matter faster away from the (anti-)neutrino emitting central PNS. Typical neutrino properties in our simulation are luminosities of \mbox{$L_{\nu_e}\eqsim 5\cdot 10^{52} \, \mathrm{erg\,s^{-1}}$}, \mbox{$L_{\bar{\nu}_e}\eqsim 6\cdot 10^{52}\, \mathrm{erg\,s^{-1}}$} and mean energies of $E_{\nu_e} \eqsim 10 \, \mathrm{MeV}$ and $E_{\bar{\nu}_e} \eqsim 13 \, \mathrm{MeV}$. When radiating matter with these neutrino properties until an equilibrium is reached, an electron fraction of $Y_e \sim 0.52$ is obtained \citep[e.g.,][]{Qian1996,Arcones2013,Martin2018,Miller2020,Just2022}. Even though matter will not come into this equilibrium state, it gives an idea that it can be beneficial for neutron-rich conditions to avoid neutrinos. The neutrino flux decrease with the radius squared. Matter that expands fast will therefore be less irradiated by neutrinos than slower expanding matter. This ultimately leads to a slightly lower electron fractions and an r-process in 2D, while the 3D model S is slightly more proton-rich and is therefore not able to host a strong r-process. With a minimum electron fraction of $Y_e\sim 0.23$ (at $7\,\mathrm{GK}$) the 3D model is however at the edge of synthesizing also a larger amount of third r-process peak elements. For this, a minimum of $Y_e\sim0.20$ would be necessary. There are  differences in abundances for $A\gtrsim70$  between the 2D models 35OC-Rs$_\mathrm{N}$ and our previous model  in \citet[][]{Reichert2021a} because of the longer evolution/simulation time. In model 35OC-Rs$_\mathrm{N}$, there is an ejection of very neutron-rich matter at the end of the simulation due to a change of the PNS morphology (see also Fig.~\ref{fig:last100ms}). This late-time effect did not develop at the final time of our previous model and will be discussed in more detail in the next section (Sect.~\ref{sct:r-process}). However, also without this effect, the 2D model shows a much lower electron fraction compared to the respective 3D counterpart (Fig.~\ref{fig:2d_3d_hydro_ye}).

Summarizing, we find no universal tendency how the dimensionality impacts the nucleosynthesis. While the 3D version of model O is more neutron-rich and synthesizes heavier elements, model S is less neutron-rich and lacks nuclei of the third r-process peak. We note that 2D models will stay an important tool to trace long-time effects on the nuclear yields (see the case of model 35OC-Rw, \citealt[][]{Reichert2021a}) as 3D models are at the moment still computational too expensive to calculate for the necessary long times. 

\subsection{The synthesis of heavy elements}
\label{sct:r-process}
\begin{figure}
 \includegraphics[width=\columnwidth]{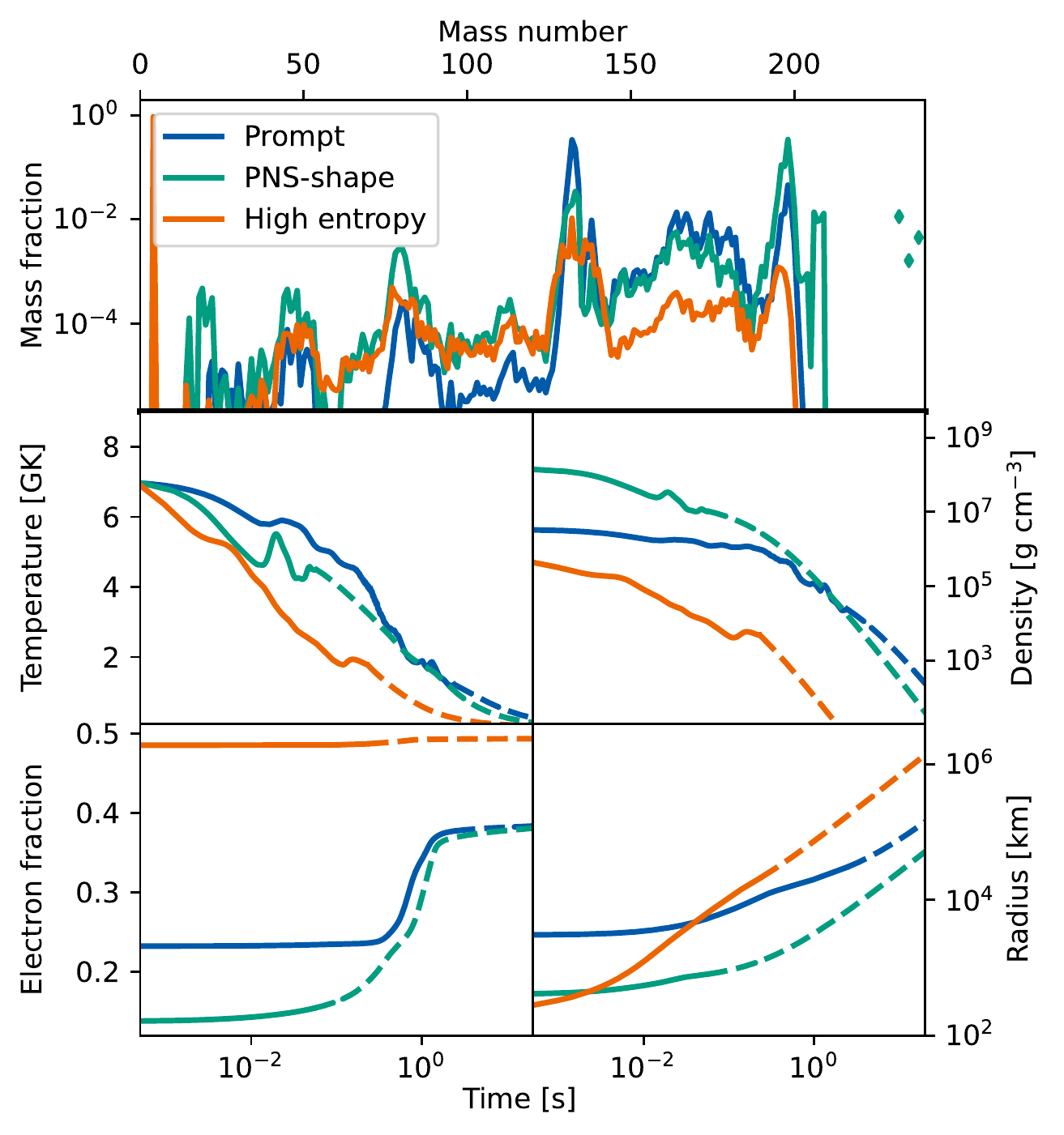}
 \caption{Mass fractions and several hydrodynamic quantities of three  representative tracer particles from simulations 35OC-Rs$_\mathrm{N}$ (prompt and PNS-shape) as well as 35OC-Rp3 (high entropy). The time is relative to the start of the nucleosynthesis calculation at $7\,\mathrm{GK}$, which corresponds to $0\,\mathrm{s}$ (prompt), $2.48\,\mathrm{s}$ (PNS-shape), and $1.58\,\mathrm{s}$ (high entropy) after bounce. All tracer particles host the conditions for a successful r-process. The corresponding neutron-to-seed ratios at $3\,\mathrm{GK}$ are $48$ (Prompt), $90$ (PNS-shape), and $73$ (high entropy). Dashed lines indicate extrapolations of the respective quantity, assuming adiabatic (i.e., constant entropy) expansion. The actinides ($A>220$) are only synthesized within the late ejection that is caused by a reconfiguration of the PNS (cyan diamonds).}
 \label{fig:tracer_case}
\end{figure}

Four of our models (35OC-Rp3, P, 35OC-Rs$_\mathrm{N}$, and S) synthesize nuclei with nuclear masses beyond the second r-process peak ($A \gtrsim 130$). All these models have the same stellar evolution pre-supernova star in common, but with a larger poloidal magnetic field strength than in the original 35OC-RO/O models. Hence they all yield powerful magneto-rotational explosions. The r-process occurs in three environments: (i) prompt ejection of matter directly after core-bounce, (ii) late ejection due to changes of the shape of the PNS, and (iii) ejection of high entropy material in the jet (for a brief discussion about proton-rich ejecta see Appendix~\ref{sct:p-rich}). As we describe in the following, the conditions and final abundances (Fig.~\ref{fig:tracer_case}) for the three cases can be very different. Indeed, not all these processes may develop in all models. Specially, process ii) may be quite stochastic and very much dependent on the evolutionary details of the PNS. Interestingly, the key for mechanism i) to yield heavy elements is that matter bouncing from the inner stellar core is not halted due to a prompt shock stagnation. This is easier to achieve if a very early magneto-rotational explosion drives a quick supernova shock expansion. If the shock stagnates relatively close to the PNS (where the temperatures still allow for NSE conditions), the intense (anti-)neutrino flux raises $Y_e$ and hampers the production of r-process elements from matter bouncing from the inner stellar core (see below). 

A critical factor for a successful r-process is the number of free neutrons available per seed nucleus ($A>4$), which can capture the neutrons. The corresponding ratio is known as neutron-to-seed ratio defined as
\begin{equation}
    r = \frac{Y_n}{\sum _{A>4}Y_{A}},
\end{equation}
where $Y_n$ is the abundance of neutrons and $Y_A$ the abundance of nuclei with mass number $A$. If matter is ejected promptly after bounce, there is short time for neutrinos to modify $Y_e$ and, thus, it stays very low, e.g., in model 35OC-Rs, we find $Y_e\sim 0.2 - 0.3$ \citep[][]{Reichert2021a}. Once NSE conditions do not hold anymore, the low $Y_e$ leads to a moderate neutron-to-seed ratio of about $\sim 50$ (at $3\, \mathrm{GK}$). The resulting r-process synthesizes matter up to the third r-process peak. However, we do not find significant production of the heaviest nuclei ($A>230$), the actinides (Fig.~\ref{fig:tracer_case}, c.f., \citealt[][]{Reichert2021a}).

\begin{figure}
 \includegraphics[width=\columnwidth]{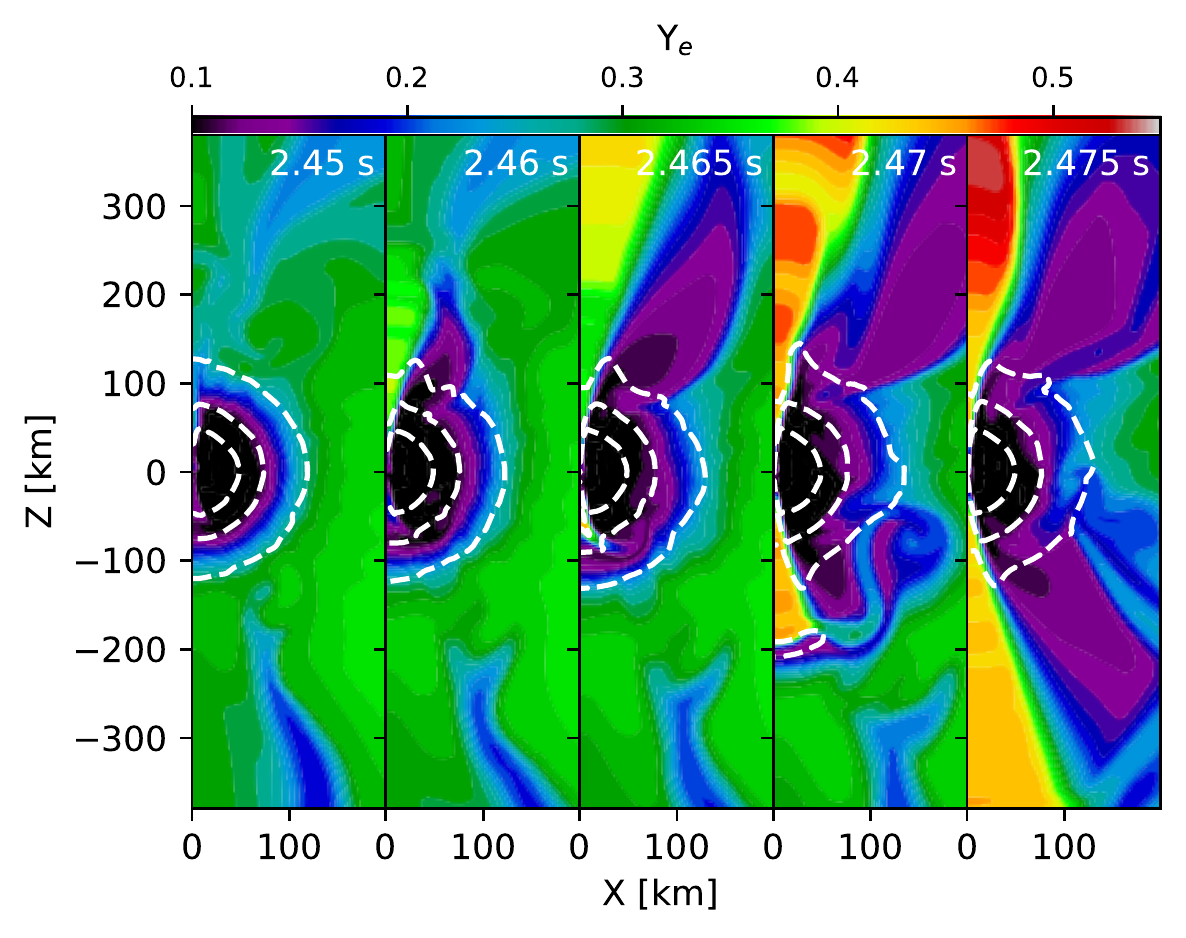}
 \caption{Colour coded is the electron fraction in the centre of model \mbox{35OC-Rs$_\mathrm{N}$} at four different times (post-bounce). The white, dashed lines are density contours at $10^9$, $10^{10}$, and $10^{11}\,\mathrm{g\,cm^{-3}}$. The reconfiguration of the PNS shape leads to the ejecta of neutron-rich material on short time-scales.}
 \label{fig:thorus_low_ye}
\end{figure}

At later times ($t_\mathrm{pb}\gtrsim 2.45\,\mathrm{s}$), even more neutron-rich matter gets ejected in model 35OC-Rs$_\mathrm{N}$. Its origin is a drastic change in the PNS shape, and it corresponds to the mechanism ii) mentioned at the beginning of the section. Around $t_\mathrm{pb}=2.5\,\mathrm{s}$ the magnetic pressure and angular-momentum redistribution transform the PNS into a torus like object with an off-centre density maximum \citepalias[see][and Fig.~\ref{fig:thorus_low_ye}]{Aloy2021}. The new configuration has a lower mass than the PNS. The mass difference of a fraction of a solar mass of neutron-rich gas is ejected within a few milliseconds. We note that this process requires special conditions. Whether it can occur also in 3D remains to be investigated. So far, the 3D counterpart of model 35OC-Rs$_\mathrm{N}$ (model S) has not been evolved long enough to corroborate the aforementioned mechanism.

Since the ejection of this matter occurs at the very end of the simulations, our results for the nucleosynthesis may depend to a significant degree on the assumptions made in the extrapolation of the hydrodynamic quantities (dashed lines in Fig.~\ref{fig:tracer_case}, see also \citealt[][]{Harris2017}), in particular the expansion velocity. We explored the impact on the final yields caused by the choice of velocities and always find a production of elements up to the third r-process peak, with only slight variations of the second to third r-process peak ratio with slightly lower masses of third peak elements for slower expansion velocities. We note that also the neutrino transport (M1) may cause uncertainties in the dynamics of this neutron-rich outflow as it is not able to treat the crossing beams at the rotational axis in an optically thin region (similar to the problems that occur when using M1 in an accretion disc hydrodynamic model; see \cite{Chan2021} for a recent discussion of the limitations of the M1 method in such a situation.) Due to the high neutron-to-seed ratios of $\sim 100$, this part of matter even reaches the heaviest synthesized nuclei, the actinides (cyan diamonds in the upper panel of Fig.~\ref{fig:tracer_case}). 

\begin{figure}
 \includegraphics[width=0.9\columnwidth]{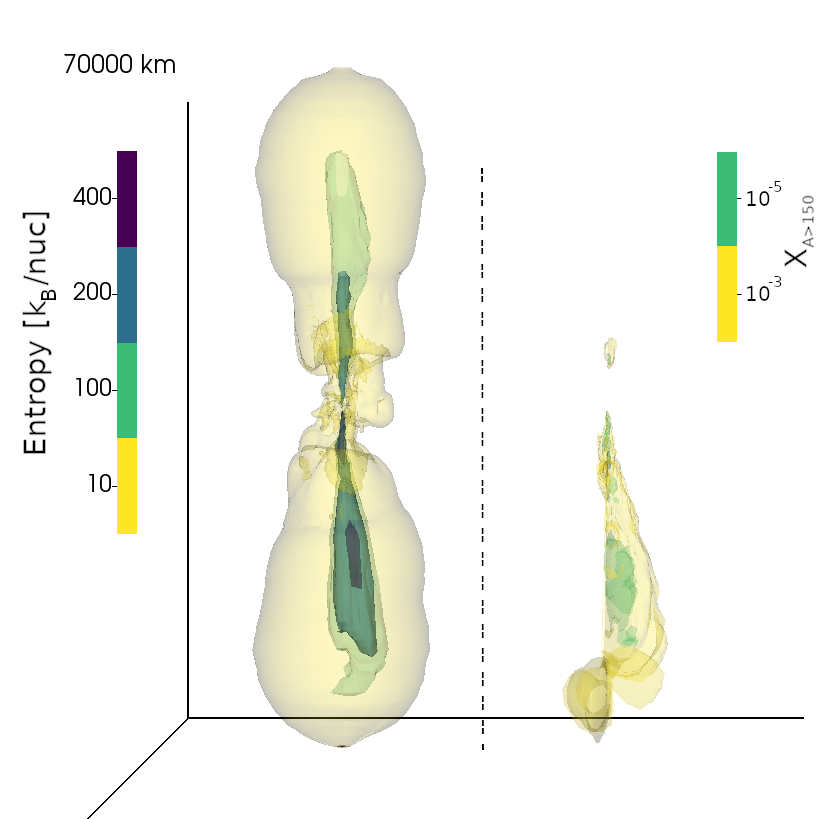} 
 \caption{Left: Contours of the entropy per baryon for model P at the end of the simulation ($t_\mathrm{pb}=1.8\,\mathrm{s}$). Right: Mass fractions of nuclei with $A>150$. In the right plot, the contours were smoothed with a Gaussian filter for improving the visualization. A clear correlation between heavier synthesized elements and high entropy regions is visible.}
 \label{fig:rp3_entr}
\end{figure}

Alternatively, high neutron-to-seed ratios can be obtained in less neutron-rich environments when the entropy is sufficiently high (mechanism iii) mentioned at the beginning of the section). In NSE, high entropy material has an excess of nucleons (protons and neutrons) compared to heavier elements. Once the temperature drops during the expansion, NSE conditions eventually break down with a high neutron-to-seed ratio \citep[][]{Meyer1994,Woosley1994,Wheeler1998,Freiburghaus1999b,Meyer2002,Thielemann2017}. In the jets of models 35OC-Rp3 and P we find $S\gtrsim 100 \, k_\mathrm{B}/\mathrm{nuc}$ (Fig.~\ref{fig:rp3_entr}). This high entropy leads to large neutron-to-seed ratios of $\sim 70$ and elements up to the third r-process peak can be synthesized (Fig.~\ref{fig:tracer_case}). The yield is thereby dominated by heavy nuclei ($A\gtrsim120$) and, in contrast to the other channels of r-process nucleosynthesis, by a high mass fraction of free protons (visible at $A=1$, upper panel of Fig.~\ref{fig:tracer_case}). We stress that this channel is only able to synthesis heavy elements for slightly neutron-rich conditions. If the electron fraction is $Y_e\ge0.5$, no heavy elements will be synthesized. The necessary high entropies are not common among our models. They can develop in the inner regions of the jet (the jet beam) when the magnetic structure of the jet is supported by a core of uniform electric current with radius $\tilde{\omega}_{\rm m}$ \citep[the magnetization radius;][]{Lind_1989ApJ...344...89,Komissarov_1999MNRAS.308.1069,Leismann_2005A&A...436..503}, and the jet beam is in approximate transverse hydromagnetic equilibrium. Under these conditions, the toroidal magnetic field grows roughly linearly with distance to the rotational axis until $\tilde{\omega}=\tilde{\omega}_{\rm m}$, and then falls as $1/\tilde{\omega}$, producing strong pinching \citep[see e.g.,][]{Leismann_2005A&A...436..503}.  The magnetic pinching yields a high pressure in a region which has lower density than the layer surrounding it, and hence rises significantly the entropy per baryon. The effect is larger when the average beam magnetization grows  and the toroidal magnetic field is stronger than the poloidal one. This explains that this channel does not develop in models S, 35OC-Rs or 35OC-Rs$_\mathrm{N}$, since in these models the poloidal field is close to equipartition with the toroidal field, and the pinching is not as effective.
The amount of r-process material ejected via this mechanism is only a small fraction of the total ejecta and adds up to a total of $10^{-8}$ to $10^{-7}\, \mathrm{M_\odot}$ for models 35OC-Rp3 and P (Fig.~\ref{fig:final_masses}). While the masses of individual packages of r-process matter ejected by this mechanism may be small, unsteady outflows can create the necessary conditions repeatedly, each episode presumably adding to the total r-process yields. In the case of model P, the total extrapolated value (see sect.~\ref{sect:upper_limit}) reaches by the end of the simulation up to about a few $10^{-6}\,\mathrm{M_\odot}$. 

\subsection{Radioactive isotopes}
\label{sct:radioactive_isotopes}
Radioactive isotopes powering  to a large degree the electromagnetic emission of SNe play a crucial role in observations related to SN. For example, to classify a SN as HNe one usually uses the explosion energy and the amount of synthesized $^{56}$Ni \citep[][]{Nomoto2006,Nomoto2013}. In addition to this radioactive isotope, we also investigate the synthesis of $^{44}$Ti whose decay is directly observable in young supernova remnants \citep[][]{INTEGRAL_Ti2012,Seitenzahl2014}. Furthermore, the radioactive elements $^{26}$Al and $^{60}$Fe can be detected in the interstellar medium \citep[see e.g.,][for recent reviews]{Diehl2021,Diehl2021b} or in sediments of the ocean crust on Earth \citep[e.g.,][]{Wallner2016,Ludwig2016}. In Tab.~\ref{tab:radioactive}, we give the yields of these radioactive elements at the end of the simulation and our estimated extrapolated values according to Sect.~\ref{sect:upper_limit}. 

\begin{figure*}
 \includegraphics[width=2\columnwidth]{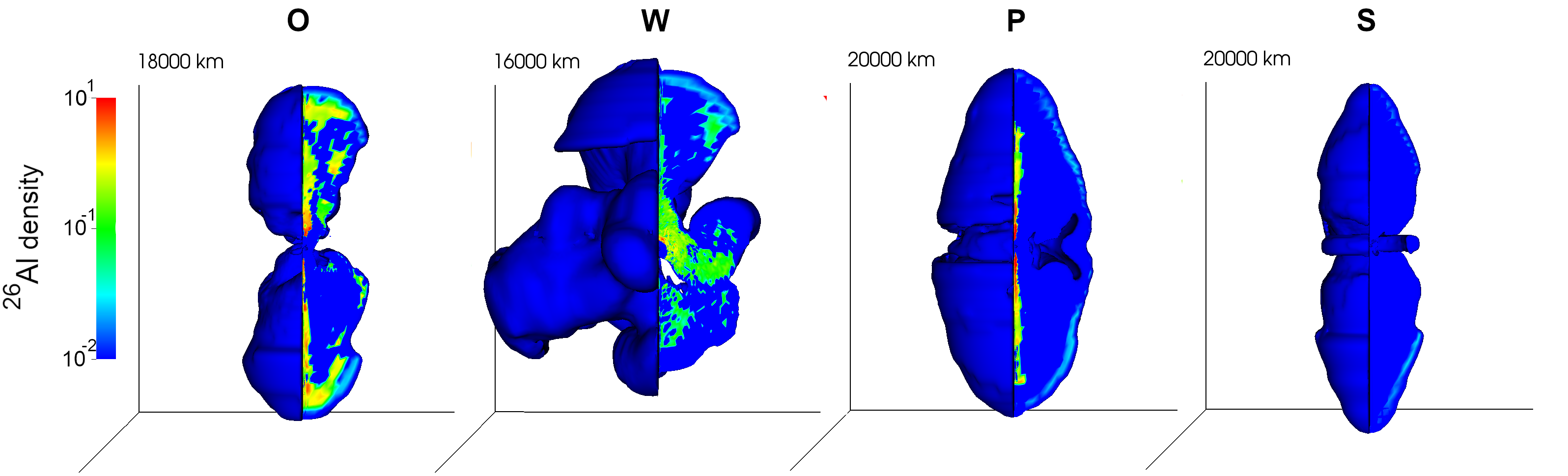}\\
 \includegraphics[width=2\columnwidth]{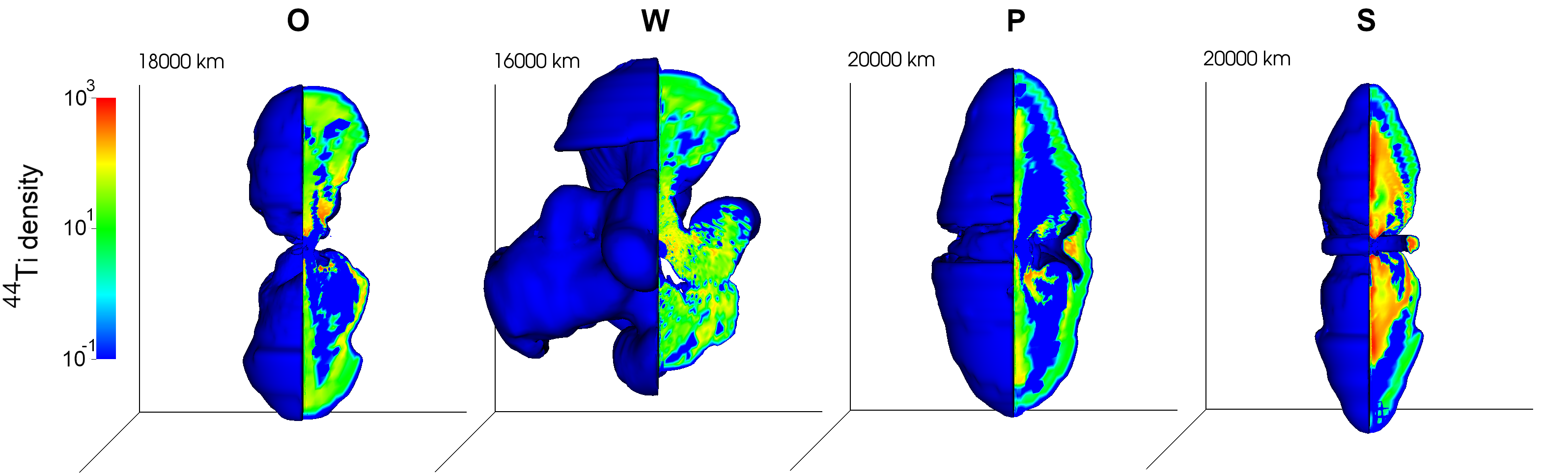}\\
 \includegraphics[width=2\columnwidth]{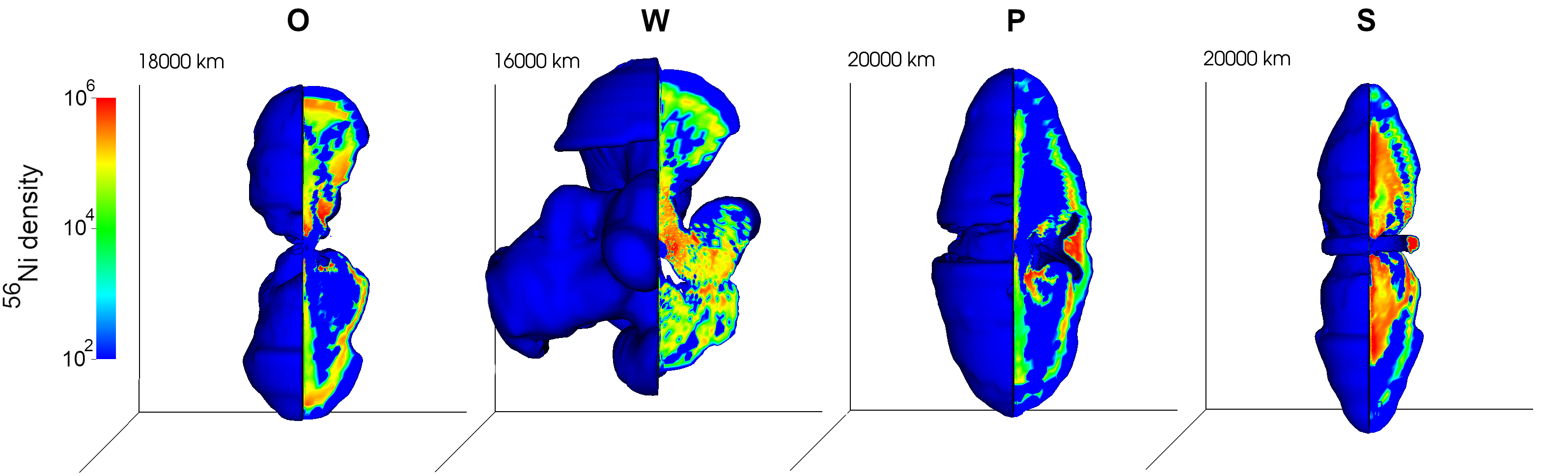}\\
 \includegraphics[width=2\columnwidth]{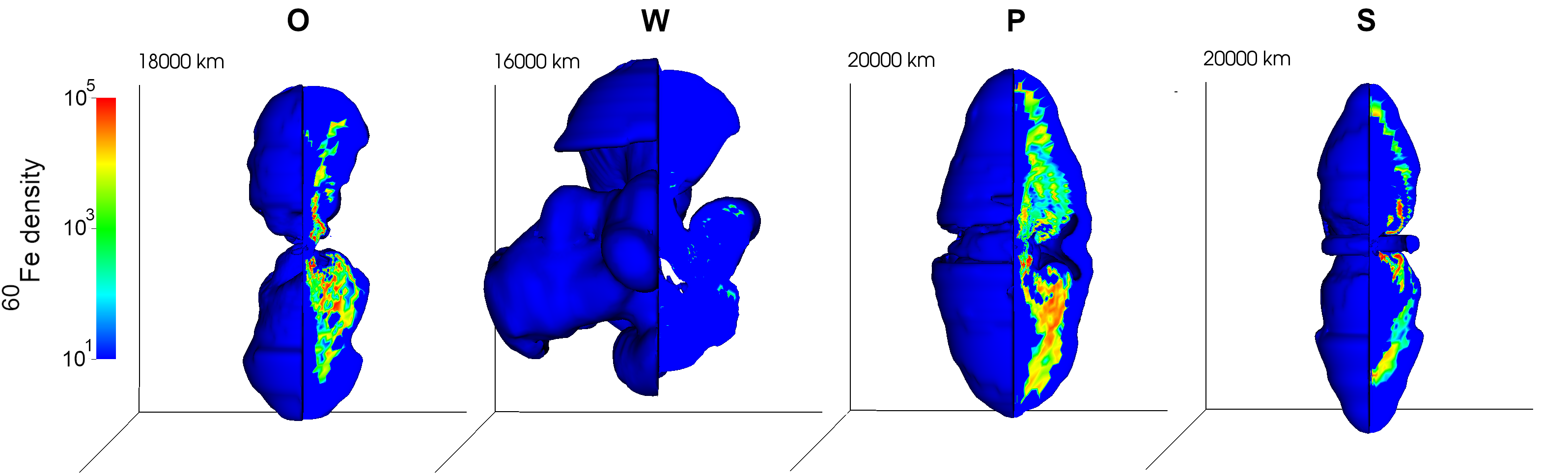}
 \caption{Spatial distribution of the unstable isotopes $^{26}$Al (first row), $^{44}$Ti (second row), $^{56}$Ni (third row), and $^{60}$Fe (last row) within unbound matter. The blue surface encloses  all unbound matter in each model up to the represented time (i.e. matter unbound up to, approximately the SN shock). The northern and southern octants are cut and a surface perpendicular to the equatorial plane is displayed to visualize the nuclear yields spatial distribution. The mass fraction of these isotopes was taken at one tenth of their half life, but the position of the tracers was taken at the simulation time  $t_\mathrm{pb}=0.80\,\mathrm{s}$, $1.13\,\mathrm{s}$, $0.63\,\mathrm{s}$, and $0.40\,\mathrm{s}$ for models O, W, P, and S, respectively.}
 \label{fig:unstables_spatial}
\end{figure*}

\begin{figure}
 \includegraphics[width=\columnwidth]{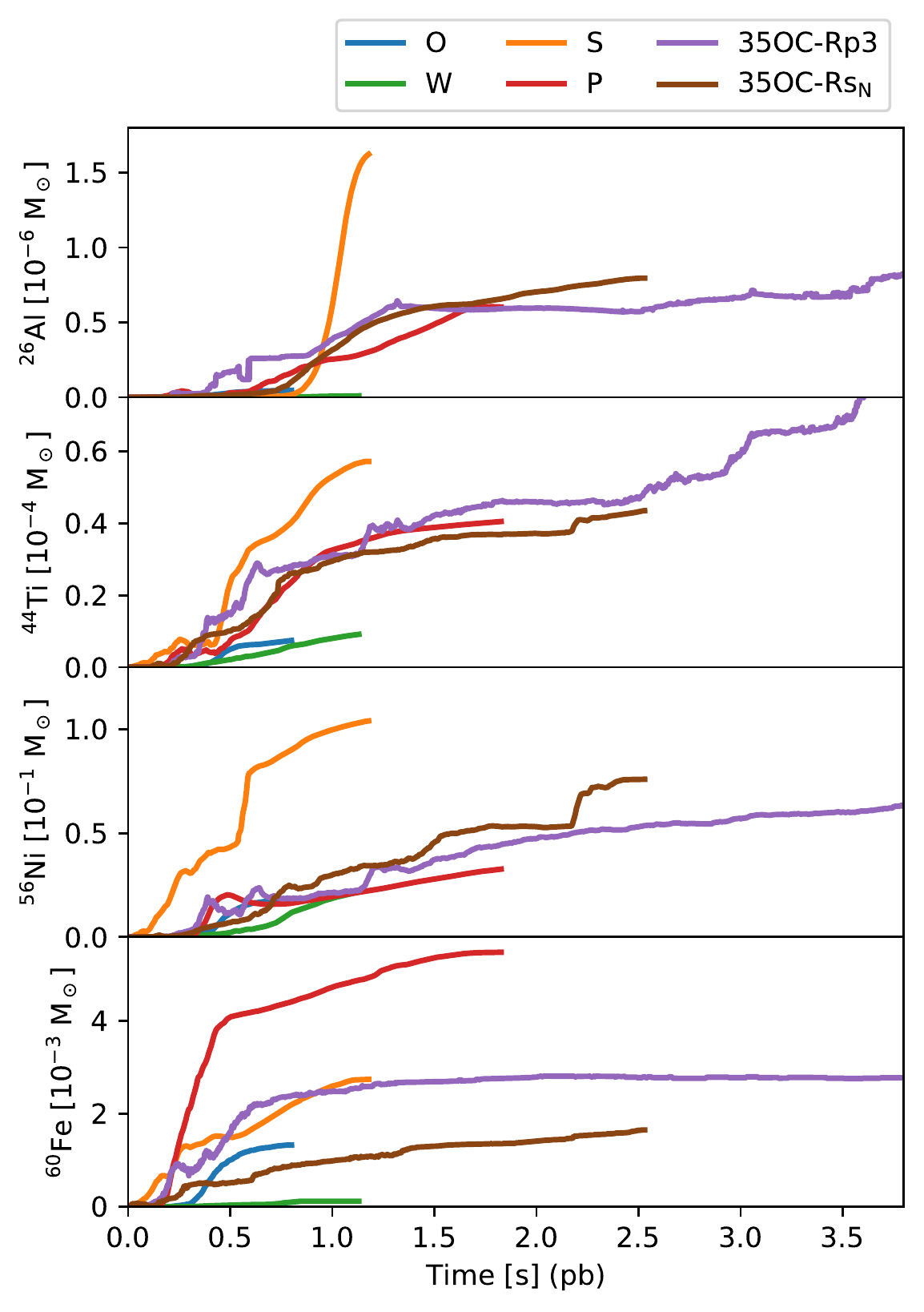}
 \caption{Mass of ejected unstable isotopes  $^{26}$Al, $^{44}$Ti, $^{56}$Ni, and $^{60}$Fe versus time after bounce. The ejecta mass of all unstable nuclei is still growing.}
 \label{fig:unstables_vs_time}
\end{figure}

The main contribution of a CC-SN to $^{26}$Al originates from hydrostatic burning in the hydrogen layer of the progenitor star before the explosion. For this production channel, the amount of $^{26}$Al may reach masses of the order of $10^{-4}$\,M$_\odot$ \citep{Karakas2010,Doherty2014,Brinkman2019,Diehl2021}. The data available for our progenitor model (35OC; \citealt{woosley-heger2006}) contain only a reduced set of abundances and does not include information about $^{26}$Al (Fig.~\ref{fig:progenitor}). Therefore, our calculated $^{26}$Al yields only reflect the contribution from explosive burning. Until the end of the simulation, none of our models reaches significant $^{26}\mathrm{Al}$ masses compared to what usually is produced during hydrostatic burning. The production of $^{26}$Al in explosive environments requires symmetric or proton-rich conditions with moderate entropy \citep[e.g.,][]{Magkotsios2011} that we mainly find in the proton-rich jet (Fig.~\ref{fig:unstables_spatial}). However, $^{26}$Al can also be synthesized during explosive burning with low peak temperatures ($2\,\mathrm{GK} \lesssim T_\mathrm{peak}\lesssim 3\,\mathrm{GK}$). The sudden ejection of $^{26}$Al at $t_\mathrm{pb}\sim0.9\,$s in model S (Fig.\ref{fig:unstables_vs_time}) is caused by explosive burning in the Oxygen-Neon layer. There, $^{26}$Al can be produced via the reaction $^{23}\mathrm{Mg}(\alpha,\mathrm{p})^{26}\mathrm{Al}$ and to smaller extends also via $^{25}\mathrm{Mg}(\mathrm{p},\gamma)^{26}\mathrm{Al}$.

$^{60}$Fe is thought to originate from neutron captures in the convective envelope during the hydrostatic burning in the progenitor, which is later ejected during the explosion, or in explosive burning inside the He- or C- shell \citep[e.g.,][]{Limongi2006,Limongi2012,woosley-heger2006,Maeder2012,Thielemann2018,Jones2019,Lawson2022}. Pre-explosion winds of the AGB stars only play a minor role for the total ejecta mass \citep{Brinkman2021,Diehl2021}. The total ejected yields of a canonical SN may reach up to $10^{-4}\,\mathrm{M_\odot}$ \citep[][]{Diehl2021}. Similar to $^{26}$Al, we only trace $^{60}$Fe that is synthesized in explosive conditions via neutron captures. Due to the neutron-rich conditions in our models, they eject a considerable amount of this radioactive element of the order of a few $10^{-3}$ M$_\odot$ \citep[c.f. to 
$\sim 10^{-5}\,\mathrm{M_\odot}$ or lower in the innermost ejecta of classical CC-SNe][]{Wanajo2018}. Because the synthesized abundances of $^{60}$Fe within our simulation are an order of magnitude higher than the ones expected by hydrostatic burning, we conclude that nucleosynthesis during explosive burning dominates the total yields for all of our models except possibly model W. Such high yields offer the potential of detecting $^{60}$Fe not only as a diffuse isotope in the ISM, but future telescopes could also detect MR-SNe as point-like sources for a galactic event \citep{Woosley1997,Diehl2021}. Remarkably, even not extremely magnetized models such as model O synthesize a considerable amount of $^{60}$Fe ($\simeq 10^{-3}$ M$_\odot$). Under NSE conditions, the synthesis of $^{60}$Fe requires a very specific neutron richness, which is approximately the proton to baryon ratio of the nucleus itself ($Y_e \approx 26/60 \approx 0.43$). Any positive or negative deviation from this ratio leads to a reduction of the $^{60}$Fe yields \citep[see also][]{Wanajo2018,Jones2019}. This is especially visible when comparing the electron fraction of model P (third panel, top row in Fig.~\ref{fig:entropy_3D}) with the corresponding $^{60}$Fe density (third panel, bottom row in Fig.~\ref{fig:unstables_spatial}). There are neutron-rich conditions in both hemispheres, north and south. However, in the northern hemisphere there are too many neutrons to synthesize $^{60}$Fe and most of the $^{60}$Fe is synthesized in the less neutron-rich southern hemisphere. The $^{60}$Fe/$^{56}$Fe ratio exceeds $10^{-2}$ for all models except for the more proton-rich model W.  This is $\sim 6$ orders of magnitude higher than the ratio in the early solar system \citep[][]{Trappitsch2018} or $\sim 3$ orders of magnitude higher than in the diffusive ISM background, which agrees with the predicted ratio of classical CC-SNe \citep[][]{Diehl2013,Sukhbold2016,Austin2017,Wang2020,Brinkman2021}.

In our models, we find total $^{44}$Ti ejecta between $\sim 10^{-5} - 10^{-4}$\,M$_\odot$ (for our lower limits) and even up to $5 \times 10^{-3}$\,M$_\odot$ for the extrapolated value of model 35OC-Rp3. The yields are in the range of reported values of SN1987A ($5.5 \times 10 ^{-5} \,\mathrm{M}_\odot$, \citealt{Seitenzahl2014}) and Cas A ($1.3 \times 10^{-4}\,\mathrm{M}_\odot$, \citealt{WangLi2016}). $^{44}$Ti is mostly produced in high entropy and symmetric or proton-rich conditions, and it is therefore located in the high entropy jet and at the shock front of the more magnetized models, while it is more uniformly spread within the ejecta of the least magnetized model W (Fig.~\ref{fig:unstables_spatial}).

\begin{figure}
 \includegraphics[width=\columnwidth]{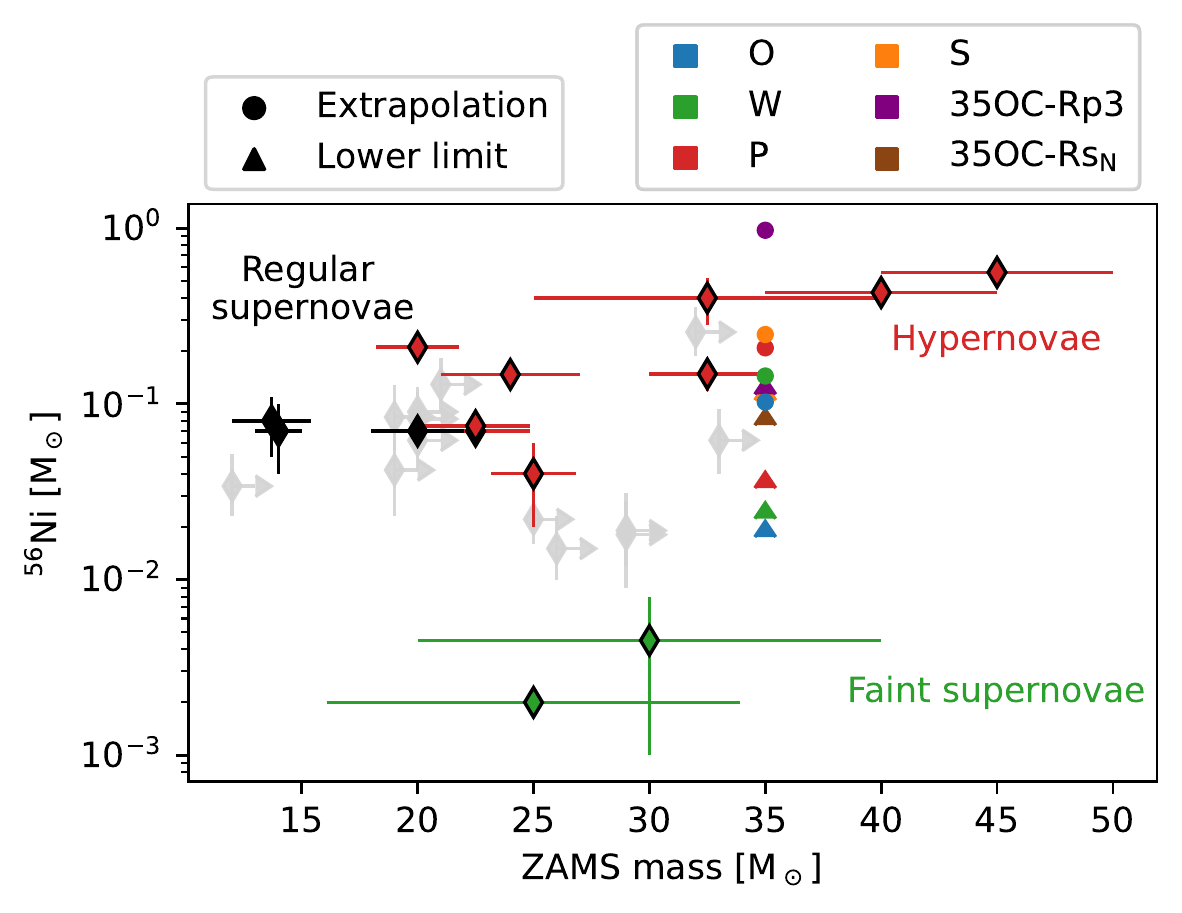}
 \caption{Estimated $^{56}$Ni masses of different supernovae versus their ZAMS mass. Hypernovae are indicated as red diamonds, faint supernovae as green diamonds, and regular supernovae as black diamonds. Data is taken from \citet[][]{Nomoto2013}. Light grey diamonds shows data by \citet[][]{Hamuy2003}. We assumed the ejected mass of the SNe given in \citet[][]{Hamuy2003} as lower limit for the ZAMS mass of the star. The uncertainty of our models is represented by the range between the extrapolated (circles) and the lower limits (triangles). The ejected $^{56}$Ni of our models is roughly compatible with observations of HNe. The plot is inspired by \citet{Nomoto2006,Nomoto2013}.}
 \label{fig:hne}
\end{figure}

Because of its half-life of $\sim6\,\mathrm{d}$ and that of its daughter isotope $^{56}$Co of $\sim77\,\mathrm{d}$, the decay of $^{56}$Ni contributes significantly to the lightcurve of SNe. Huge amounts of $^{56}$Ni (\mbox{$\sim 0.1 - 1 \, \mathrm{M}_\odot$} for a $35\, \mathrm{M}_\odot$ ZAMS mass progenitor like ours) may even produce HNe \citep[see Fig.~\ref{fig:hne} and][]{Nomoto2006,Nomoto2013}. 
Estimated $^{56}$Ni masses of SNe associated to long gamma-ray bursts are of comparable magnitude ($0.18\pm 0.01\,\mathrm{M}_\odot$; \citealt[][]{Izzo2019}). At the end of the respective simulation, only models 35OC-Rp3, 35OC-Rs$_\mathrm{N}$, and S reach comparable and large yields of $^{56}$Ni, namely $\sim 0.08 - 0.11\,$M$_\odot$ (see Tab.~\ref{tab:radioactive}). However, nucleosynthesis is still ongoing in all our models. In Tab.~\ref{tab:radioactive} we provide both the yield masses computed until the end of the neutrino-MHD evolution and their extrapolated values according to the methodology of sect.~\ref{sect:upper_limit}. Noteworthy, for some of the models  mentioned above (35OC-Rp3 and S), the extrapolated $^{56}$Ni is $\sim 0.25 - 1\,$M$_\odot$. Even the weakly magnetized models W and O may produce sufficient $^{56}$Ni to be possible HN candidates (Fig.~\ref{fig:hne}) according to our extrapolation method. The late-time synthesis of $^{56}$Ni may cease, whenever a BH forms, which occurs (or it is expected to happen) in the simulations of \citetalias{Aloy2021} of the 2D counterparts of models O (35OC-RO) and W (35OC-Rw). Whether or not it also happens in 3D requires longer simulation times than we could afford in the present models. In the case of black-hole formation, the final yields would be much closer to the values computed without extrapolating the neutrino-MHD results (see Tab.~\ref{tab:radioactive}). The low mass yields are compatible with regular (i.e., no HNe) CC-SNe (e.g., $7\times 10^{-2}\,\mathrm{M}_\odot$ for SN1987A, \citealt{Seitenzahl2014} or between $5.8 - 16 \times 10^{-2}\, \mathrm{M}_\odot$ for Cas A, \citealt{Eriksen2009}). However, the $^{56}$Ni yields are already too high to be classified as faint SN (Fig~\ref{fig:hne}). The spatial distribution of $^{56}$Ni is similar to the one of $^{44}$Ti (Fig.~\ref{fig:unstables_spatial}), i.e., it is located within the jet and at the shock front. The amount of $^{56}$Ni that got liberated through the funnel of the jet hereby contributes at maximum to only $\lesssim 15\%$ of all the created Nickel. Consequently, most of $^{56}$Ni is synthesized at the shock front. Therefore, the amount of $^{56}$Ni broadly correlates with the explosion energy also in MR-SNe, extending the relation found for ordinary supernovae \citep[e.g.,][]{Maeda2003,Nomoto2013,Chen2017,Nomoto2017,Suwa2019,Grimmett2021}.

\begin{table*}
 \caption{Yields of radioactive nuclei for different models. The subscript $l$ denotes a lower limit as calculated within our models and the subscript $e$ denotes our estimate of the final yields as outlined in Sect.~\ref{sect:upper_limit}. The yields are taken at one-tenth of the half life of the radioactive nucleus. }
 \label{tab:radioactive}
 \begin{tabular}{lccccccccc}
  \hline
  Model & M($^{26}$Al)$_l$ & M($^{26}$Al)$_e$ & M($^{44}$Ti)$_l$ & M($^{44}$Ti)$_e$ & M($^{56}$Ni)$_l$ & M($^{56}$Ni)$_e$& M($^{60}$Fe)$_l$ & M($^{60}$Fe)$_e$ & M($^{56}$Ni)$_l$/M($^{44}$Ti)$_l$\\
   & [$10^{-7}$ M$_{\sun}$] & [$10^{-7}$ M$_{\sun}$] & [$10^{-5}$ M$_{\sun}$] & [$10^{-5}$ M$_{\sun}$]&[$10^{-2}$ M$_{\sun}$] &[$10^{-2}$ M$_{\sun}$]&[$10^{-3}$ M$_{\sun}$]&[$10^{-3}$ M$_{\sun}$]& $[\times 10^3]$\\
  \hline
  35OC-Rp3              & $21.0$ &$31.4$  & $20.0$& $521.7$ & $11.3$ & $97.3$ & $3.1$ & $208.5$ & $0.57$\\
  35OC-Rs$_\mathrm{N}$  & $8.1$  & - & $4.3$ & - & $7.5$  & - & $1.7$ & - & $1.74$ \\
  P      & $6.2$  & $15.6$ & $4.2$ & $37.4$ & $3.3$  & $20.8$ & $5.5$ & $11.8$ & $0.79$ \\
  O      & $0.4$  & $5.3$ & $0.9$ & $4.9$  & $1.7$  & $10.2$ & $1.2$ & $5.6$ & $1.89$ \\
  W      & $0.1$  & $2.5$ & $1.1$ & $4.6$  & $2.2$  & $14.4$ & $0.1$ & $0.1$ & $2.00$\\
  S      & $16.5$ & $16.5$ & $5.9$ & $17.1$  & $10.5$ & $24.8$ & $2.8$ & $8.4$ & $1.78$ \\
  \hline
 \end{tabular}
\end{table*}
The ratio of $^{44}$Ti to $^{56}$Ni has been proposed as a diagnostic of the entropy in SNe \citep[][]{Nagataki1997,Nagataki1998,Vance2020,Sato2021}. High entropy environments are characterized by a larger fraction of matter undergoing an $\alpha$-rich freezeout and, hence, a larger amount of $^{44}$Ti is indicative of high entropy conditions. Our models span a wide range of entropy, with the higest values for models 35OC-Rp3 and P. This is directly reflected in their low M($^{56}$Ni)/M($^{44}$Ti)$<10^{3}$ ratios (see Tab.~\ref{tab:radioactive}), while all other models have larger ratios ($\gtrsim 1.7 \times 10^3$).

\subsection{The nucleosynthesis of zinc}
\label{ssct:zinc}
\begin{figure*}
 \includegraphics[width=2\columnwidth]{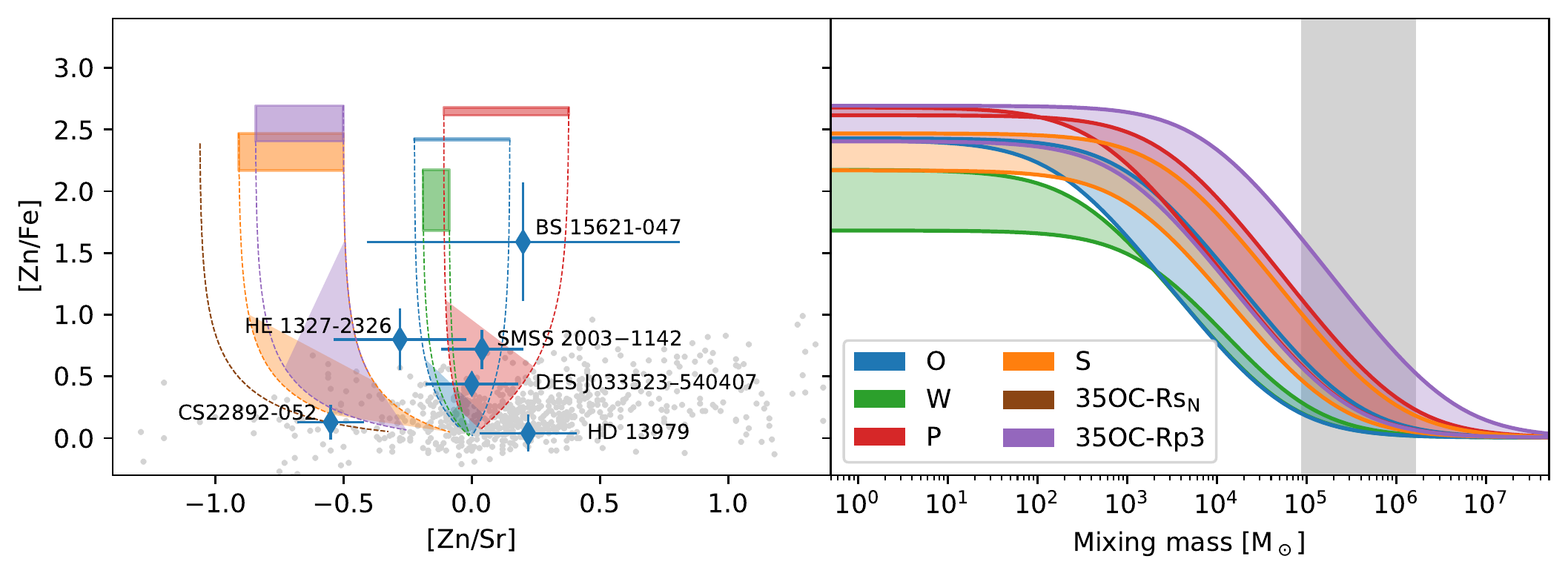}
 \caption{The left-hand panel shows [Zn/Fe] versus [Zn/Sr]. Every grey dot is a star taken from \citet{Westin2000,Hill2002,Aoki2002,Cowan2002,Ivans2003,Christlieb2004,Barbuy2005,Aoki2005b,Barklem2005,Sivarani2006,Cohen2006,Masseron2006,Honda2007,Aoki2007,Aoki2008,Lai2008,Cohen2009,Bonifacio2009,Mashonkina2010,Frebel2010,Roederer2010,Cohen2010,Honda2011,Hollek2011,Allen2012,Kirby2012,Cui2013,Cohen2013,Gilmore2013,Ishigaki2013,Frebel2014,Roederer2014,Placco2015,Hansen2015,Jacobson2015,Li2015,Siqueira-Mello2015,Skuladottir2015,Placco2016,Kirby2017,Mishenina2017,Roriz2017,Hansen2017,Holmbeck2018,Sakari2018,Bandyopadhyay2018,Ji2018,Gull2018,Cain2018,Mardini2019,Purandardas2019}, and \citet[][]{Ji2019}, accessed via the SAGA database \citep[][]{SAGA}. Additionally, we added stars from \citet{Reichert2020}. Blue diamonds are chosen r-process enhanced stars namely He 1327-2326 \citep[][]{Ezzeddine2019}, BS 15621-047 \citep[][]{Allen2012}, SMSS 2003-1142 \citep[][]{Yong2021}, DES J033523-540407 \citep[][]{Ji2018}, HD 13979 \citep[][]{Roederer2014}, and CS22892-052 \citep[][]{Sneden2003}. Coloured boxes show the range of ratios for the individual MR-SNe models using the lower and estimated ejected yields into account (see sect.~\ref{sect:upper_limit}). Dashed lines indicate the ratio when solar scaled material gets mixed into the ejecta. Here, solar scaled material is material of originally solar abundances in which we reduce all abundances but hydrogen by the same factor and add the excluded mass to hydrogen to conserve the total mass until we reach $\mathrm{[Fe/H]=-1}$. The coloured shapes indicate the range of ratios when mixing $\sim 10^5-10^6\,\mathrm{M_\odot}$ of solar scaled gas into the ejecta. The right-hand panel illustrates [Zn/Fe] ratios in dependence of different mixing masses. The grey region shows the expected mixing masses that we obtain for assumed explosion energies between $10^{52}$ and $10^{53}\,\mathrm{erg}$. Our models are able to explain stars with an excess of $\mathrm{[Zn/Fe]}$.}
 \label{fig:zn_fe_sr}
\end{figure*}

In addition to their large explosion energy of $\sim 10^{52}\,\mathrm{erg}$ and their high ejected Ni mass of $\mathrm{M}(^{56}\mathrm{Ni})> 0.1\,\mathrm{M}_\odot$ many studies suggest that HNe may also eject a substantial amount of zinc \citep[e.g.,][]{Umeda2002,Kobayashi2006,Tominaga2007,Barbuy2015,Nishimura2017,daSilveira2018,Hirai2018,Tsujimoto2018,Ezzeddine2019,Grimmett2020,Grimmett2021,Yong2021}. 
Our models have $1/10$th of the solar metallicity and thus differ from the interesting case of very metal-poor environments. Nevertheless, the fact that Zn is predominantly produced in hot environments that reach NSE conditions and thus lose memory of the progenitor composition makes our results applicable for the low metallicity case, too.
As a consequence, we investigate if our models can explain a high $[\mathrm{Zn}/\mathrm{Fe}]$ ratio similar to what can be observed in the atmosphere of old stars. The ejected amount of Zn lies between \mbox{$1\times 10^{-2}\,\mathrm{M}_\odot \lesssim \mathrm{M(Zn}) \lesssim 9 \times 10^{-1}\,\mathrm{M}_\odot$}. It is mainly synthesized in slightly neutron-rich conditions, and the spatial distribution of Zn is similar to the one of $^{60}$Fe (Fig.\ref{fig:unstables_spatial}). The nucleosynthetic pathway to Zn differs among our models. While most of them dominantly synthesize the slightly neutron-rich $^{66}$Zn, model W synthesizes more $^{64}$Zn via the decay of $^{64}$Ge due to the less neutron-rich conditions therein. The amount of Fe lies between $1\times 10^{-1}\,\mathrm{M}_\odot \lesssim \mathrm{M(Fe}) \lesssim 1.1 \,\mathrm{M}_\odot$ (including the extrapolation as outlined in sect.~\ref{sect:upper_limit}). Besides Zn and Fe, we also look at the fraction between Zn and first r-process peak elements such as Sr which synthesized about $2\times 10^{-4}\,\mathrm{M}_\odot \lesssim \mathrm{M(Sr}) \lesssim 7 \times 10^{-2} \,\mathrm{M}_\odot$. This leads to values of $1.5 \lesssim \mathrm{[Zn/Fe]}\lesssim 3 $ and $-1 \lesssim \mathrm{[Zn/Sr]}\lesssim 0.5 $ (upper rectangles in the left-hand panel of Fig.~\ref{fig:zn_fe_sr}).
This exceeds typical values of $\mathrm{[Zn/Fe]} \sim 0.7$ of observed stars that are proposed to carry signatures of HNe. 

However, when comparing the yields of a specific MR-SN to the composition of an individual star, we have to take into account that the ejecta of the explosion did not exclusively form the successor star. Instead they were mixed with the surrounding ISM  (see also, e.g., \citealt[][]{Reichert2021b}), in the process diluting the high yields of elements like Zn or Fe. We therefore estimate the amount of mass the ejecta will mix into. This mixing mass is approximately given by the Sedov-Taylor blast wave solution \citep[e.g.,][]{Ryan1996,Magg2020},
\begin{equation}
\label{eq:dillmass} 
M_\mathrm{gas} = 1.9 \times 10^4 \mathrm{M}_\odot E_{51}^{0.96}n_0^{-0.11}, 
\end{equation}
where $n_0$ is the ambient number density, which we assume to be $1\,\text{cm}^{-3}$. For the explosion energy in units of $10^{51}\,$erg using the diagnostic explosion energy is not adequate, because the latter is still growing by the end of the computed neutrino-MHD evolution. Instead, we assume typical explosion energies in a energy range of observed HNe \citep[$E_{51}=5-100$;][]{Nomoto2006,Nomoto2013}. These explosion energies lead to mixing masses of $10^5\lesssim M_\mathrm{gas} \lesssim 10^6\,\mathrm{M}_\odot$ (grey region in the right-hand panel of Fig.~\ref{fig:zn_fe_sr}). Assuming a metallicity of $\mathrm{[Fe/H]=-1}$ and mixing this amount of gas with the ejecta of our models indeed leads to the necessary amounts of $\mathrm{[Zn/Fe]}$ to explain the high ratios of some stars (lower coloured regions in the left-hand panel of Fig.~\ref{fig:zn_fe_sr}, the dashed lines correspond to ratios for lower mixing masses). We note that a high $[\mathrm{Zn/Fe}]$ is obtained in all of our models and not only in the most energetic model explosions. Even in the least magnetized models that are closer in terms of explosion energy to regular CC-SNe we obtain an excess in Zn \citep[also compare to similar possible high fractions in other regular multidimensional CC-SNe models][]{Eichler2017,Wanajo2018,Sieverding2020,Sandoval2021}. The fact that the $\mathrm{[Zn/Fe]}$ ratio is low for HNe in 1D models may be the result of missing physics such as downflows \citep[c.f., also to][who were unable to reproduce large $\mathrm{[Zn/Fe]}$ within 1D HNe models]{Grimmett2018}. Nevertheless, when including mixing, our models can also explain $\mathrm{[Zn/Fe]}$ ratios up to $\sim 1.5\,\mathrm{dex}$ in agreement with the typical values suggested for HNe. 

\subsection{Candidates for superluminous supernovae?}
\label{ssct:lightcurve}
MR-SNe have also been suggested as explanations for SLSNe \citep[e.g.,][]{Quimby2011,Gal-Yam2012,Inserra2013b,Soker2017,Gal-Yam2019,Nicholl2021,Soker2022}. This is because MR-SNe bring the promise of producing larger amounts of $^{56}$Ni than ordinary CCSNe, and they may also host protomagnetars, which may act as central engines releasing energy in the ejecta that decisively contribute to the overall luminosity of the explosion.

We apply an extended version of the simplified model for supernova light curves of \citet{Dado2015} to the results of our simulations. It is based on the assumption of a spherical cloud of hot gas dominated by photon pressure and expanding with a given velocity into the surrounding medium. The thermal energy of the gas changes due to adiabatic expansion, the emission of photons diffusing out of the cloud, and radioactive heating. For the latter effect, while the original prescription only accounted for the decay chain ${}^{56}\mathrm{Ni} \to  {}^{56}\mathrm{Co} \to {}^{56}\mathrm{Fe}$, we include a larger set of isotopes with abundances obtained by our detailed nucleosynthesis calculations. 

We assume that photons diffuse out of the cloud on a time-scale $t_\mathrm{diff}\approx t_\mathrm{r}^2/t$ with $t_r$ being dominated by Compton scattering
\begin{equation}
    t_{\rm r} \approx \sqrt{ \frac{3 M_{\rm ej} f_{\rm e} \sigma_{\rm T}}{8\pi m_{\rm p} c V_{\rm ej}} }.
\end{equation}
Here, $M_{\rm ej}$ is the total ejected mass, $f_{\rm e}$ the fraction of free electrons ($\sim 0.3$; see \citealt{Dado2015}), $\sigma_{\rm T} \approx 6.5 \times 10^{-25} \, \mathrm{cm}^2$ the Thomson cross section, $c$ the speed of light, and $V_{\rm ej}$ the ejecta velocity. We estimate the velocity of the models by assuming that the total explosion energy and ejecta mass, obtained from the final state of the neutrino-MHD simulations, will be converted into kinetic energy, i.e.,
\begin{equation}
    V_\mathrm{ej} = \sqrt{ 2 \frac{E_\mathrm{ej}}{M_\mathrm{ej}}}, 
\end{equation}
which leads to velocities of the order of a few $10^4\, \mathrm{km\,s^{-1}}$. The bolometric luminosity is given by (Eq.~(4) in \citealt{Dado2015})
\begin{equation}
    L = \frac{e^{-t^2/2t_\mathrm{r}^2}}{t_r^2}\int _0 ^t t\,e^{-t^2/2t_\mathrm{r}^2} \dot{E}\, \mathrm{d}t.
\end{equation}
The heating rate $\dot{E}$ is computed as sum over all nuclear decay channels (here for an individual nucleus $N$) according to
\begin{equation}
    \dot{E}(N) = \left(\sum _i \lambda_i \, Y_{N} \, Q_i \right) \times M
\end{equation}
with the decay constant $\lambda_i$, the corresponding Q-value of the reaction $Q_i$ and the abundance of the parent nucleus $Y_{N}$. The energy released in the decays is split into (i) photons, $\dot{E}_\gamma$, (ii) positrons or electrons, $\dot{E}_{e^\pm}$, (iii) $\alpha$-particles, $\dot{E}_\alpha$, and (iv) neutrinos, $\dot{E}_{\nu_e,\bar{\nu}_e}$. Additionally, we assume that the energy released in electron captures escapes entirely via neutrinos. We have \mbox{$\dot{E}_\mathrm{tot}(N) = \dot{E}_\gamma+ \dot{E}_{e^\pm}$+$\dot{E}_\alpha$+$\dot{E}_{\nu_e,\bar{\nu}_e}$} for each nucleus $N$. We take into account that a part of the photons leaves without depositing energy in the gas and estimate the fraction of photons that thermalize in the ejecta (Eq.~(7) in \citealt{Dado2015}),
\begin{equation}
    A_\gamma \approx 1-e^{-\tau_\gamma}.
\end{equation}
Here, $\tau_\gamma$ is the optical depth given by
\begin{equation}
    \tau_\gamma = \frac{3M_\mathrm{ej}\sigma_\mathrm{t}}{8\pi m_\mathrm{p} V_\mathrm{ej} t^2}.
\end{equation}
The so called Klein-Nishina energy transfer cross-section $\sigma_\mathrm{t}$ is dependent on the average photon energy $\bar{E}_\gamma$ of the decay\footnote{From ENSDF database as of 1/12/21. Version available at http://www.nndc.bnl.gov/ensarchivals/} and calculated via (e.g., \citealt{Attix1987})
\begin{align}
    \sigma_\mathrm{t} (\bar{E}_\gamma) = 2\pi r_\mathrm{e}^2 &\left[ \frac{2\times (1+x)^2}{x^2\times (1+2x)} - \frac{1+3x}{(1+2x)^2} \right. \\ \nonumber
                         &- \frac{(1+x)\times (2x^2-2x-1)}{x^2\times (1+2x)^2} 
                         - \frac{4x^2}{3\dot (1+2x) ^3} \\ \nonumber
                         &-\left. \left(\frac{1+x}{x^3} - \frac{1}{2x}+\frac{1}{2x^3}\right)\times \ln{\left( 1+2x\right)} \right]
\end{align}
with the electron radius $r_\mathrm{e}$ and the photon energy in units of the electron rest energy $x = \bar{E}_\gamma/m_\mathrm{e} c^2$, where $m_\mathrm{e}$ is the electron mass. Because $\sigma_\mathrm{t}$ and the fractions of energy released as photons ($f_\gamma$), positrons/electrons ($f_{e ^\pm}$) and $\alpha$-particles ($f_\alpha$) differ between decay reactions, we obtain the contributing energy as sum over all decaying nuclei,
\begin{equation}
\label{eq:lightcurve_energy}
    \dot{E} = \sum _N \left( A_\gamma (N) f_\gamma (N) + f_{e ^\pm}(N) + f_{\alpha}(N) \right)\times  \dot{E}_\mathrm{tot}(N).
\end{equation}
We neglect the impact of neutrinos on the supernova light curve and assume that they are radiated away without any further interaction. Trivially, \mbox{$f_\gamma + f_{e^\pm} + f_\alpha +f_{\nu_e,\bar{\nu}_e}=1$} holds. In eq.~\ref{eq:lightcurve_energy} we also neglect energy from pair annihilation.
 
\begin{table}
\caption{$\beta$-decay properties of selected nuclei. The table contains the name of the parent nucleus, half-lives (note that $\lambda=\ln{2}/T_{1/2}$), released energies, fraction of released energy in form of photons, average photon energy per decay, fraction of energy in form of electrons or positrons, and the fraction of energy in form of neutrinos. Half-lives and Q-values are taken from the JINA Reaclib database \citep[][]{Cyburt2010}, while all other properties are taken from the ENSDF database \citep[][]{Brown2018}.}
 \label{tab:decay_props}
 \centering
 \begin{tabular}{lrccccc}
  \hline
  Parent &$T_{1/2}$& Q  & $f_\gamma$ & $\bar{E}_\gamma$  & $f_{e^\pm}$ & $f_{\nu_e,\bar{\nu}_e}$  \\
    & [d]& [MeV] & & [MeV] & &  \\
  \hline
  $^{56}$Co &$77.2$ & $4.57$ & $0.787$& $1.13$ & $0.028$ & $0.185$ \\
  $^{57}$Co &$272.0$& $0.84$ & $0.154$& $0.07$ & $0.017$ & $0.829$ \\
  $^{56}$Ni &$6.1$  & $2.13$ & $0.798$& $0.48$ & $0.004$ & $0.198$ \\
  $^{57}$Ni &$1.5$  & $3.26$ & $0.595$& $0.86$ & $0.049$ & $0.356$ \\
  $^{66}$Ni &$2.3$  & $0.25$ & $0.000$& -      & $0.291$ & $0.709$ \\
  $^{66}$Cu &$0.004$& $2.64$ & $0.037$& $1.03$ & $0.404$ & $0.559$ \\
  $^{72}$Ga &$0.6$  & $4.00$ & $0.692$& $1.21$ & $0.118$ & $0.190$ \\
  \hline
 \end{tabular}
\end{table}

For early times ($\sim 1$\,d), most of the energy released by decays thermalizes, heats the gas, and thus contributes to the light curve.
\begin{figure}
 \includegraphics[width=\columnwidth]{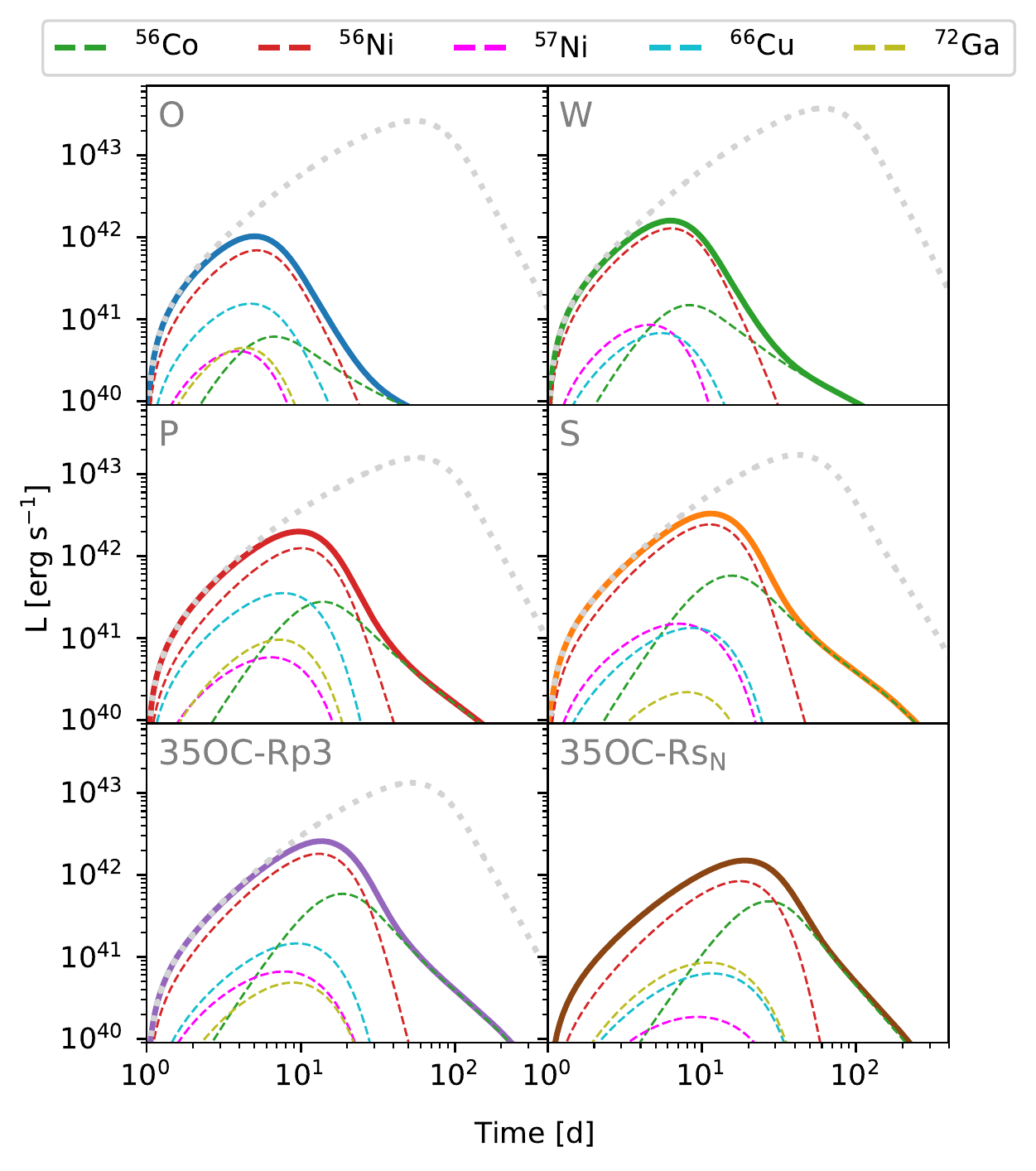}\\
 \caption{Light curve as estimated with ejected masses and yields at the end of the simulation (coloured solid lines). Coloured dashed lines show the individual light curves due to the radioactivity of the most contributing nuclei. The light-grey dotted line indicates the light curve using the extrapolated values of ejected matter described in sect.~\ref{sect:upper_limit}.}
 \label{fig:light_curve}
\end{figure}
The light curve is dominated by the radioactive decay of $^{56}$Ni initially and of $^{56}$Co later on. However, the decay of other nuclei also slightly contributes to the light curve, increasing the peak luminosities by $38\%$, $13\%$, $37\%$, $13\%$, $13\%$, and $24\%$ for models O, W, P, S, 35OC-Rp3, and 35OC-Rs$_\mathrm{N}$, respectively. 

The luminosity of all models is increased by at least $13\%$ in comparison to a model that considers the decay chain of $^{56}$Ni only. An additional boost up to a $38\%$ higher peak luminosity can come from the decay chain $^{66}$Ni $\to ^{66}$Cu $\to ^{66}$Zn. This decay chain contributes predominantly to heating of the gas via the decay of the short-lived $^{66}$Cu (due to the nuclear properties, see Tab.~\ref{tab:decay_props}, c.f. to the nuclear properties given in~\citealt{Nadyozhin1994,Wu2019,Shingles2020}), powered by the longer lived $^{66}$Ni. 
For the synthesis of $^{66}$Ni the neutron-richness plays a dominant role. We find a major contribution to the light curve if more matter drops out of NSE ($T = 7 \, \mathrm{GK}$) with $0.42 < Y_e < 0.47$ than with $Y_e>0.47$, which is the case for models O, P, and 35OC-Rs$_\mathrm{N}$. 

At times $100 \lesssim t\lesssim 1000$\,d the light curve is dominated by the decay of $^{56}$Co emitting positrons that subsequently thermalize \citep[][]{Seitenzahl2009,Dado2015}. For even later times, the assumptions of our simple model may break down as the ejecta become optically thin.

The peak luminosities lie within $10^{42}-4\times10^{42}\,\mathrm{erg\, s^{-1}}$ (Fig.~\ref{fig:light_curve}) in a range which is expected for moderate luminous supernovae \citep[][]{Inserra2013a}. Superluminous supernovae usually exceed $10^{43}\,\mathrm{erg\, s}$ \citep[see e.g.,][for a recent review]{Gal-Yam2019}. Indeed, when using the extrapolated value of the ejected mass and nickel masses outlined in sect.~\ref{sect:upper_limit}, we obtain peak luminosities exceeding $10^{43}\,\mathrm{erg\, s^{-1}}$ (grey dashed lines in Fig.~\ref{fig:light_curve}).

However, so far we only investigated a simplified light curve model that assumes spherical symmetry. The impact of a viewing angle dependence of non-spherical models on the luminosity depends on the shape and $^{56}$Ni distribution. The variation may lie between $\sim 10 - 40 \%$, having a higher luminosity at the equator compared to the poles \citep[][]{Wollaeger2017,Barnes2018}. Furthermore, we observe a certain structure in the distribution of $^{56}$Ni (Fig.~\ref{fig:unstables_spatial}), while our light-curve model assumes that all radioactive sources are located in the centre. In model S, most of the $^{56}$Ni is even behind a fairly opaque lanthanide enriched shell (see the shell of low electron fraction in Fig.~\ref{fig:entropy_3D}). The structure possesses some analogy to the 'lanthanide curtain' found in models of neutron star mergers (NSM) by, e.g., \cite{Perego_2014MNRAS.443.3134}. This in-homogeneity will introduce further variations in the light-curve. A better mixing can shift the peak of the light-curve to an earlier time \citep[][]{Taddia2016,Taddia2019}. 

Additionally, we note that the light-curve model does not include any luminosity enhancing effects that could arise from an interaction with the circumstellar matter and that could be significant. It has been pointed out previously that these interactions could be responsible for the occurrence of SL-SNe \citep[e.g.,][and references therein]{Jerkstrand2020}.

\begin{figure}
 \includegraphics[width=\columnwidth]{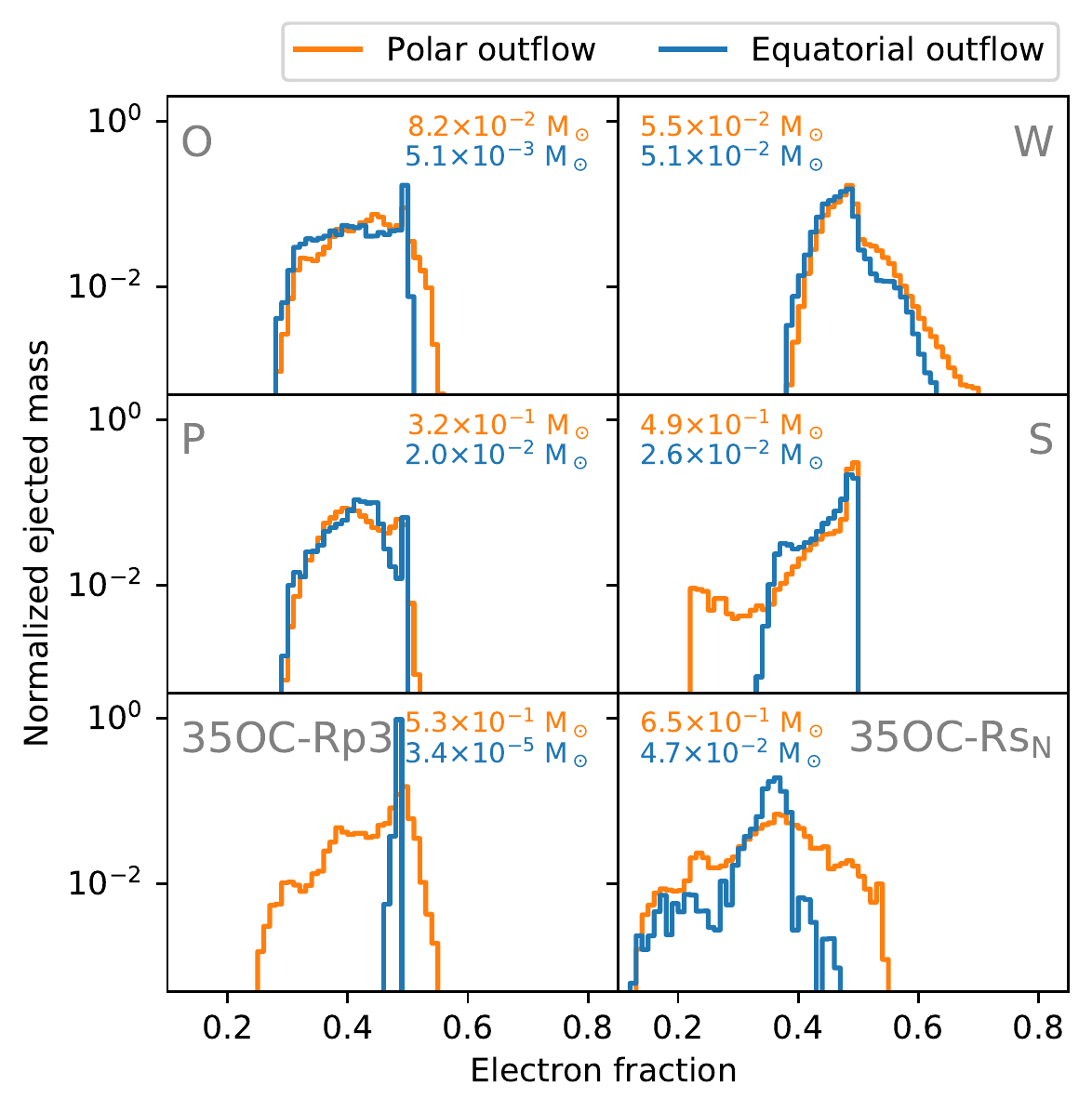}
 \caption{Histograms of the electron fraction at $T=7\,\mathrm{GK}$ normalized to the ejected mass in the equator and polar region, respectively. The outflow is divided into a polar one (orange) defined as $0\le \theta \le 60$ and $120\le \theta \le 180$ as well as an equatorial one (blue) defined as $60< \theta <120$. The histogram is extracted at the end of the simulations (Tab.~\ref{tab:sim_properties}). The coloured text indicates the total ejected mass in each of the components. We note that only ejecta with $T\ge7\,\mathrm{GK}$ is considered here.}.
 \label{fig:ye_mass_hist}
\end{figure}

We note that there could also be an interesting dependence of the colours of the light-curved due to the quite different conditions involved when looking at ejecta at the equator and at the polar region (see Fig.~\ref{fig:ye_mass_hist}). The
distribution of the electron fraction differs in the ejecta along the polar and along the equatorial regions due to the slower expansion of the latter ejecta, which allows for a more efficient neutrino heating in the equatorial plane. Even though the ejected mass in the equatorial region is rather small in all jet-driven models, we observe differences in the different directions. Model W is an exception. This model is close to what one would expect in a regular CC-SNe and explodes rather spherical and the explosion is not driven by jets (see Appendix~\ref{sct:p-rich}). The distribution of the electron fraction is therefore more similar in the equatorial and jet direction. The distinct distributions of the electron fraction may yield the key to distinguish usual CC-SNe and jet-driven MR-SNe in the future.

Summarized, the models here may reach the necessary peak luminosities to be classified as SLSNe even only accounting for the radioactive contribution to the light curve \citepalias[i.e. not even including the potential contribution of the central engine as in][]{Obergaulinger2022}. However, the general shape of the light curve is dominated by the decay chain of $^{56}$Ni. As pointed out by several authors \citep[e.g.,][]{Inserra2013b,Gal-Yam2019} the slope of the decaying light curve of observed SLSNe is often incompatible with a slope that is inferred by the decay of $^{56}$Ni and $^{56}$Co. For these cases, the accretion on a central BH or the spin-down of a magnetar in the centre of the SLSN has been proposed as energy source of the extremely high luminosities \citep[][]{Kasen2010,Woosley2010,Dessart_et_al__2012__mnras__Superluminoussupernovae:$56$Nipowerversusmagnetarradiation,Inserra2013b,Chatzopoulos2013,Mazzali_et_al__2014__mnras__Anupperlimittotheenergyofgamma-rayburstsindicatesthatGRBs/SNearepoweredbymagnetars,Nicholl2014,Nicholl2015,Metzger2015,Obergaulinger2022}. \citet{Nicholl2017} and \citet{De_Cia2018} find typical required magnetic field strengths of $B\sim 10^{14}\,\mathrm{G}$ and spin periods of $P\sim 2\,\mathrm{ms}$ of the magnetar to match light curves of SLSNe. For those of our models that avoid BH formation, i.e., most likely models O, P, S, 35OC-Rp3, and 35OC-Rs$_\mathrm{N}$, the PNS does indeed fulfil these conditions. The investigation of the impact of a central engine is beyond the scope of this work, but we refer the reader to the simplified light curve models obtained in \citetalias{Obergaulinger2022}. In these models, the central engine contribution is dominant, and allows for the possibility of extremely bright events. The obtained radioactive peak luminosities here are higher than the ones for the models presented in \citetalias{Obergaulinger2022}, because of the various additional radioactive energy sources (there only Ni and Co where considered) and, specially, because of the extrapolation of the radioactive yields following the methodology of sect.~\ref{sect:upper_limit}. 

\section{Discussion and conclusions}
\label{sct:conclusion}
We presented the detailed nucleosynthesis of long-time 2D models as well as 3D models of MR-SNe evolved to final times of the order of at least $1$~s for the 3D models and up to $9$~s for one 2D model. For some of the new 3D models, axisymmetric versions were presented in our previous article \mbox{\cite{Reichert2021a}}. The simulations were performed with a sophisticated energy-dependent M1 neutrino transport \citep[][]{Just2015}, which allows for a reliable determination of nucleosynthesis-relevant properties such as the electron fraction of the ejecta. The set of models includes one case of an explosion driven mostly by neutrino heating and several magnetorotational explosions with polar jets.

All the models in this study, regardless of their dimensionality (2D or 3D), are variations of a single magnetized, fast-rotating, massive stellar progenitor. These variations are broadly compatible with the uncertainties carried by 1D stellar evolution models. Here, we have extended our previous results \citep{Reichert2021a} to 3D models and to longer 2D evolutionary times. That has allowed us to show that not only the dynamics (\citealt{ObergaulingerAloy2017}, \citetalias{Obergaulinger2021}), the explosion mechanism \citep{Bugli2020, Obergaulinger2020}, and the compact remnant type \citepalias[][]{Aloy2021} are critically dependent on (order of magnitude) variations of the rotational rate and magnetic field strength, but also the nucleosynthetic yields depend significantly on these initial conditions. 

We computed the detailed nucleosynthesis by applying the large nuclear network \textsc{WinNet} with 6545 isotopes to Lagrangian tracer particles following the dynamics of the ejecta. For this purpose, we sampled the unbound material at the final time of the simulations with tracers, and evolved them backward in time according to the velocity field of the neutrino-MHD models. To accurately sample the ejecta, of the order of a solar mass, the tracers have different masses with an upper limit of $10^{-4}\,\mathrm{M_\odot}$, leading to a total number of tracers of $\mathcal{O}(10^4)-\mathcal{O}(10^6)$. We reduce the computational effort by binning the tracers into groups with similar physical properties.

Despite the long simulation times, the ejection of mass has not been finished when the models were terminated. We developed a method to estimate the final ejected mass fractions. For this, we use an extrapolation procedure to approximately determine the conditions relevant for nucleosynthesis. While the uncertainties of this method are considerable, it allows us to obtain estimates of the yields of several elements that would otherwise be underestimated. 

We compared the nucleosynthetic results of the 3D models with their respective 2D versions. The dynamics \citepalias[see][]{Obergaulinger2021}, but also the calculated yields can differ significantly. Consistently, there is no universal trend in neutron-richness between 2D and 3D. On the one hand, the nucleosynthetic imprints of long time effects that are visible in the long simulated 2D models cannot be investigated in their 3D version due to the shorter simulation times. On the other hand, the different dynamics of the 3D models can leave a fingerprint in the nucleosynthetic yields. Therefore both, 2D and 3D models, will stay important tools for the investigation of MR-SNe in the future. 

We find three major mechanisms to synthesize r-process elements. The first one is connected with an early and fast ejection of neutron-rich matter. This matter will later be located in a cocoon around the jet as reported already in earlier studies \citep[][]{Winteler2012,Nishimura2015,Nishimura2017,Moesta2018,Reichert2021a}. This production mechanism may be less favorable in 3D given the less neutron-rich conditions of model S \citepalias[][]{Obergaulinger2022}. However, model S is nevertheless very neutron-rich with minimum electron fractions of $Y_\mathrm{e,min}\sim 0.23$ (at $T=7\,\mathrm{GK}$) which is only slightly larger than the boundary of $0.20$ that have been found to be sufficient for the production of third r-process peak elements. Given this marginal difference, a slight change in the setup (either numerical or physical) may enable a full r-process by this mechanism also in 3D. The long simulation times enabled us to investigate a second mechanism. If the magnetic field is strong enough, it can have a strong impact on the shape of the PNS. In model 35OC-Rs$_\mathrm{N}$ this even lead to a transition of the PNS to a toroidal configuration in the centre. This process ejects very neutron rich material from the PNS itself on short time-scales \citepalias[see also][]{Aloy2021}, which ultimately lead to an r-process synthesizing heavy elements that include the actinides. This ejection mechanism is not unique among our models as we found a similar behaviour already in model 35OC-Rw presented in \citealt[][]{Reichert2021a}. However, it has to be shown that this effect can also occur in full 3D simulations and over a broader set of initial models (including different masses, and mechanisms for angular momentum transport). Determining the conditions in the massive collapsing star that yield to the actinide production is highly uncertain with only a few models. But the model in which this distinctive feature has happened harbours a strong ($\sim 10^{12}\,$G), large-scale, poloidal magnetic field in equipartition with the toroidal magnetic field. Although this field is not directly obtained from a consistent stellar evolution, the uncertainties still remaining in the secular (1D) modelling of massive stars, would likely allow for the realization of a quantitatively similar case \citepalias[see][for a more detailed discussion, and \citet{Griffiths_2022arXiv220400016} for the potential action of MRI in the topology of the magnetic field]{Aloy2021}. The last mechanism is driven by high entropies rather than very neutron-rich conditions. Two of our models, \mbox{35OC-Rp3} and P, showed high entropies (exceeding $S>200\,\mathrm{k_B}$) in the beam of the magnetically driven jets. Under these conditions the neutron-to-seed ratio can exceed $>70$ and is therefore high enough to perform a full r-process \citep[e.g.,][]{Woosley1994,Wheeler1998,Freiburghaus1999b,Meyer2002,Thielemann2017}. 

Thus, summarizing the r-process results, we find two aspects within our 3D models: (i) The strongest pre-collapse magnetic fields lead to the strongest r-process ejecta in a cocoon around the jet, resulting in an early and fast ejection (consistent with earlier investigations). Whether such strong pre-collapse magnetic fields result from realistic stellar evolution calculations has to be verified in the future.
(ii) While the hope that an r-process can occur for very moderate $Y_e$-values, resulting from earlier neutrino interactions with matter during the explosion phase, has been raised \citep[e.g., by][]{Takahashi1994,Woosley1994}, if high entropies are obtained, later CCSN simulations never supported such conditions. In the present paper we have found such conditions in some models during the later phase of the explosion. Should these conditions be recreated in nature, they would be a fingerprint of the existence of a collimated/jetted ejecta inside the star, since they require magnetized jet beams where the toroidal magnetic field is larger than the poloidal one, thus allowing for a strong beam pinching. However, the resulting nucleosynthetic pattern is probably not one of a full solar-type r-process.

Even though we have shown that MR-SNe are able to host a small portion of matter undergoing an r-process also in 3D, there are some caveats when considering MR-SNe as dominant astrophysical production sides of the r-process. If we consider a typical mixing mass of $10^{5}\,\mathrm{M_\odot}$ and assume that the ejecta is homogeneously mixed into pure hydrogen we also obtain a minimum metallicity of the remnant composition. Taking the considerable amount of iron given for our extrapolated final yields, this would lead to remnant metallicities with $\mathrm{[Fe/H]\gtrsim-2.5}$ for the possible r-process candidate models S and P. However, newly born stars will not form from the remnant composition, but might rather have only a 10 or 1\% admixture from a nearby event. Therefore, extending the mixing mass to quite extreme $10^{6}\,\mathrm{M_\odot}$ would also extend the limit to $\mathrm{[Fe/H]\gtrsim-3.5}$, which becomes consistent with low-metallicity observations. A (very extreme) explosion energy of $10^{53}\,\mathrm{erg}$ would already lead to such a low metallicity in the remnant. 
Similar arguments can be applied for collapsars. For these events, model O and W are possible candidates. If a collapsar (or BH) forms, our iron amount will be closer to our lower limit, i.e., as obtained by our tracer particles. A further iron contribution that we here do not account for will come from ejecta of the later forming accretion disc. For model O and W, a likely maximum mixing mass of $10^5\,\mathrm{M_\odot}$ result in $\mathrm{[Fe/H]}\eqsim-3.5$ as lower metallicity limit, indicating similar conditions for describing very metal-poor r-process enhanced stars with collapsars. Therefore, MR-SNe and collapsars may occur early in galactic history, earlier than merging neutron stars,permitting with their iron ejecta to describe very metal-poor r-process enhanced stars down to $\mathrm{[Fe/H]}\eqsim-3.5$. Additionally, we notice that our extrapolations of iron also limit the ratio $\mathrm{[Eu/Fe]}$. While the strongest magnetized 3D model S reaches values of $\mathrm{[Eu/Fe]}\eqsim 1.5$ that fit with very r-process enhanced stars (r-II stars, $\mathrm{[Eu/Fe]}>1$, \citealt[][]{Beers2005}), model P only reaches $\mathrm{[Eu/Fe]}\eqsim0.7$ and a successor star would therefore be at most categorized as less enriched r-I star ($\mathrm{0.3< [Eu/Fe]}<1$; \citealt[][]{Beers2005}). We stress that these are upper limits that are unlikely obtained as frequently occurring CC-SNe could further contribute to iron, but not to europium thus lowering $\mathrm{[Eu/Fe]}$. Whether these values are sufficient to describe the evolution of europium in the early Universe has to be addressed by more complex galactic chemical evolution models \citep[e.g.][]{Schoenrich2019,Kobayashi2020,VandeVoort2020,Cavallo2021,VandeVoort2022}, combined with progenitors with lower metallicity, and/or longer simulations in the future.

Additionally, it seems to be challenging to synthesize a considerable amount of actinides within our 3D models except via high entropy conditions. Model P could reach $\mathrm{[Th/Eu]}\eqsim 0.37$ (after $1\,\mathrm{Gyr}$); however, this value is based on our extrapolation and involves therefore large uncertainties. On the other hand, model S reaches $\mathrm{[Th/Eu]}\eqsim -3.9$ only. This value might be less uncertain as we expect that the ratio is not modified significantly with ongoing simulation time. This is a difference compared to a confirmed r-process site, NSM, which are expected to robustly eject larger amounts of actinides \citep[see e.g.,][for recent reviews]{Horowitz2019,Cowan2021,Wu2022}. 
Judging from our models, stars with a high amount of actinides \citep[as, e.g., reported in][]{Yong2021} are extremely challenging to describe with a dominant contribution of MR-SNe. On the other hand, a certain variability in the actinides has be observed in form of so called "actinide boost" stars \citep[][]{Roederer2009,Mashonkina2014,Holmbeck2018,Holmbeck2019a,Eichler2019,Farouqi2021}. While the existence of actinide boost stars may be explained with different contributions of the NSM disc and dynamical ejecta only \citep[][]{Eichler2019,Holmbeck2019a} an additional contributing source can not be excluded. This is furthermore underlined by the observation of an "actinide deficient" star \citep[][]{Ji2018} which could possibly be explained by our model S. For this group of stars, there is only one representative observed so far and it is therefore a very rare class of stars. As a consequence, MR-SNe may only give a small contribution to the total r-process content in our Universe. On the other hand, the sparsity of these stars can also be an observational bias as deficiencies are harder to detect than elemental enhancements, especially for the actinides that are challenging to detect. We can, however, not fully neglect a later (after the simulation ended) contribution to the actinides by the outflow of a possibly forming collapsar. Whether these outflows are able to contribute to the synthesis of actinides is still a matter of ongoing investigations \citep[][]{Siegel2019,Miller2020, Just_2022arXiv220514158} and goes beyond the scope of this work.

We studied the viability of MR-SNe as candidates for HNe or SLSNe. Two of our models, model 35OC-Rs$_\mathrm{N}$ and S, reach explosion energies that are compatible with those of observed HNe \citep[][]{Nomoto2013}. We therefore investigated the amount of unstable nuclei such as $^{26}$Al, $^{44}$Ti, $^{56}$Ni, and $^{60}$Fe that are also observed in HNe. The ejected amount of $^{44}$Ti lies with $10^{-5}-10^{-4}\,\mathrm{M_\odot}$ in the range of regular CC-SNe. While only two models reach $^{56}$Ni masses of $>0.1\, \mathrm{M_\odot}$ necessary for HNe \citep[][]{Nomoto2013}, the extrapolated values of all other models indicate that such masses are also possible if the simulations were carried out for a longer time. Within our models there is no visible trend between the neutron-richness of the models and the amount of ejected $^{56}$Ni \citep[as found, e.g., in][]{Nishimura2017}. Rather than the neutron-richness, a more important factor is the explosion energy at the end of the simulations which correlates for our models to the ejected mass of $^{56}$Ni. We obtain the relation $M(^{56}\mathrm{Ni}) \eqsim \left(0.72\times E_{51} + 2.31\right) \times 10^{-2} \mathrm{M_\odot}$, which is well in agreement with the observed correlation found within CC-SNe \citep[see e.g.,][]{Nomoto2006,Nomoto2013}. Except for the weakest magnetized model W, all our models show an exceptional high amount of $^{60}$Fe exceeding $10^{-3}\,\mathrm{M_\odot}$ and possibly even growing to $10^{-1}\,\mathrm{M_\odot}$ originating in the moderately neutron-rich conditions with the highest values found in model 35OC-Rp3. Such large values could even be visible as a point source for future telescopes if the event was galactic \citep[][]{Woosley1997,Diehl2021b}. 

We stress that the possibility that MR-SNe are directly connected to HNe is not in conflict with the coincident observations of lGRBs and HNe \citep[][]{Nomoto2006,Nomoto2013}, since lGRBs may not only be produced by collapsars (i.e., by central engines hosting a BH). An alternative scenario in the case of MR-SNe, is that the spin down of the protomagnetar acts as central engine of lGRBs (\citealt[e.g.,][]{Metzger2018}, \citetalias{Aloy2021}).
A direct connection to HNe would infer an event rate of $\sim 10^{-5}\,\mathrm{yr^{-1}}$ in an average galaxy or one MR-SNe every $\sim 700$ regular CC-SNe \citep[][]{Podsiadlowski2004}. Given this rate and $M_\mathrm{A\geq90} \sim 1\times 10^{-2}-4\times 10^{-2} \,\mathrm{M_\odot}$ (for model S, P, and 35OC-Rp3), this fits with the observational determined properties of the dominant r-process site \citep[][]{Hotokezaka2018}. However, we note that our models synthesize a much larger second to third r-process peak ratio than the solar r-process pattern.

Besides a large amount of $^{56}$Ni and a large explosion energy, a high $\mathrm{[Zn/Fe]}$ ratio has been reported as signature of HNe, which our models reproduce. We applied a simple model for the mixing of the SN ejecta into the ISM to compare these ratios to the observation of metal-poor stars. The obtained ratios are in agreement with observations when assuming a mixing mass around $\sim 10^5-10^6\,\mathrm{M_\odot}$ inferred by typical HNe explosion energies of $10^{52}-10^{53}\,\mathrm{erg}$ \citep[][]{Nomoto2006,Nomoto2013}. On the other hand, a high $\mathrm{[Zn/Fe]}$ ratio is not exclusively obtained by our most magnetized models, but also by the less magnetized ones that are more similar to regular CC-SNe. Indeed also other multidimensional CC-SNe simulations without magnetic fields can obtain high $\mathrm{[Zn/Fe]}$ ratios \citep[e.g.,][]{Eichler2017,Wanajo2018,Sieverding2020,Sandoval2021}. Low $\mathrm{[Zn/Fe]}$ ratios may therefore be an indication of missing physics in 1D CC-SNe models (see also \citealt[][]{Sieverding2020} for a similar conclusion).  

Finally, we applied a simplified light curve model based on the spherical expansion of the ejecta and the energy input by radioactive decays to show whether our MR-SN models can reproduce peak luminosities compatible with SLSNe. The model uses the masses of several radioisotopes produced in the explosion. Thus, the resulting light curves depend on whether we use the yields obtained at the final time of the neutrino-MHD simulations, or the (higher) masses obtained by the extrapolation of our results. In the former case, the peak luminosities are in the range of $10^{42}\,\mathrm{erg \, s^{-1}}$ for all models. In the latter case, the luminosity peaks are much broader and with $\sim 10^{43}\,\mathrm{erg \, s^{-1}}$ also brighter. Hence, our results suggest that peak luminosities of the dimmest SLSNe may be produced by the radioactive decay of a blend of isotopes generated explosively during the SN. Larger peak luminosities ($\sim 10^{44}\,\mathrm{erg \, s^{-1}}$ or higher) may require an extra energy release of the central engine \citepalias[see e.g.][]{Obergaulinger2022}, or the interaction of the SN ejecta with the circumstellar medium.
We note that the presence of radioactive nuclei different from $^{56}$Ni and $^{56}$Co can increase the peak luminosity by $10\%-40\%$. This increase is mostly powered by the synthesis of neutron-rich $^{66}$Cu and $^{66}$Ni. However, all our models are dominantly powered by the nuclear decay chain of $^{56}$Ni and therefore also the tail of the light curve follows the slope of this decay. 
This has been shown to be not the case for all SLSNe. The discrepancy can be explained by a central engine as, e.g., the spin-down of a magnetar or the accretion on a central BH. Estimating the effects of such an engine was, however, beyond the scope of our work and leaves room for future investigations. 

An investigation of the effect of higher numerical resolution of the neutrino-MHD models would be desirable. A low resolution leads to enhanced numerical diffusion and can, in our environment, smear out the neutron-rich features. Furthermore, the dynamics within the simulation may change significantly \citep[e.g.,][]{Nagakura2019}. It is therefore an interesting question to investigate if model S is able to host a more neutron-rich environment when applying a higher resolution. Additionally, longer simulated 3D models would enable the investigation of long time effects and therefore get more realistic models. 

As a final note, we stress that we investigated the nucleosynthesis of some of the most advanced neutrino-MHD models of MR-SNe to date. Our results reinforce the existing possibility stating that MR-SNe are viable candidates for HNe and possibly for SLSNe as well. We furthermore have shown that MR-SNe remain as important candidates for the synthesis of r-process elements in the early Universe.

\section*{Acknowledgements}
We would like to thank Marta Molero and Athanasios Psaltis for fruitful discussions. This work has been supported by the Spanish Ministry of Science, Education and Universities (PGC2018-095984-B-I00, PID2021-127495NB-I00) and the Valencian Community (PROMETEU/2019/071). This article benefited from the 'ChETEC' COST Action (CA16117). MG acknowledges support through the Generalitat Valenciana via the grant CIDEGENT/2019/031. AA acknowledges support from the European Research Council under grant EUROPIUM-667912, and from the Deutsche Forschungsgemeinschaft (DFG, German Research Foundation) - Projektnummer 279384907 - SFB 1245 and the State of Hessen within the Research Cluster ELEMENTS (Project ID 500/10.006). MO acknowledges support from the Spanish Ministry of Science via the Ram\'on y Cajal programme (RYC2018-024938-I).

\section*{Data Availability}
Yield tables are available in the electronic form of the manuscript online. Other data will be made available upon reasonable requests made to the authors.



\bibliographystyle{mnras}
\bibliography{mnras} 



\appendix
\section{Tracer integration}
\label{app:tracer_integration}
The tracer particles are calculated backwards in time using snapshots of the neutrino-MHD simulations. These snapshots are available every millisecond. To integrate the trajectories backwards in time, we employ the so called Runge-Kutta-Fehlberg method (RKF45; \citealt[][]{fehlberg1969}). This method employs a fourth order Runge-Kutta algorithm with a fifth order error estimate, allowing for an automatized step size control. Our implementation includes an individual step size for every tracer and, if necessary, interpolate linearly in time between the snapshots of the simulations. In the following, we briefly discuss the impact of the time resolution of the checkpoint files.

\begin{figure}
 \includegraphics[width=\columnwidth]{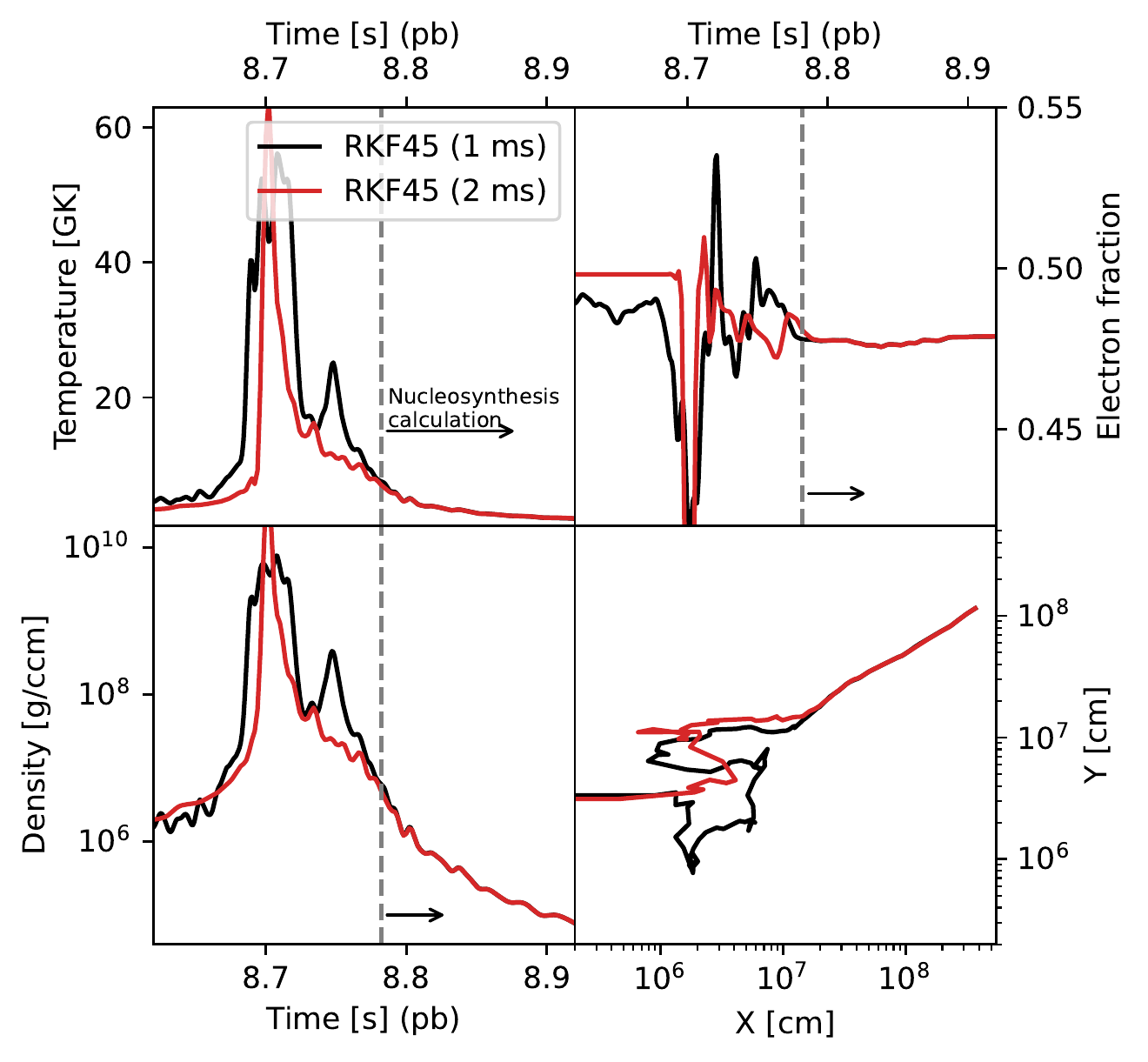}
 \caption{Impact on different properties (see vertical labels) of the time resolution for the checkpoint files, as well as different integration schemes. The dashed line shows the point in time when the tracer falls below $7\,\mathrm{GK}$. The lower right-hand panel shows the projection of the trajectories on the $XY$-plane. The largest differences in the trajectory happen relatively close to the compact remnant. Above a few $10^7\,$cm the trajectories run in parallel.}
 \label{fig:tr_time_res}
\end{figure}

To test if the time resolution of the checkpoint files is sufficiently high, we computed tracers based on the simulation data with a time sampled every millisecond and every two milliseconds. In between two checkpoint files, the velocity field is linearly interpolated. In regions close to the centre of the exploding star or in turbulent flows with rapidly varying velocity fields, this can lead to diverging trajectories. The impact can exemplary be seen in Fig.~\ref{fig:tr_time_res}, where we have picked a representative tracer of model 35OC-Rp3.  

We emphasize that the integration is performed backward in time with the initial conditions set at $t \approx 8.96 \, \mathrm{s}$. The two versions of the tracer (black and red lines) agree very well for $t \ge 8.78 \, \mathrm{s}$, which corresponds to the time after they are ejected from the vicinity of the PNS (distances from the centre of $\ge 200 \, \mathrm{km}$). Before that point, the tracers are in a region where the velocity field varies strongly with time and position causing the positions and physical properties of the two calculations to disagree considerably. In practice, this disagreement has little impact on the nucleosynthesis because during the entire period both versions of the tracer possess sufficiently high temperatures, $T > 7 \, \mathrm{GK}$, for NSE to apply. Both drop out of NSE after this phase and at almost the same positions and with almost identical density, temperature, and $Y_e$. Since the subsequent nucleosynthesis is insensitive to the history of a tracer prior to leaving NSE, the final yields of both tracer calculations are the same. 
To exploit this insensitivity, it is important to integrate the tracers backward in time. The forward integration of a tracer starting in (or passing through) the turbulent velocity field would introduce a sensitivity of the post-NSE conditions to the precise initial positions and prevent the same good agreement.

\begin{figure}
 \includegraphics[width=\columnwidth]{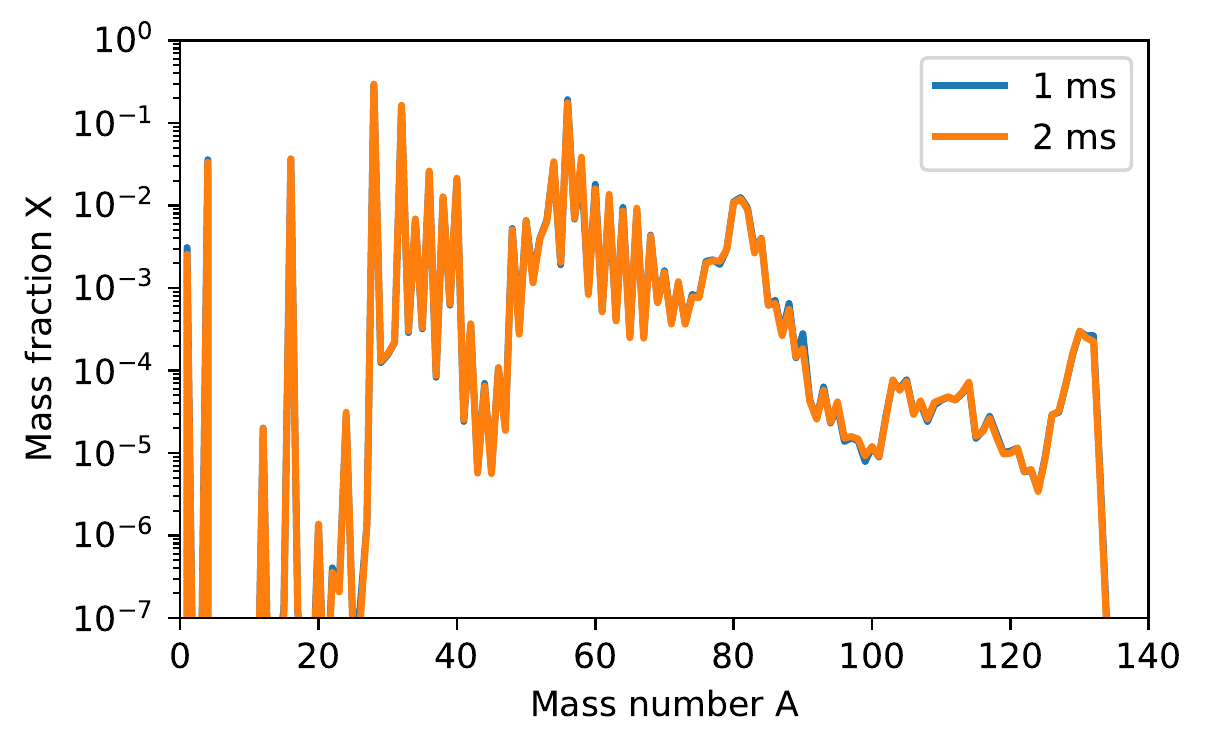}
 \caption{Final nucleosynthetic yields for model O with varied checkpoint resolution of the neutrino-MHD simulation.}
 \label{fig:RO_2ms_1ms}
\end{figure}

To test the impact on the integrated yields, we calculate model O twice, one time with a snapshot time resolution of $1\, \mathrm{ms}$ and once with a resolution of $2\,\mathrm{ms}$ (Fig.~\ref{fig:RO_2ms_1ms}). There are small differences visible, most dominant around $A\sim 90$. However, these differences are negligible and smaller than other typical errors associated to uncertainties of nuclear reaction rates or of astrophysical origin. 

\begin{figure}
 \includegraphics[width=\columnwidth]{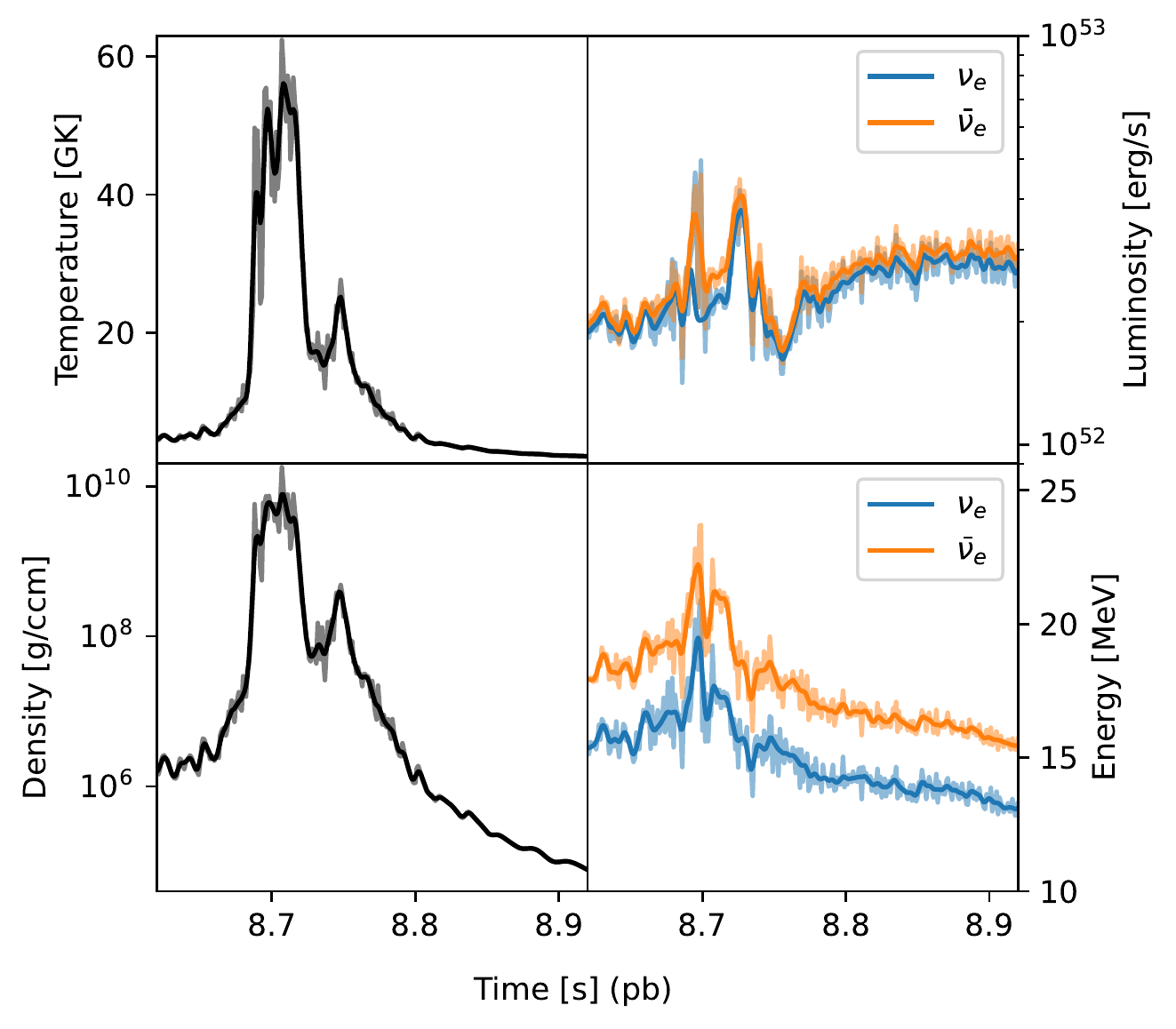}
 \caption{Impact of the applied low-pass filter on the temperature (upper left), density (lower left), neutrino luminosities (upper right), and neutrino energies (lower right) of an example trajectory. The thick solid lines correspond to the filtered data, and the (background) softer and more variable lines to the unfiltered ones.}
 \label{fig:tracer_lowpass}
\end{figure}

In order to speed up the nucleosynthesis calculation, we reduce the noise of the temperature, density, and neutrino quantities. This noise stems from interpolation artifacts and small uncertainties of the advection scheme, especially in the central region close to the PNS. We apply a low-pass filter on all tracked quantities. We choose the threshold frequency of the low-pass filter fairly high and, as a consequence, we mainly smooth the neutrino luminosities and energies (see Fig.~\ref{fig:tracer_lowpass}).


\section{Tracer selection}
\label{app:tracer_selection}
\begin{figure}
 \includegraphics[width=\columnwidth]{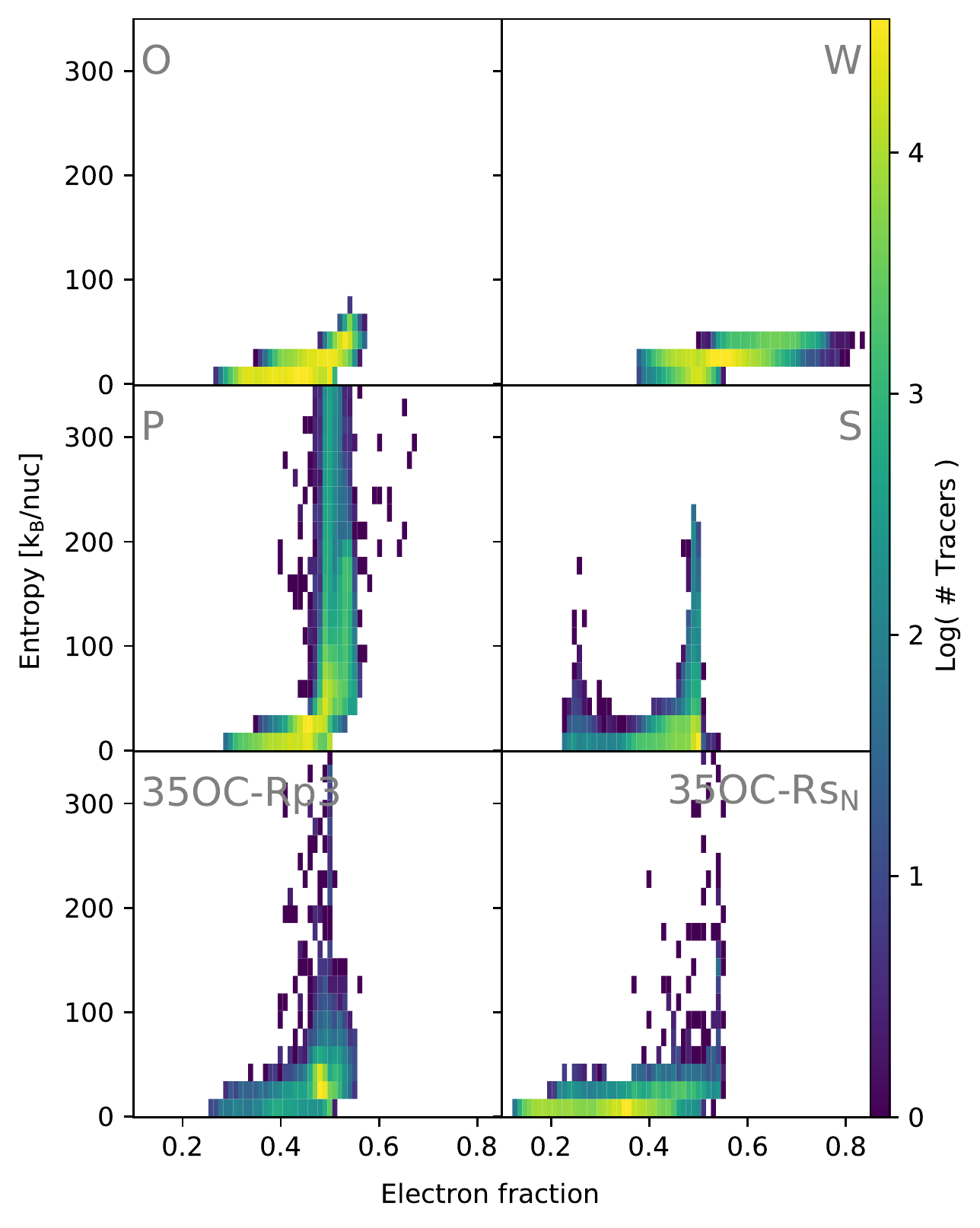}
 \caption{Entropy versus electron fraction at $7\, \mathrm{GK}$. Each panel corresponds to a different model (see legends). The colour indicates the number of tracer particles in the individual bin. Notice that the mass distribution will differ because the tracers have different masses.}
 \label{fig:ye_entr_all_3D}
\end{figure}

Our neutrino-MHD models are the first ones that combine accurate neutrino transport, full 3D, and long evolution of the order of seconds. As a consequence of the long evolution times, also the amount of ejected mass is fairly large ($\sim 10^{-1}-1 \, \mathrm{M_\odot}$; Tab.~\ref{tab:sim_properties}). This is one to two orders of magnitude larger than previous 3D simulations (c.f., $\sim 7\times 10^{-3}\, \mathrm{M_\odot}$ in \citealt{Winteler2012} or $\sim 3\times 10^{-2}\, \mathrm{M_\odot}$ in \citealt{Moesta2018}) and poses a computational challenge, since for an accurate estimate of the nuclear yields around one million tracer particles per model are necessary (Tab.~\ref{tab:sim_properties}). The average computation time of the nucleosynthesis of a single tracer particle lies around $\sim 20\,\mathrm{min}$ on one standard computing core ($22\,\mathrm{min}$ for model 35OC-Rp3). An adequate resolution of tracer particles in our 3D models requires $\sim 10^6$ tracer particles (Tab.~\ref{tab:sim_properties}). The calculation of one model would therefore result in $\sim 3\times 10^5$ core hours. This is less than the neutrino-MHD calculation of the 3D models itself ($\sim 10^7$ core hours). Anyway, the reported numbers show that a nucleosynthesis calculation of all tracer particles would be extremely computational demanding, and we therefore select representative conditions and only calculate a subset of tracer particles. 

Tracer particles that reach maximum temperatures of at least $7\, \mathrm{GK}$ attain nuclear statistical equilibrium (NSE).
In a first step, we therefore divide all tracer particles in hot and cold ones, depending respectively on whether their  maximum temperatures reach $T_\mathrm{max} \ge 7\, \mathrm{GK}$ or not. 
Since the entropy and electron fraction dominantly influence the composition in the equilibrium for Lagrangian tracers with $T_\mathrm{max} \ge 7\, \mathrm{GK}$, it is useful to group them in bins of $S$ and $Y_e$ for a fixed reference temperature ($7$\,GK in our case; see e.g., \citealt{Freiburghaus1999b,Thielemann2017,Moesta2018,Reichert2021a}). The bin size and the number of calculated tracer particles per bin was chosen from the experience gained with model 35OC-Rp3 ($0.01$ in electron fraction and $15\,k_\mathrm{B}/\mathrm{nuc}$ in specific entropy). In a Cartesian tessellation of the $(S,Y_e)$ phase space, most of the elements of the partition do not contain any tracer (white areas in Fig.~\ref{fig:ye_entr_all_3D}). For the chosen partition size, only $N_{\rm b}=198$ bins contain hot tracers in the case of model 35OC-Rp3, as can be seen in the coloured rectangles of Fig.~\ref{fig:ye_entr_all_3D}. Other models display different thermodynamic conditions, noticeable in the diversity of morphologies in the corresponding entropy-electron fraction plane.

\begin{figure*}
 \includegraphics[width=0.9\linewidth]{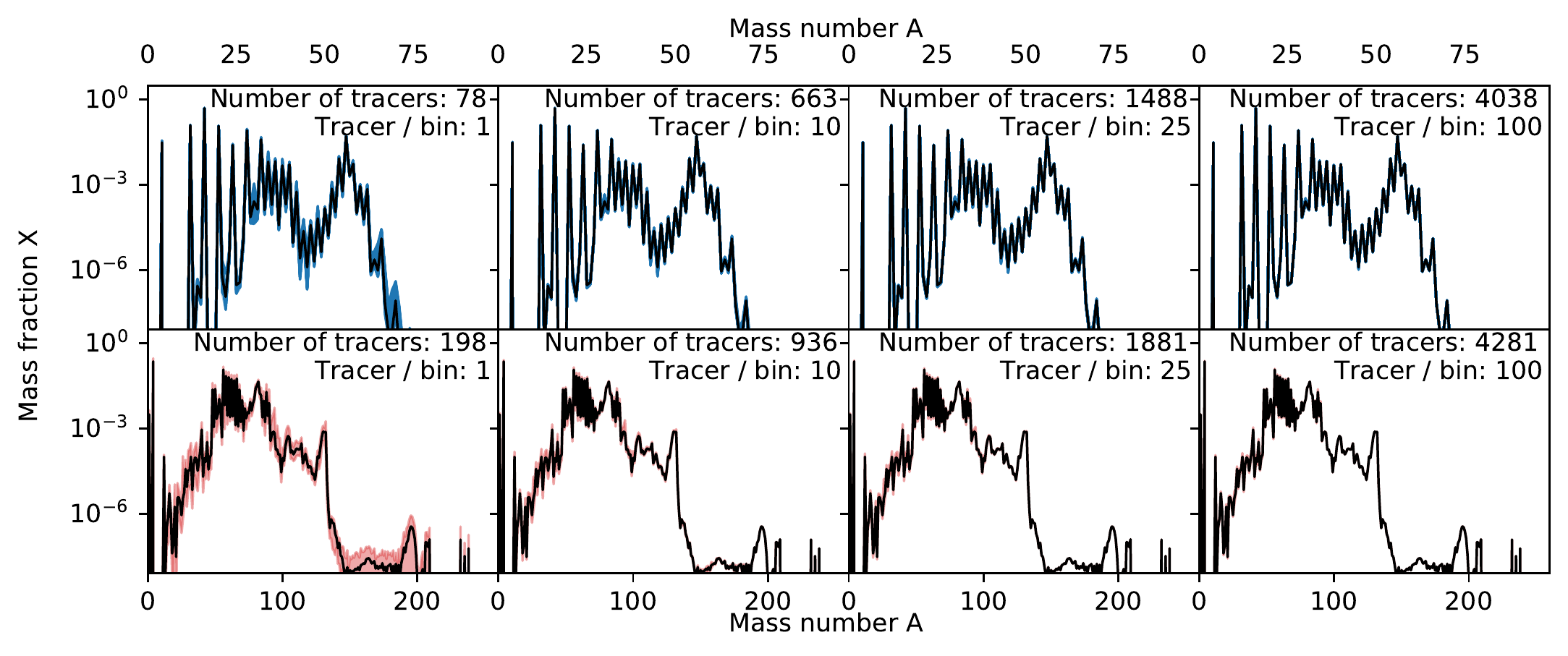}
 \caption{Obtained nucleosynthetic yield for different number of tracers per bin in model 35OC-Rp3. In the upper row, we consider the cold subset of tracers (with a maximum temperature below $7\,$GK) and, in the lower row, the hot subset (with a maximum temperature above $7\,$GK). The total number of bins containing tracers in the $(S,Y_e)$ tessellation of the phase space are \mbox{$N_{\rm b}=78$} and $N_{\rm b}=198$ for the cold and hot subsets, respectively. From the left to the right column we randomly pick an increasing number of \emph{representative} tracers in each bin, $N_{\rm rep}$. We recall that some bins may contain a number of Lagrangian tracers smaller than $N_{\rm rep}$. To quantify the variability in the random choice of tracers inside a bin, we repeat the random choice of tracers within each bin 100 times. For each of the repetitions, we compute the mass fraction of the nucleosynthethic yields as a function of the mass number. The minimum and maximum mass fractions obtained by the former procedure at each mass number are registered, and shown by bands for cold tracers in blue (upper panel) and hot tracers in red (lower panel). The mass-fraction including all available tracer particles of the model is displayed with a solid black line. A good agreement can be obtained when selecting $25$ tracer particles per bin.}
 \label{fig:yields_rp3_2d}
\end{figure*}

In order to reduce the total number of tracers to be processed, we randomly select a number $N_{\rm rep}$ of \emph{representative} Lagrangian markers per $(S,Y_e)$ bin at $T=7\,$GK. In Fig.~\ref{fig:yields_rp3_2d} we show the obtained nucleosynthetic yields for various choices of $N_{\rm rep}$ for model 35OC-Rp3. For $N_{\rm rep}=25$ we find a good trade-off between employing the minimum possible total number of tracers and reducing the scattering of the nucleosynthetic yields prediction induced by the random selection of Lagrangian particles inside each bin. In most $(S,Y_e)$ bins, the number of tracers is significantly larger than $N_{\rm rep}$. However, if the bin contains less than $N_{\rm rep}$ tracers, we calculate the nucleosynthetic yields for all tracers in the bin. This leads to a sufficient agreement between the overall yields and our representatives in model 35OC-Rp3 (lower panel of Fig.~\ref{fig:yields_rp3_2d}). Employing this procedure, there is a reduction by a factor $\gtrsim 100$ of the processing time and memory/storage requests of hot tracers.

\begin{figure}
 \includegraphics[width=\columnwidth]{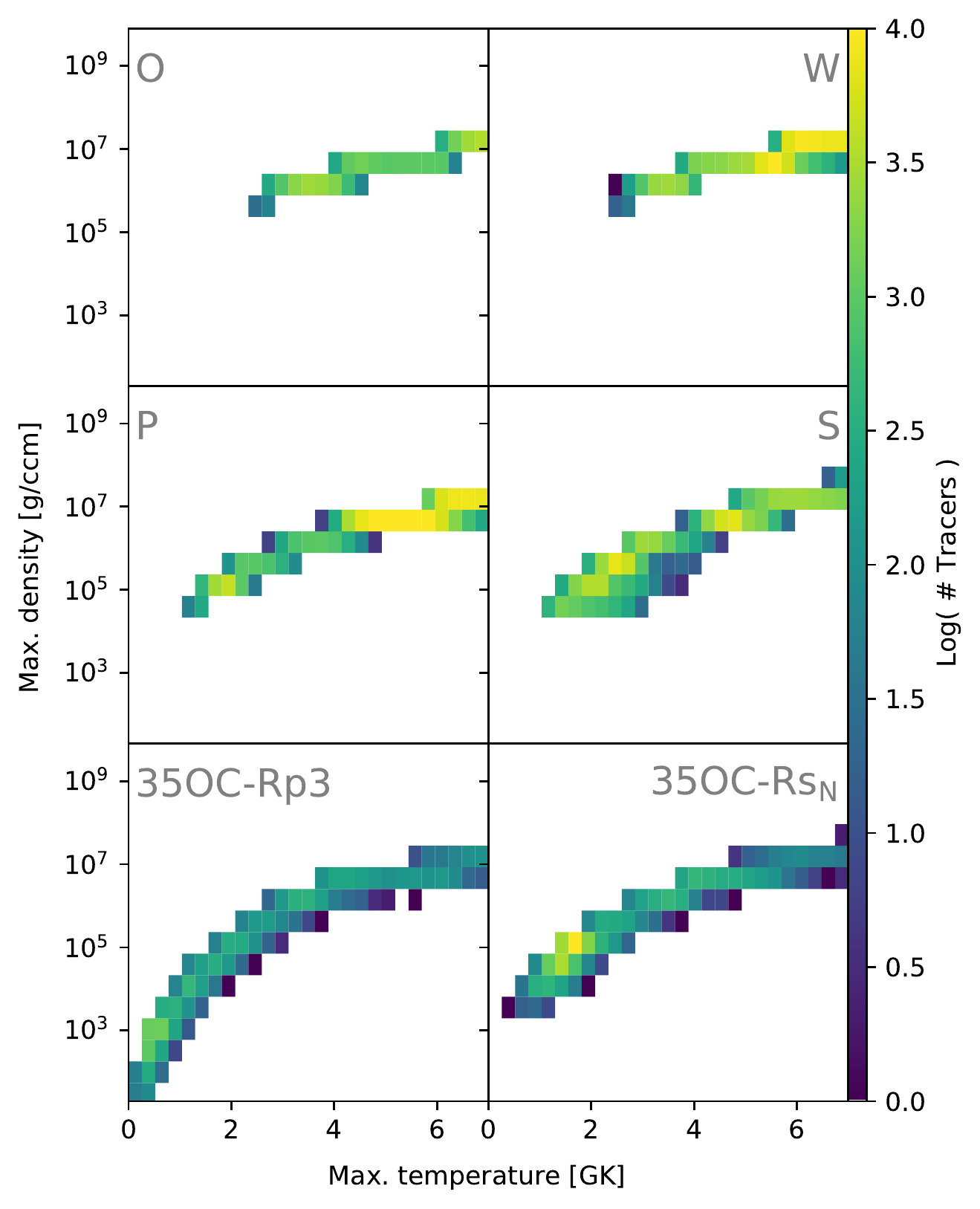}
 \caption{Maximum density versus maximum temperature distribution of tracer particles colder than $7\,$GK. Notice that the mass distribution will differ because the tracers have different masses.}
 \label{fig:cold_tracers_bin}
\end{figure}

For tracers with $T_{\rm max} <7$\,GK the assumption of NSE does not necessarily hold. Instead, the maximum temperature and density is a good indicator of the location in the progenitor and the final nucleosynthetic pattern (see e.g., \citealt{Vance2020} or \citealt{Reichert2021a}). Similarly to the case of the hot tracers, we tessellate the $(T_{\rm max},\rho_{\rm max})$ phase space into bins. Out of this tessellation, only $N_{\rm b}=78$ bins contain tracers (coloured rectangles in Fig.~\ref{fig:cold_tracers_bin}). Analogously to hot tracers, for cold ones, we have experienced picking only $N_{\rm rep}$ per bin in model 35OC-Rp3, again finding that $N_{\rm rep}=25$ suffices for a convergent estimate of the nucleosynthetic yields (Fig.~\ref{fig:yields_rp3_2d}; upper panels).  

Also for the cold tracers, the selection of only a few representatives for individual conditions decreases the necessary amount of calculated tracer particles and, therefore, the computational cost by a factor of around $500$ (Tab.~\ref{tab:sim_properties}) without loosing accuracy (upper panel of Fig.~\ref{fig:yields_rp3_2d}).

\begin{figure}
 \includegraphics[width=\columnwidth]{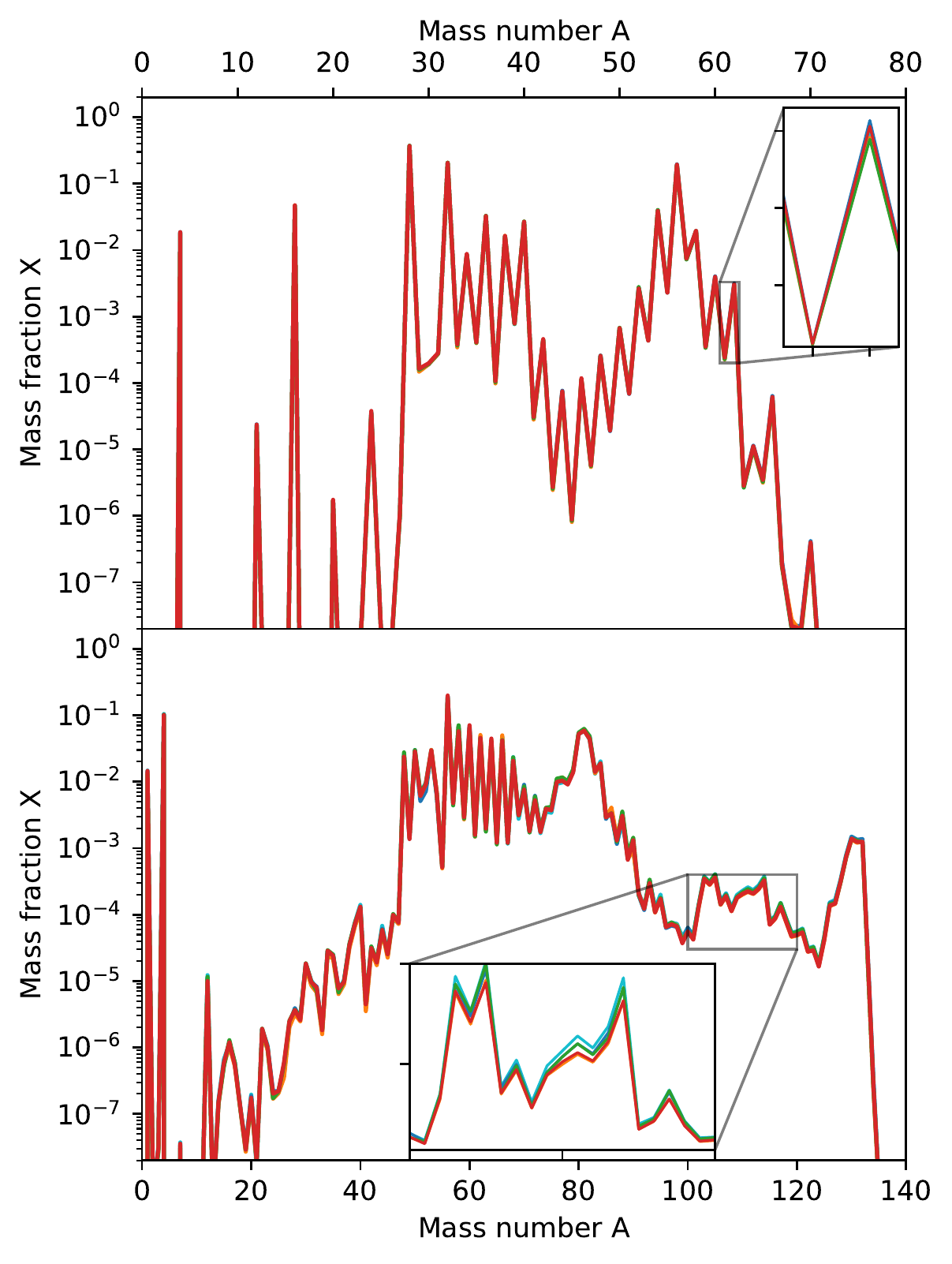}
 \caption{Different nucleosynthetic result for five times randomly chosen trajectories of the individual bins. The upper panel shows the result for cold trajectories, while the lower one shows the result for hot trajectories. Insets show magnifications of some regions, displaying the small scattering of the nucleosynthetic predictions among different random realizations of the choice of representative tracers in model O.}
 \label{fig:select_test}
\end{figure}

We tested the selection criterion explained above also in the 3D model O. We have calculated it five times, always choosing randomly different representative tracers. The result agrees very well with negligible deviations (Fig.~\ref{fig:select_test}).

\section{Proton-rich outflow}
\label{sct:p-rich}
While studies of MR-SNe usually focus on the outflow of neutron-rich material and the synthesized heavy elements, we take the opportunity to shortly discuss also the proton-rich (i.e., $Y_e>0.5$) outflow of our models. This proton-rich outflow is a novelty of our studies (see also \citealt[][]{Reichert2021a}) and has not yet been observed in other nucleosynthesis studies \citep[c.f.,][]{Nishimura2006,Winteler2012,Nishimura2015,Nishimura2017,Moesta2018}. One of the main differences between the underlying hydrodynamic simulations here and in other studies is the more reliable M1 neutrino transport scheme and the longer simulation times \citep[][]{ObergaulingerAloy2017,Obergaulinger2020,Obergaulinger2021,Aloy2021}. With the exception of model W, the proton-rich ejecta is located in the centre of the jet (Figs.~\ref{fig:entropy_3D} and \ref{fig:2d_3d_hydro_ye}). In total, the ejected mass of proton-rich material is smaller than $<8\times 10^{-2}\,\mathrm{M_\odot}$. However, this can make up around to $7$, $11$, $0.8$, $5$, $0.03$, and $0.8\,\%$ of the total ejected mass of models O, W, P, 35OC-Rp3, S, and 35OC-Rs$_\mathrm{N}$, respectively. The stronger the magnetic field of the model, the lower is the fraction of proton-rich matter in the ejecta. Furthermore, we note that a more oblate PNS leads to more extreme values of the electron fractions in the jet. This phenomena is more common within the 2D axisymmetric models.
The synthesized elements in proton-rich ejecta are distributed around a peak of $^{56}$Ni, but can also contribute to the synthesis of so called p-nuclei such as $^{74}$Se, $^{78}$Kr, $^{84}$Sr, or $^{92}$Mo. However, p-nuclei are only synthesized in a non-negligible amount in model W, the model that is most close to a regular CC-SNe. Additionally, given the expected rareness of MR-SNe in contrast to regular SNe, the galactic contribution to p-nuclei in the lower mass range ($A<100$) from the proton-rich outflow of MR-SNe may therefore be negligible.
Therefore, we focus only shortly on model W in the following. For this model, we find proton-rich nuclei up to $A\sim100$ \citep[c.f.,][]{Bliss2018,Eichler2017}, heavier ones may get formed by the p/$\gamma$-process \citep[see e.g.,][and references therein]{Rayet1995,Arnould2003,Pignatari2016,Travaglio2018,Choplin2022}. However, the composition within our progenitor is not detailed enough to investigate this process.

\begin{figure}
 \includegraphics[width=\columnwidth]{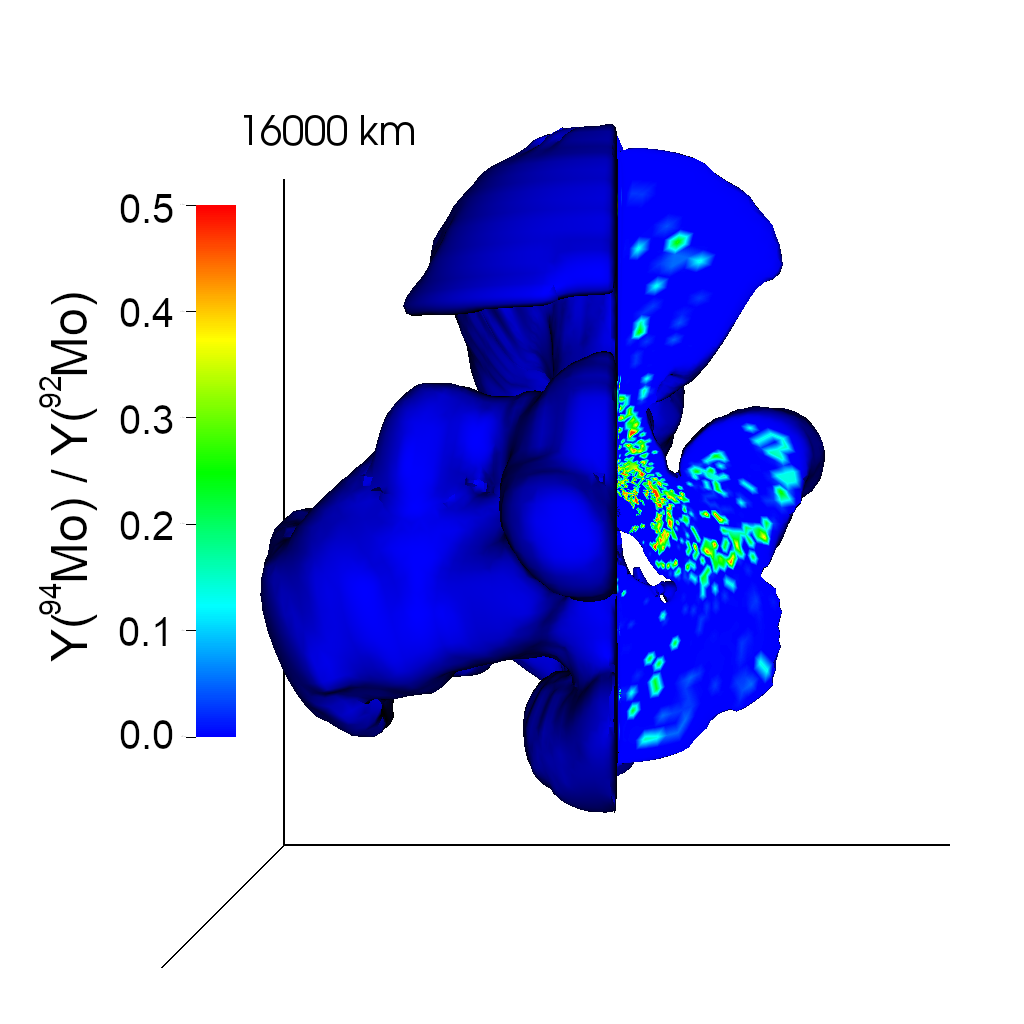}
 \caption{Spatial distribution of the Y($^{94}$Mo)/Y($^{92}$Mo) ratio at the end of the simulation in model W. Regions with $Y(^{92}\mathrm{Mo})=0$ or $Y(^{94}\mathrm{Mo})=0$ are shown as blue colours.}
 \label{fig:mo94mo92}
\end{figure}

Despite the total ejected mass of p-nuclei, isotopic ratios can give interesting clues about the formation of elements. An interesting isotopic ratio of a proton-rich isotope is given by Y($^{94}$Mo)/Y($^{92}$Mo). This ratio can be determined within meteorites, so called SiC grains of type X which are thought to be formed from the ejecta of CC-SNe. The measured ratios of the grains and also of the sun lie within $\sim 0.46-0.74$ \citep[][]{Pellin2006,Eichler2017,Bliss2018}. Considering all ejecta from model W, the isotopic ratio is $\sim 0.014$ if only tracers ejected at the end of the simulation are used or $\sim 0.176$ if also the extrapolated contribution of the outer layers is taken into account. This range of values falls far below that of the grains but is in agreement with \citet[][]{Eichler2017} who found an integrated value of $Y(^{94}\mathrm{Mo})/Y(^{92}\mathrm{Mo})\approx 0.06$ for both of their modelled CC-SNe models. However, small clumps can locally reach Y($^{94}$Mo)/Y($^{92}$Mo)$>0.5$ (Fig.~\ref{fig:mo94mo92}) and could thus be the origin of the SiC X grains with the aforementioned ratios. Whether or not a contribution of the p/$\gamma$-process could possibly fill the gap between observations and theoretical modelling is beyond the scope of our work. 

\section{Yield tables}
\label{app:yields}
The yield tables are given for model O, W, S, P, 35OC-Rp3, and 35OC-Rs$_\mathrm{N}$ in Tab.~\ref{tab:ro_yields}, \ref{tab:rw_yields}, \ref{tab:rs_yields}, \ref{tab:rp3_yields}, \ref{tab:rp3_2D_yields}, and \ref{tab:rs_2D_yields}, respectively. They are separated into different contributions as outlined in Sect.~\ref{sect:upper_limit}. The contributions are given by the tracer particles, the significantly heated matter, the slightly shocked progenitor material, and the stellar wind. Furthermore, we tabulate the total extrapolated mass which is the sum of all contributions.
\clearpage
\onecolumn
\begin{longtable}{lrrrrrrrr}
\caption{Yield table for model O after $1\,\mathrm{Gyr}$. The first four columns give information of the nucleus as the name, atomic number, neutron number, and mass number. The column M$_\mathrm{l}$ indicates the yield in M$_\odot$ obtained from the tracer particles, M$_\mathrm{h}$ the yield from significantly heated matter, M$_\mathrm{s}$ the yield of slightly shocked progenitor material after the simulation has ended, M$_\mathrm{w}$ the approximated yield of the stellar wind with solar abundances according to \citet[][]{Lodders2009}, and M$_\mathrm{e}$ as sum of all previous yields. Nuclei with masses lower than $10^{-10}\mathrm{M}_\odot$ for M$_\mathrm{e}$ are excluded from the table.}
\endfirsthead
\caption* {\textbf{Table~\ref{tab:ro_yields}} Continued}
\endhead
\label{tab:ro_yields}\\
\hline
     name &  Z &   N &   A &  M$_{l}$ [M$_\odot$] &  M$_{h}$ [M$_\odot$] &  M$_{s}$ [M$_\odot$] &  M$_{w}$ [M$_\odot$] &  M$_{e}$ [M$_\odot$] \\
\hline
   $^{1}$H &  1 &   0 &   1 &             3.05e-04 &             2.95e-03 &             0.00e+00 &             4.99e+00 &             5.00e+00 \\
   $^{2}$H &  1 &   1 &   2 &             1.29e-10 &             2.10e-09 &             0.00e+00 &             1.94e-04 &             1.94e-04 \\
  $^{3}$He &  2 &   1 &   3 &             5.72e-10 &             6.76e-09 &             0.00e+00 &             2.40e-04 &             2.40e-04 \\
  $^{4}$He &  2 &   2 &   4 &             6.47e-03 &             5.84e-02 &             3.50e-01 &             1.93e+00 &             2.34e+00 \\
  $^{6}$Li &  3 &   3 &   6 &             0.00e+00 &             0.00e+00 &             0.00e+00 &             5.12e-10 &             5.12e-10 \\
  $^{7}$Li &  3 &   4 &   7 &             1.73e-11 &             2.91e-10 &             0.00e+00 &             7.27e-09 &             7.58e-09 \\
\hline
\end{longtable}

\newpage
\begin{longtable}{lrrrrrrrr}
 \caption{Same as Tab.~\ref{tab:ro_yields}, but for model W.}
\endfirsthead
\caption* {\textbf{Table~\ref{tab:rw_yields}} Continued}
\endhead
 \label{tab:rw_yields}\\
\hline
      Nucleus &  Z &   N &   A &  M$_\mathrm{l}$ [M$_\odot$] &  M$_\mathrm{h}$ [M$_\odot$] &  M$_\mathrm{s}$ [M$_\odot$] &  M$_\mathrm{w}$ [M$_\odot$] &  M$_\mathrm{e}$ [M$_\odot$] \\
\hline
   $^{1}$H &  1 &   0 &   1 &             1.66e-03 &             4.58e-02 &             0.00e+00 &             4.99e+00 &             5.04e+00 \\
   $^{2}$H &  1 &   1 &   2 &             4.12e-09 &             1.50e-07 &             0.00e+00 &             1.94e-04 &             1.94e-04 \\
  $^{3}$He &  2 &   1 &   3 &             1.53e-08 &             5.37e-07 &             0.00e+00 &             2.40e-04 &             2.40e-04 \\
  $^{4}$He &  2 &   2 &   4 &             1.55e-02 &             1.53e-01 &             3.51e-01 &             1.93e+00 &             2.45e+00 \\
  $^{6}$Li &  3 &   3 &   6 &             0.00e+00 &             0.00e+00 &             0.00e+00 &             5.12e-10 &             5.12e-10 \\
  $^{7}$Li &  3 &   4 &   7 &             1.23e-09 &             4.46e-08 &             0.00e+00 &             7.27e-09 &             5.31e-08 \\
\hline
\end{longtable}

\newpage
\begin{longtable}{lrrrrrrrr}
 \caption{Same as Tab.~\ref{tab:ro_yields}, but for model P.}
 \endfirsthead
\caption* {\textbf{Table~\ref{tab:rp3_yields}} Continued}
\endhead
 \label{tab:rp3_yields}\\
\hline
      Nucleus &  Z &   N &   A &  M$_\mathrm{l}$ [M$_\odot$] &  M$_\mathrm{h}$ [M$_\odot$] &  M$_\mathrm{s}$ [M$_\odot$] &  M$_\mathrm{w}$ [M$_\odot$] &  M$_\mathrm{e}$ [M$_\odot$] \\
\hline
   $^{1}$H &  1 &   0 &   1 &             4.10e-05 &             1.81e-03 &             0.00e+00 &             4.99e+00 &             5.00e+00 \\
   $^{2}$H &  1 &   1 &   2 &             9.84e-10 &             2.47e-07 &             0.00e+00 &             1.94e-04 &             1.94e-04 \\
  $^{3}$He &  2 &   1 &   3 &             1.43e-10 &             1.93e-08 &             0.00e+00 &             2.40e-04 &             2.40e-04 \\
  $^{4}$He &  2 &   2 &   4 &             2.40e-02 &             5.45e-01 &             3.53e-01 &             1.93e+00 &             2.85e+00 \\
  $^{6}$Li &  3 &   3 &   6 &             0.00e+00 &             0.00e+00 &             0.00e+00 &             5.12e-10 &             5.12e-10 \\
  $^{7}$Li &  3 &   4 &   7 &             1.19e-11 &             1.17e-09 &             0.00e+00 &             7.27e-09 &             8.45e-09 \\
\hline
\end{longtable}

\newpage
\begin{longtable}{lrrrrrrrr}
 \caption{Same as Tab.~\ref{tab:ro_yields}, but for model S.}
  \endfirsthead
\caption* {\textbf{Table~\ref{tab:rs_yields}} Continued}
\endhead
 \label{tab:rs_yields}\\
\hline
      Nucleus &  Z &   N &   A &  M$_\mathrm{l}$ [M$_\odot$] &  M$_\mathrm{h}$ [M$_\odot$] &  M$_\mathrm{s}$ [M$_\odot$] &  M$_\mathrm{w}$ [M$_\odot$] &  M$_\mathrm{e}$ [M$_\odot$] \\
\hline
   $^{1}$H &  1 &   0 &   1 &             1.32e-07 &             1.50e-07 &             0.00e+00 &             4.99e+00 &             4.99e+00 \\
   $^{2}$H &  1 &   1 &   2 &             0.00e+00 &             0.00e+00 &             0.00e+00 &             1.94e-04 &             1.94e-04 \\
  $^{3}$He &  2 &   1 &   3 &             0.00e+00 &             0.00e+00 &             0.00e+00 &             2.40e-04 &             2.40e-04 \\
  $^{4}$He &  2 &   2 &   4 &             7.29e-02 &             1.93e-01 &             3.54e-01 &             1.93e+00 &             2.55e+00 \\
  $^{6}$Li &  3 &   3 &   6 &             0.00e+00 &             0.00e+00 &             0.00e+00 &             5.12e-10 &             5.12e-10 \\
  $^{7}$Li &  3 &   4 &   7 &             0.00e+00 &             0.00e+00 &             0.00e+00 &             7.27e-09 &             7.27e-09 \\
\hline
\end{longtable}

\newpage

\begin{longtable}{lrrrrrrrr}
 \caption{Same as Tab.~\ref{tab:ro_yields}, but for model 35OC-Rs$_\mathrm{N}$. Since we did not attempt to extrapolate this model, the table contains missing values marked with a minus (see sect.~\ref{sect:upper_limit} for details). }
  \endfirsthead
\caption* {\textbf{Table~\ref{tab:rs_2D_yields}} Continued}
\endhead
 \label{tab:rs_2D_yields}\\
\hline
      Nucleus &  Z &   N &   A &  M$_\mathrm{l}$ [M$_\odot$] &  M$_\mathrm{h}$ [M$_\odot$] &  M$_\mathrm{s}$ [M$_\odot$] &  M$_\mathrm{w}$ [M$_\odot$] &  M$_\mathrm{e}$ [M$_\odot$] \\
\hline
   $^{1}$H &  1 &   0 &   1 &             1.12e-03 &                    - &             0.00e+00 &             4.99e+00 &                    - \\
   $^{2}$H &  1 &   1 &   2 &             3.37e-09 &                    - &             0.00e+00 &             1.94e-04 &                    - \\
  $^{3}$He &  2 &   1 &   3 &             6.91e-09 &                    - &             0.00e+00 &             2.40e-04 &                    - \\
  $^{4}$He &  2 &   2 &   4 &             3.78e-02 &                    - &             3.66e-01 &             1.93e+00 &                    - \\
  $^{6}$Li &  3 &   3 &   6 &             0.00e+00 &                    - &             0.00e+00 &             5.12e-10 &                    - \\
  $^{7}$Li &  3 &   4 &   7 &             5.43e-10 &                    - &             0.00e+00 &             7.27e-09 &                    - \\
\hline
\end{longtable}
\newpage
\begin{longtable}{lrrrrrrrr}
 \caption{Same as Tab.~\ref{tab:ro_yields}, but for model 35OC-Rp3.}
  \endfirsthead
\caption* {\textbf{Table~\ref{tab:rp3_2D_yields}} Continued}
\endhead
 \label{tab:rp3_2D_yields}\\
\hline
      Nucleus &  Z &   N &   A &  M$_\mathrm{l}$ [M$_\odot$] &  M$_\mathrm{h}$ [M$_\odot$] &  M$_\mathrm{s}$ [M$_\odot$] &  M$_\mathrm{w}$ [M$_\odot$] &  M$_\mathrm{e}$ [M$_\odot$] \\
\hline
   $^{1}$H &  1 &   0 &   1 &             1.62e-03 &             8.16e-04 &             0.00e+00 &             4.99e+00 &             5.00e+00 \\
   $^{2}$H &  1 &   1 &   2 &             3.43e-10 &             8.98e-11 &             0.00e+00 &             1.94e-04 &             1.94e-04 \\
  $^{3}$He &  2 &   1 &   3 &             1.30e-09 &             2.57e-10 &             0.00e+00 &             2.40e-04 &             2.40e-04 \\
  $^{4}$He &  2 &   2 &   4 &             1.52e-01 &             1.80e+00 &             2.99e-01 &             1.93e+00 &             4.17e+00 \\
  $^{6}$Li &  3 &   3 &   6 &             0.00e+00 &             0.00e+00 &             0.00e+00 &             5.12e-10 &             5.12e-10 \\
  $^{7}$Li &  3 &   4 &   7 &             5.53e-11 &             1.16e-11 &             0.00e+00 &             7.27e-09 &             7.34e-09 \\
\hline
\end{longtable}

\clearpage
\twocolumn

\bsp	
\label{lastpage}
\end{document}